\documentclass[a4paper,fleqn]{cas-sc}

\usepackage[numbers,sort&compress]{natbib}
\usepackage{graphicx}
\usepackage{multirow}
\usepackage{amsmath,amssymb,amsfonts}
\usepackage{amsthm}
\usepackage{braket}
\usepackage{mathtools}
\usepackage{subcaption}
\usepackage{mathrsfs}
\usepackage{xcolor}
\usepackage{textcomp}
\usepackage{booktabs}
\usepackage{algorithm}
\usepackage{algorithmicx}
\usepackage{algpseudocode}
\usepackage{listings}

\begin{document}
\let\WriteBookmarks\relax
\def\floatpagepagefraction{1}
\def\textpagefraction{.001}

\shorttitle{Noisy-QSMOTE Robustness Analysis}
\shortauthors{Patel et al.}

\title[mode=title]{Noise-Robust Quantum-Inspired Synthetic Minority Oversampling for Condition Monitoring and Fault Classification in Industrial and Energy Systems}

\author[1]{Amit S. Patel}

\ead{aspatel.mh@ddu.ac.in}

\author[2]{Himanshukumar R. Patel}
%\cormark[2]
\ead{himanshupatel.ic@ddu.ac.in}

\author[3,4]{Bikash K. Behera}
\ead{bikas.riki@gmail.com}

\author[5]{Pranav Mehta}
\ead{pranavmehta.mech@ddu.ac.in}

\author[6,7]{David Bassir}
\cormark[1]
\ead{david.bassir@ens-paris-saclay.fr}

\author[8,9]{Ghanshyam G. Tejani}
\ead{gtejani@saturn.yzu.edu.tw}

\affiliation[1]{organization={Department of Mechanical Engineering, Dharmsinh Desai University},
            addressline={College Road},
            city={Nadiad},
            postcode={387001},
            state={Gujarat},
            country={India}}

\affiliation[2]{organization={Department of Instrumentation \& Control Engineering, Dharmsinh Desai University},
            addressline={College Road},
            city={Nadiad},
            postcode={387001},
            state={Gujarat},
            country={India}}

\affiliation[3]{organization={Bikash's Quantum (OPC) Pvt. Ltd.},
            addressline={Balindi},
            city={Mohanpur},
            postcode={741246},
            state={West Bengal},
            country={India}}

\affiliation[4]{organization={Universit\`a degli Studi di Cagliari},
            addressline={Via Is Mirrions},
            city={Cagliari},
            postcode={09123},
            country={Italy}}

\affiliation[5]{organization={Department of Mechanical Engineering, Dharmsinh Desai University},
            city={Nadiad},
            state={Gujarat},
            postcode={387001},
            country={India}}

\affiliation[6]{organization={Smart Structural Health Monitoring and Control Laboratory, DGUT-CNAM, Dongguan University of Technology},
            city={Dongguan},
            country={China}}

\affiliation[7]{organization={ENS-Paris-Saclay University, Centre Borelli, UMR CNRS 9010},
            city={Gif-sur-Yvette},
            postcode={91190},
            country={France}}

\affiliation[8]{organization={Department of Research Analytics, Saveetha Dental College and Hospitals, Saveetha Institute of Medical and Technical Sciences, Saveetha University},
            city={Chennai},
            postcode={600077},
            country={India}}

\affiliation[9]{organization={Department of Industrial Engineering and Management, Yuan Ze University},
            city={Taoyuan},
            postcode={320315},
            country={Taiwan}}

\cortext[1]{Corresponding author}
%\cortext[2]{Corresponding author}

\begin{abstract}
Class imbalance remains a major challenge in industrial condition monitoring and fault classification, where minority fault patterns can be poorly represented during machine learning. This work develops a noise-aware quantum-inspired oversampling framework and systematically evaluates its effectiveness and robustness for industrial and energy-system fault diagnosis. The principal artificial-intelligence contribution is a noise-aware extension of the Quantum Synthetic Minority Oversampling Technique (QSMOTE), in which quantum-inspired perturbations are introduced directly into the compact-swap-test-based similarity estimation used during synthetic minority-sample generation. Consequently, noise propagates through overlap estimation and angle computation to the generated samples, enabling direct assessment of the robustness of the oversampling mechanism rather than perturbing only the downstream classifier. The engineering application is evaluated on four public datasets covering solar-panel faults, bearing condition monitoring, engine failures, and industrial fault detection. Five classical classifiers are used to compare performance on the original imbalanced data, after QSMOTE-based balancing, and under noisy-QSMOTE. The results show substantial classifier-dependent benefits from QSMOTE, including improvements of up to 170\% on the engine-failure dataset and accuracy exceeding 0.99 on the industrial fault-detection dataset. Robustness experiments across six quantum-inspired noise channels further demonstrate that perturbations in similarity estimation can produce markedly different downstream effects depending on the noise model and classifier. The proposed framework therefore provides a reproducible approach for jointly assessing class-imbalance mitigation and noise robustness of quantum-inspired oversampling in practical industrial artificial-intelligence applications.
\end{abstract}


\maketitle

\section{Introduction}\label{Sec1}

\subsection{Context and Motivation}

Modern industrial, manufacturing, and energy infrastructures are increasingly equipped with large-scale sensing, monitoring, and control systems that continuously generate multimodal data for predictive maintenance and fault diagnosis. As these systems evolve into highly automated and interconnected Industry 4.0 environments, the need for reliable, noise-resilient, and imbalance-aware learning frameworks has become increasingly important \cite{lee2015industrial,li2017intelligent}. Real-world condition-monitoring datasets, including the Solar Panel Image Dataset (SPID), CWRU Bearing Dataset (CWRUBD), Engine Failure Detection Dataset (EFDD), and Industrial Fault Detection Dataset (IFDD), often exhibit severe class imbalance, with fault conditions significantly underrepresented relative to normal operating states. Such an imbalance can lead to biased decision boundaries, poor minority-class representation, and reduced classification performance across both classical and advanced machine-learning (ML) models \cite{chawla2002smote,he2009learning}.

Industrial systems such as rotating machinery, induction motors, wind turbines, and industrial pumps frequently operate in noisy environments affected by sensor degradation, electromagnetic interference, environmental disturbances, and hardware faults \cite{qin2019review,zhang2020deep}. These conditions motivate the development of learning frameworks that remain effective under structured perturbations. Furthermore, the emergence of quantum sensing, quantum communication, and hybrid quantum-classical technologies has increased interest in quantum-inspired noise models that emulate realistic perturbation mechanisms beyond conventional stochastic corruption \cite{preskill2018quantum}. Understanding how such perturbations influence data processing and imbalance-mitigation mechanisms is therefore essential for the development of robust next-generation fault-diagnosis systems.

The Synthetic Minority Oversampling Technique (SMOTE) is one of the most widely adopted approaches for addressing class imbalance by generating synthetic minority samples through feature-space interpolation \cite{chawla2002smote}. Although SMOTE has demonstrated effectiveness across domains such as cybersecurity, healthcare, and industrial fault diagnosis, it does not explicitly exploit the geometric structure of quantum-state representations or account for robustness during synthetic sample generation. These limitations motivate the exploration of quantum-inspired oversampling methods that leverage quantum-state geometry for more expressive minority-sample synthesis \cite{Mohanty2025QuantumSMOTE,schuld2015introduction,biamonte2017quantum}. Recently, Mohanty \textit{et al.} proposed Quantum SMOTE (QSMOTE), a quantum-inspired oversampling framework that leverages compact-swap-test-based similarity estimation to generate synthetic minority samples while better preserving the geometric structure of the feature space \cite{Mohanty2025QuantumSMOTE}. The reported results demonstrated the effectiveness of QSMOTE for addressing class imbalance across multiple benchmark datasets. However, the robustness of QSMOTE under quantum-inspired perturbations and the impact of noise-corrupted similarity estimation on synthetic sample generation remain unexplored.
At the same time, advances in quantum machine learning (QML), including Quantum Boltzmann Machines (QBMs), quantum kernels, and quantum generative models, have demonstrated strong potential for learning complex data distributions and enhancing sample generation capabilities \cite{gao2017quantum,kieferova2017tomography}.

Motivated by these developments, this work investigates the effectiveness of Quantum SMOTE (QSMOTE) for imbalance mitigation and, more importantly, examines how quantum-inspired perturbations introduced into compact-swap-test-based similarity estimation affect synthetic sample generation and subsequently influence downstream fault-classification performance. By jointly analyzing baseline classification, QSMOTE-based balancing, and noisy-QSMOTE robustness across multiple industrial datasets, this study provides a systematic framework for evaluating quantum-inspired oversampling under realistic perturbation scenarios and contributes toward the development of robust fault-diagnosis systems for industrial and energy applications.

\subsection{Gap Analysis}

Despite substantial progress in machine-learning-based fault diagnosis and
class-imbalance mitigation, important research gaps remain in the development
of robust learning pipelines for industrial and energy-system condition
monitoring. Existing fault-diagnosis studies commonly evaluate learning
performance using datasets obtained under nominal or controlled acquisition
conditions \cite{qin2019review,liu2018fault}. In practical industrial
environments, however, data acquisition and processing may be affected by
disturbances associated with mechanical vibration, temperature variation,
electromagnetic interference, sensor degradation, and other sources of
uncertainty. Although the influence of data quality on fault diagnosis is
well recognized, considerably less attention has been given to perturbations
that affect intermediate data-processing and imbalance-mitigation mechanisms.
In particular, the effect of structured quantum-inspired perturbations, such
as bit-flip (BF), depolarizing (DP), and amplitude damping (AD), on
similarity-driven synthetic sample generation remains insufficiently
understood. Consequently, it is important to determine how errors introduced
during similarity estimation propagate through an oversampling procedure and
ultimately influence downstream fault-classification performance.

A second gap concerns the treatment of class imbalance itself. Classical
oversampling approaches, particularly SMOTE and its variants, remain widely
used to increase the representation of minority classes
\cite{chawla2002smote,blagus2013smote}. However, interpolation-based
oversampling can generate synthetic samples in ambiguous or overlapping
regions of the feature space, potentially altering class boundaries and
producing over-generalization \cite{he2009learning,fernandez2018learning}.
Most investigations therefore concentrate on the classification improvement
obtained after balancing the training data, whereas comparatively limited
attention is given to the robustness of the mechanism responsible for
determining relationships between minority samples. This becomes particularly
relevant for similarity-driven oversampling, where errors in the estimated
relationship between samples can directly modify the location and
distribution of the generated synthetic observations.

A third gap arises from the growing use of quantum and quantum-inspired
principles in machine learning. Quantum machine learning has introduced
alternative approaches to data representation, distribution learning, and
generative modelling \cite{schuld2015introduction,biamonte2017quantum}.
Quantum generative approaches, including quantum Boltzmann machines and
quantum generative adversarial networks, have further demonstrated the
potential of quantum models for learning and generating complex data
distributions \cite{gao2017quantum,kieferova2017tomography}. Nevertheless,
the robustness of quantum-inspired imbalance-mitigation mechanisms has
received considerably less attention, particularly when perturbations occur
within the similarity-estimation process used to construct synthetic minority
samples. For industrial fault diagnosis, it therefore remains necessary to
investigate not only whether quantum-inspired oversampling improves
classification performance, but also how reliably the oversampling mechanism
behaves when its similarity estimates are perturbed.

Finally, noise has been extensively considered in quantum circuits,
parameterized quantum models, and related quantum-learning settings
\cite{mcclean2016theory,benedetti2019parameterized}. However, these studies
do not directly resolve the robustness question associated with
compact-swap-test-based QSMOTE, where the estimated similarity between
minority samples participates directly in synthetic sample generation. In
this setting, perturbations can propagate sequentially from similarity
estimation to overlap-derived quantities, synthetic sample construction, and
eventually classifier training. A systematic assessment is therefore needed
to distinguish the benefit obtained from QSMOTE-based class balancing from
the performance changes caused by perturbations within its
quantum-inspired similarity-estimation mechanism.

Accordingly, the literature considered in this study leaves three closely
related questions insufficiently addressed:

\begin{itemize}

\item[1)] How do perturbations in compact-swap-test-based similarity
estimation affect the generation of synthetic minority samples in QSMOTE?

\item[2)] How do perturbations introduced during QSMOTE-based synthetic
sample generation propagate to downstream fault-classification performance
across different classifier families?

\item[3)] How does sensitivity to noisy-QSMOTE-generated samples vary across
industrial datasets, classifier families, and structured quantum-inspired
noise models?

\end{itemize}

The present work addresses these questions through a unified empirical
framework comprising four industrial and energy-system fault-diagnosis
datasets, five classical machine-learning classifiers, six quantum-inspired
noise channels, and QSMOTE-based imbalance mitigation. Rather than injecting
perturbations directly into the downstream classifiers, noise is introduced
into the compact-swap-test-based similarity-estimation process used during
QSMOTE synthetic sample generation. This experimental design enables the
effects of perturbations to be traced from similarity and overlap estimation,
through synthetic sample construction, to downstream classification
performance, while separately evaluating the benefit of class-imbalance
mitigation and the robustness of the quantum-inspired oversampling mechanism.

\subsection{Novelty and Contributions}\label{Sec1.4}

The present work extends the QSMOTE framework introduced in
\cite{Mohanty2025QuantumSMOTE}. While the original framework demonstrated
the effectiveness of compact-swap-test-based quantum-inspired oversampling
for class-imbalance mitigation, the robustness of its similarity-estimation
mechanism under structured quantum-inspired perturbations has not been
systematically investigated. This issue is particularly relevant because
similarity estimation directly determines the interpolation used to generate
synthetic minority samples; perturbations at this stage can therefore
propagate through the oversampling process and alter the data subsequently
used for classifier training.

To address this gap, we develop a noise-aware extension of QSMOTE in which
quantum-inspired noise is introduced directly into the
compact-swap-test-based similarity-estimation process during synthetic
minority-sample generation. The principal artificial-intelligence
contribution of this work is therefore not merely an evaluation of
classifiers under noisy conditions, but a systematic framework for
investigating the robustness of the quantum-inspired oversampling mechanism
itself. Specifically, perturbations introduced during similarity estimation
affect the estimated overlap, the corresponding angular representation, and
the resulting synthetic samples, allowing their downstream effect on
classification performance to be quantified.

From an engineering perspective, the proposed framework is investigated for
condition monitoring and fault classification across four industrial and
energy-system datasets. The study follows three complementary experimental
scenarios: (i) classification using the original imbalanced data to establish
baseline performance, (ii) classification after QSMOTE-based minority-class
oversampling to quantify the effect of imbalance mitigation, and
(iii) classification after synthetic samples are generated using noisy
compact-swap-test-based similarity estimates. This design separates the
benefits of oversampling from the effects introduced by perturbations in the
quantum-inspired component of the learning pipeline.

To the best of our knowledge, no previous study has provided a unified
investigation combining QSMOTE-based class-imbalance mitigation with
systematic noise analysis at the level of compact-swap-test-based similarity
estimation across multiple industrial fault-diagnosis applications. The
primary contributions of this work are summarized as follows:

\begin{itemize}

\item[1)] We develop a noise-aware QSMOTE framework that introduces
quantum-inspired perturbations directly into compact-swap-test-based
similarity estimation during synthetic minority-sample generation, enabling
the robustness of the oversampling mechanism itself to be systematically
investigated.

\item[2)] We establish a three-stage evaluation methodology comprising
baseline learning on the original imbalanced data, QSMOTE-based imbalance
mitigation, and noisy-QSMOTE robustness analysis, thereby separating the
effects of oversampling from those of perturbations in the quantum-inspired
similarity-estimation process.

\item[3)] We evaluate the framework across four multi-class industrial and
energy-system fault-diagnosis datasets (SPID, CWRUBD, EFDD, and IFDD),
covering distinct condition-monitoring and fault-classification scenarios.

\item[4)] We systematically investigate six quantum-inspired noise channels,
namely bit-flip (BF), phase-flip (PF), bit-phase-flip (BPF), depolarizing
(DP), amplitude damping (AD), and phase damping (PD), and analyze how these
perturbations propagate through overlap estimation, angle computation, and
synthetic minority-sample generation.

\item[5)] We quantify the downstream consequences of noise-aware synthetic
sample generation using multiple classifier families and determine how the
benefits of QSMOTE and sensitivity to noisy similarity estimation vary across
classifiers, datasets, and noise models.

\item[6)] We identify dataset-, classifier-, and noise-dependent robustness
patterns that provide practical insight into when quantum-inspired
oversampling remains effective under perturbed similarity estimation and
where performance degradation becomes significant.

\item[7)] The resulting analysis provides practical guidance for the design
and evaluation of noise-aware quantum-inspired imbalance-learning methods
for condition monitoring and fault classification in industrial and energy
systems.

\end{itemize}

\begin{figure}
    \centering
    \includegraphics[width=\linewidth]{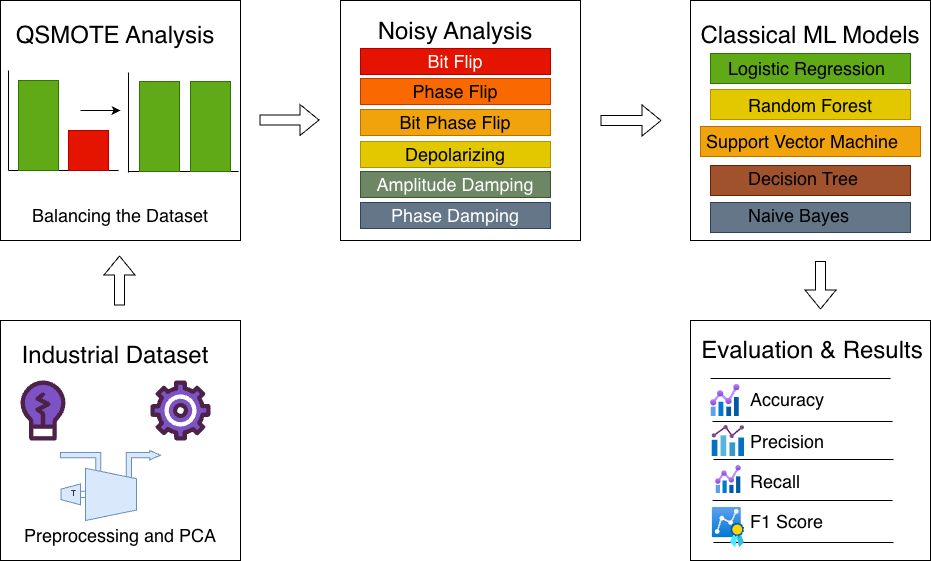}
    \caption{Overall experimental workflow of the proposed framework. Industrial datasets are first preprocessed and balanced using QSMOTE. In the noisy-QSMOTE setting, quantum-inspired noise channels (BF, PF, BPF, DP, AD, and PD) are injected directly into the compact-swap-test-based similarity estimation process used for synthetic sample generation. The resulting datasets are subsequently evaluated using classical machine-learning models (LR, RF, SVM, DT, and NB). Performance is quantified using accuracy, precision, recall, and F1-score to analyse baseline classification behaviour, the impact of QSMOTE-based imbalance mitigation, and the robustness of noisy-QSMOTE-generated samples.}
    \label{fig:schematic}
\end{figure}

Fig. \ref{fig:schematic} illustrates the overall experimental framework used in this study. Industrial and energy-system datasets are first preprocessed through normalization, encoding, and dimensionality reduction where applicable. QSMOTE is then employed to address class imbalance using compact-swap-test-based quantum-inspired similarity estimation. The study consists of three stages: (i) baseline evaluation on the original imbalanced datasets, (ii) evaluation after QSMOTE-based balancing, and (iii) noisy-QSMOTE evaluation. In the third stage, six quantum-inspired noise channels, namely bit-flip (BF), phase-flip (PF), bit-phase-flip (BPF), depolarizing (DP), amplitude damping (AD), and phase damping (PD), are injected directly into the compact swap test used for similarity estimation and synthetic sample generation. The resulting datasets are evaluated using LR, RF, SVM, DT, and NB classifiers. Performance is assessed using accuracy, precision, recall, and F1-score to analyze the impact of QSMOTE and the robustness of noisy-QSMOTE-generated samples.

\subsection{Organization}
The rest of the paper is organized as follows. Section \ref{Sec2} discusses the existing works related to class imbalance and ML models for fault diagnosis. Then, in Section \ref{Sec3}, a detailed explanation of the proposed methodology is provided. Next, Section \ref{Sec4} provides the details on the dataset, preprocessing steps, hyperparameters, evaluation metrics and experimental results with a discussion. Finally, in Section \ref{Sec5}, the conclusion is made with the future direction of research.

\section{Related Works}\label{Sec2}

Research on class imbalance handling, industrial fault diagnosis, and noise-robust ML has evolved significantly over the past decade. This section reviews the most relevant literature across three major themes: (i) imbalance mitigation and resampling techniques, (ii) classical ML models for condition monitoring, and (iii) robustness under noise and quantum-inspired corruption models.

\subsection{Imbalance Mitigation and SMOTE-Based Resampling}

Class imbalance is a well-documented challenge in industrial condition-monitoring datasets, where failure samples are rare compared to normal operation. Traditional oversampling and undersampling methods, such as random oversampling, Tomek links, or cluster-based undersampling, often struggle to preserve minority-class structure \cite{chawla2002smote, he2009learning}. The introduction of the SMOTE marked a pivotal advancement by generating synthetic minority samples through linear interpolation between feature-space neighbors \cite{chawla2002smote}. Variants such as Borderline-SMOTE \cite{han2005borderline}, Adaptive Synthetic Sampling (ADASYN) \cite{he2008adasyn}, and SMOTE-Tomek Links \cite{batista2004study} further refine interpolation quality or perform noise-aware resampling. In fault diagnosis, SMOTE is frequently used to correct imbalance in vibration, acoustic, and multi-sensor health-monitoring datasets \cite{xu2021novel, zhang2020fault}. Mohanty \textit{et al.} introduced Quantum SMOTE (QSMOTE), a quantum-inspired oversampling technique that employs compact-swap-test-based similarity estimation for synthetic minority sample generation and demonstrated its effectiveness for imbalanced learning problems \cite{Mohanty2025QuantumSMOTE}. 

Recently, Behera \textit{et al.} proposed QSMOTE-PGM/kPGM, which combines multiple QSMOTE variants with Pretty Good Measurement (PGM) and kernelized PGM classifiers for imbalanced dataset classification, demonstrating improved minority-class detection and balanced classification performance through quantum-inspired oversampling and classification mechanisms \cite{behera2025qsmotepgmkpgm}. However, the study focused on classification effectiveness and did not investigate the robustness of QSMOTE under quantum-inspired perturbations or the impact of noise-corrupted similarity estimation on synthetic sample generation. Existing research, however, focuses mostly on performance improvements after balancing, rather than investigating how oversampled datasets behave under noise-corrupted assessment or in multi-class industrial contexts. This study contributes to the field by examining SMOTE's impact not only on classification metrics but also on robustness against structured quantum-inspired noise.

\subsection{Classical Machine Learning Models for Fault Diagnosis}

Numerous classical models have been successfully deployed for industrial fault classification. Linear classifiers such as LR provide interpretable decision boundaries but often struggle with nonlinear relations common in mechanical and electrical systems \cite{hosseini2016review}. Probabilistic models like NB are computationally efficient yet sensitive to feature correlation \cite{rish2001empirical}. SVMs remain widely used in fault diagnosis due to their strong margin maximization properties and robustness to high-dimensional data \cite{vapnik1998statistical}. Ensemble methods, including RF, have proven particularly effective for multi-class industrial datasets due to their ability to model complex nonlinear patterns \cite{breiman2001random}. While these models show strong baseline performance, prior work primarily evaluates them under noise-free conditions or simple stochastic corruptions (e.g., Gaussian noise). This paper extends prior studies by benchmarking models across six quantum-inspired noise channels, enabling a more realistic assessment of robustness for next-generation cyber-physical and quantum-touched diagnostic systems.

\subsection{Noise Robustness and Quantum-Inspired Perturbation Models}

ML robustness under various perturbation models has been explored in adversarial learning \cite{szegedy2014intriguing}, random corruption studies \cite{hendrycks2019benchmark}, and sensor-noise simulations for industrial monitoring \cite{wang2016noise}. However, quantum-inspired noise originating from fundamental quantum channels such as BF, PF, DP, and AD/PD has recently gained interest due to the growing integration of quantum sensing and quantum communication in industrial IoT systems \cite{nielsen2002quantum}. Prior works primarily analyze these channels in QML settings, studying their effect on quantum kernels, variational circuits, or entangled state discrimination \cite{schuld2019quantum, zhang2022noise}. Yet, no existing work systematically evaluates how noise-corrupted quantum-inspired similarity estimation affects synthetic sample generation and subsequently influences downstream classification performance on multi-class industrial datasets. This manuscript provides a cross-dataset and cross-noise analysis of how six quantum-inspired noise channels affect compact-swap-test-based similarity estimation within QSMOTE and how the resulting noisy-QSMOTE-generated datasets influence LR, RF, SVM, NB, and DT performance.

\subsection{Positioning and Novelty of This Study}

To the best of our knowledge, no prior study simultaneously examines multi-class industrial datasets (SPID, CWRUBD, EFDD, IFDD), QSMOTE-based balancing effects, five classical classifier families, and quantum-inspired noise robustness across six channels. Existing literature typically examines these aspects independently, whereas this study investigates them within a common evaluation setting. Such an analysis is becoming increasing relevant as industrial systems begin to adopt quantum communication technologies, quantum sensors, and hybrid classical-quantum architectures. In contrast to studies where perturbations are applied directly to datasets or classifiers, the noise considers here affects the compact-swap-test-based similarity estimation process used during QSMOTE-based synthetic sample generation.

\section{Methodology}\label{Sec3}
\subsection{Problem Formulation and Overall Framework}
\label{sec:problemformulation}
In imbalanced classification problems, the training dataset 
$X = \{x_i \in \mathbb{R}^d\}_{i=1}^{n}$ with labels $y_i \in \mathcal{Y} = \{1,2,\dots,K\}$ 
contains classes with highly unequal sample sizes. Let $N_\ell = |\{i : y_i = \ell\}|$ denote the number of samples in class $\ell$, and define
\begin{equation}
N_{\max} = \max_{\ell \in \mathcal{Y}} N_\ell, 
\qquad 
N_{\min} = \min_{\ell \in \mathcal{Y}} N_\ell,
\end{equation}
where typically $N_{\min} \ll N_{\max}$.  
The objective is to construct an augmented dataset 
$(X_{\text{out}}, y_{\text{out}})$ 
that restores approximate balance across all classes while preserving the intrinsic manifold structure of the feature space.
%\textbf{Overall Framework.}
The proposed \emph{QSMOTE} utilizes a three-stage hybrid pipeline that blends classical clustering and quantum-inspired geometric reasoning:

\begin{enumerate}
\item {Cluster Formation:}  
    The $K$-means algorithm partitions the feature matrix $X$ into $K$ clusters, resulting in centroids $\{\mu_j\}_{j=1}^{K}$.  
    Each sample $x$ is assigned to its nearest centroid $c(x)$, which offers a local geometric framework for creating synthetic samples.

    \item {Quantum similarity estimation:}  
    The normalized feature representations of each minority instance $x$ and centroid $c(x)$ are treated as amplitude-encoded quantum states.  
    The \emph{compact swap test} is used to calculate the quantum overlap or inner product $\braket{M|C}$ of these states.  
    The \emph{quantum-inspired angle} $\alpha(x,c)$ quantifies the alignment between $x$ and $c(x)$ in a normalized feature space.
    
    \item{Synthetic Sample Generation:}  
    The angle $\alpha(x,c)$ determines the step size and direction for creating a new synthetic instance $\widetilde{x}$ along the vector connecting $x$ and $c(x)$.  
    Larger angular deviations result in correspondingly larger synthetic displacements, allowing adaptive interpolation based on local cluster geometry.  
    Repeating this method for all minority points results in an oversampled dataset that is consistent with the majority class distribution while minimizing redundancy.
\end{enumerate}

The QSMOTE framework combines clustering-based information from the data with compact swap test-based similarity estimation. 
This integration introduces an angle-based mechanism for generating synthetic data that is both geometrically meaningful and statistically balanced. Unlike conventional SMOTE variants that rely solely on Euclidean interpolation, the proposed method leverages quantum overlap information to adaptively regulate the synthetic sampling process, ensuring improved diversity, cluster consistency, and minority-class representation in the resulting balanced dataset.

\subsection{Compact Swap Test}

\begin{figure}
    \centering
    \includegraphics[width=\linewidth]{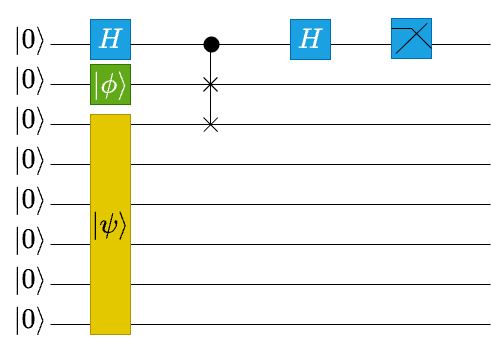}
    \caption{Quantum circuit illustrating the compact swap test.}
    \label{fig:compactswaptest}
\end{figure}

The compact swap test is performed using two data points, centroid ($C$) and minority ($M$). Their quantum states can be written as,
\begin{eqnarray}
    \ket{C}&=&\frac{1}{|C|}\sum_i c_i\ket{i},\nonumber\\
    \ket{M}&=&\frac{1}{|M|}\sum_i m_i\ket{i},
\end{eqnarray}

where, $|C|$ and $|M|$ are the normalization constants, and $c_i$ and $m_i$ are the feature values for data points $C$ and $M$ respectively. The inner product between $\ket{C}$ and $\ket{M}$ is calculated using the compact swap test circuit shown in Fig. \ref{fig:compactswaptest}. This is a resource-efficient method to estimate the overlap between two quantum states \( |C\rangle \) and \( |M\rangle \). First, the quantum states $|\psi\rangle$ and $|\phi\rangle$ are defined as follows:

\begin{eqnarray}
|\psi\rangle&=&\frac{|0\rangle \otimes|C\rangle+|1\rangle \otimes|M\rangle}{\sqrt{2}},  \nonumber \\
|\phi\rangle&=&\frac{|C||0\rangle-|M||1\rangle}{\sqrt{Z}}=\frac{C|0\rangle-M|1\rangle}{\sqrt{Z}}, \nonumber \\
Z&=&|C|^2 + |M|^2.
\end{eqnarray}

The top qubit, initialized in \( |0\rangle \), is put into superposition using a Hadamard gate and acts as a control for the subsequent operations involving the states \( |\phi\rangle \) (green box) and \( |\psi\rangle \) (yellow box). Through controlled operations, the ancilla qubit effectively compares the two states, and a second Hadamard followed by measurement encodes the overlap information into the ancilla's outcome. The initial state of the quantum circuit is given as,

\begin{eqnarray}
&&\ket{+}\ket{\phi}\ket{\psi}%=\frac{1}{\sqrt{2}}(\ket{0}+\ket{1})\ket{\phi}\ket{\psi}
\nonumber\\
&=&\frac{1}{\sqrt{2}}(\ket{0}\ket{\phi}\ket{\psi}+\ket{1}\ket{\phi}\ket{\psi})\nonumber\\
&=&\frac{1}{\sqrt{2}}(\ket{0}\left(\frac{C|0\rangle-M|1\rangle}{\sqrt{Z}}\right)\left(\frac{|0\rangle \otimes|C\rangle+|1\rangle \otimes|M\rangle}{\sqrt{2}}\right)\nonumber\\
&&+\ket{1}\left(\frac{C|0\rangle-M|1\rangle}{\sqrt{Z}}\right)\left(\frac{|0\rangle \otimes|C\rangle+|1\rangle \otimes|M\rangle}{\sqrt{2}}\right))
\end{eqnarray}

After CSWAP operation, the expression becomes,

\begin{eqnarray}
&&\xrightarrow{CSWAP}\frac{1}{2\sqrt{Z}}((C\ket{0}\ket{0}\ket{0}\ket{C}+C\ket{0}\ket{0}\ket{1}\ket{M}\nonumber\\
&&-M\ket{0}\ket{1}\ket{0}\ket{C}-M\ket{0}\ket{1}\ket{1}\ket{M})\nonumber\\
&&+(C\ket{1}\ket{0}\ket{0}\ket{C}+C\ket{1}\ket{1}\ket{0}\ket{M}\nonumber\\
&&-M\ket{1}\ket{0}\ket{1}\ket{C}-M\ket{1}\ket{1}\ket{1}\ket{M}))  
\end{eqnarray}

After Hadamard application, the expression can be written as,

\begin{eqnarray}
    &&\xrightarrow{H_1}\frac{1}{2\sqrt{Z}}((C\ket{+}\ket{0}\ket{0}\ket{C}+C\ket{+}\ket{0}\ket{1}\ket{M}\nonumber\\
&&-M\ket{+}\ket{1}\ket{0}\ket{C}-M\ket{+}\ket{1}\ket{1}\ket{M})\nonumber\\
&&+(C\ket{-}\ket{0}\ket{0}\ket{C}+C\ket{-}\ket{1}\ket{0}\ket{M}\nonumber\\
&&-M\ket{-}\ket{0}\ket{1}\ket{C}-M\ket{-}\ket{1}\ket{1}\ket{M}))
\end{eqnarray}

The probability of measuring the ancilla qubit in the `0' state is calculated as,
\begin{eqnarray}
    P_0&=&\frac{1}{8Z}\Big|C\ket{0}\ket{0}\ket{C}+C\ket{0}\ket{1}\ket{M}\nonumber\\
&&-M\ket{1}\ket{0}\ket{C}-M\ket{1}\ket{1}\ket{M}\nonumber\\
&&+C\ket{0}\ket{0}\ket{C}+C\ket{1}\ket{0}\ket{M}\nonumber\\
&&-M\ket{0}\ket{1}\ket{C}-M\ket{1}\ket{1}\ket{M}\Big|^2
\end{eqnarray}

After simplifying the above expression, finally, the inner product between the centroid and minority data point is found to be,
\begin{eqnarray}
\braket{M|C}&=&\frac{(3-4P_0)Z}{2|C|.|M|}
\end{eqnarray}

Alternatively, the probability of obtaining \( |0\rangle \) on the ancilla is 
\begin{eqnarray}
P(0) = \frac{1}{4}\Big(3-\frac{2|C|.|M|\braket{M|C}}{Z}\Big),
\label{Eq8}
\end{eqnarray} 
from which the overlap \( \langle M | C \rangle \) can be extracted. This compact variant of the SWAP test is particularly useful in NISQ-era devices, as it reduces the number of qubits required while still enabling reliable comparison of quantum states for applications in verification, quantum communication, and ML. The compact swap test method shown in algorithm \ref{Algo-CompactSwapTest} computes a quantum-inspired angle between two vectors $c$ and $m$ by encoding their normalized forms into a quantum circuit. It first prepares an ancilla qubit in a superposition state using a Hadamard gate, then encodes a rotation based on the relative norms of the vectors. The vectors themselves are concatenated, normalized, and initialized on the remaining qubits. A controlled-SWAP operation followed by another Hadamard on the ancilla allows interference that encodes the inner product between $c$ and $m$. Measuring the ancilla and calculating the probability of outcome `0' enables estimation of the inner product, which is clipped to the valid range and converted to an angle using the $\arccos$ function. This angle reflects the similarity or quantum-inspired distance between the two input vectors. In the noisy-QSMOTE experiments, the considered quantum-inspired noise channels (BF, PF, BPF, DP, AD, and PD) are injected directly into the compact swap-test circuit before ancilla measurement. Consequently, the measured probability $p_0$, overlap estimation, similarity score, and the angle used for synthetic sample generation are all affected by the noise process. The resulting synthetic samples are therefore generated from noisy similarity estimates, enabling direct investigation of the robustness of the QSMOTE mechanism under quantum-inspired perturbations.

\begin{algorithm}
\caption{Compact Swap Test}
\label{Algo-CompactSwapTest}
\textbf{Input:} Vectors $c, m$ \\
\textbf{Output:} Quantum-inspired angle between $c$ and $m$
\begin{algorithmic}[1]
\State Convert $c$ and $m$ to numpy arrays of type float64
\State Compute norms: $norm\_c \gets ||c||$, $norm\_m \gets ||m||$
\State Compute normalization factor: $Z \gets norm\_c^2 + norm\_m^2$
\State Set number of qubits $n \gets \log_2(\text{len}(c)) + 3$
\State Initialize quantum circuit $qc$ with $n$ qubits and 1 classical bit

\State Apply Hadamard gate on qubit 0 (ancilla)

\State Compute angle $\theta \gets 2 \cdot \arctan(norm\_m / norm\_c)$
\State Apply single-qubit rotation $u(\theta, \pi, 0)$ on qubit 1

\State Normalize vectors: $c \gets c / norm\_c$, $m \gets m / norm\_m$
\State Concatenate to form $\psi \gets [c, m]$
\State Normalize $\psi \gets \psi / ||\psi||$

\State Initialize $\psi$ on qubits 2 to $n-1$ of $qc$
\State Apply controlled-SWAP gate with control qubit 0 and target qubits 1 and 2
\State Apply Hadamard gate on qubit 0
\State Measure qubit 0

\State Run $qc$ on AerSimulator with 1000 shots
\State Get counts for outcome `0': $c_0$
\State Compute probability $p_0 \gets c_0 / 1000$
\State Compute inner product $inner\_product \gets (3 - 4p_0) \cdot Z / (2 \cdot norm\_c \cdot norm\_m)$
\State Clip value: $final \gets \text{clip}(inner\_product, -1.0, 1.0)$
\State Compute angle $\phi \gets \arccos(final)$

\State \Return $\phi$
\end{algorithmic}
\end{algorithm}

\subsection{Quantum SMOTE}
%\section{Theory of Synthetic Sample Generation in Quantum SMOTE}
\label{sec:qsmote-theory}

%\paragraph{Setting and notation.}
Let $X=\{x_i\in\mathbb{R}^d\}_{i=1}^n$ be the feature set with labels $y_i\in\mathcal{Y}$ and let $\mathcal{L}_{\min}\subset\mathcal{Y}$ denote the set of minority labels.
We assume features are standardized (zero mean, unit variance per coordinate) and write
\[
N_\ell \coloneqq \bigl|\{i:\, y_i=\ell\}\bigr|,\qquad 
M \coloneqq \max_{\ell\in\mathcal{Y}} N_\ell .
\]
A $K$-means model fitted on $X$ yields centroids $\{\mu_j\}_{j=1}^{K}$.
Each point $x$ is associated to its nearest centroid
\begin{equation}
\label{eq:cluster-id}
\mathrm{cid}(x) \;\coloneqq\; \arg\min_{1\le j\le K}\,\|x-\mu_j\|_2,
\qquad 
c(x) \;\coloneqq\; \mu_{\mathrm{cid}(x)}.
\end{equation}
For each $\ell\in\mathcal{L}_{\min}$, the number of synthetic samples to be generated is
\begin{equation}
\label{eq:need}
\mathrm{need}_\ell \;=\; M - N_\ell, \qquad \mathrm{if}\ \mathrm{need}_\ell>0.
\end{equation}

%\paragraph{Amplitude encoding and quantum similarity.}
Given a nonzero vector $v\in\mathbb{R}^d$, define its amplitude-encoded quantum state
\begin{equation}
\label{eq:amp-enc}
\ket{\psi_v}
\;\coloneqq\;
\frac{1}{\|v\|_2}\sum_{m=1}^{d} v_m \ket{m},
\qquad v\neq 0.
\end{equation}
For a minority sample $x$ and its assigned centroid $c=c(x)$, the (squared) state overlap is
\begin{equation}
\label{eq:overlap}
s(x,c) \;\coloneqq\; \braket{\psi_x\,|\,\psi_c}
%\;=\; \left(\frac{\langle x,c\rangle}{\|x\|_2\,\|c\|_2}\right)^{\!2}
\;=\; {\braket{M|C}}
\;=\; \cos{\alpha(x,c)},
\end{equation}
where $\alpha(x,c)$ is the geometric angle between $x$ and $c$.
A compact swap test provides an unbiased estimator of $s(x,c)$ via the ancilla-0 probability $p_0$ given in Eq. \eqref{Eq8}.
We define the \emph{quantum-inspired angle}
\begin{equation}
\label{eq:q-angle}
\alpha(x,c) \;\coloneqq\; \cos^{-1}({s(x,c)}\bigr),
\end{equation}
%so that larger misalignment ($x$ less aligned with $c$) yields a larger $\theta$.
For $x$, define the unit direction pointing from $x$ to its centroid:
\begin{equation}
\label{eq:direction}
\hat{d}(x,c) \;\coloneqq\; \frac{c-x}{\|c-x\|_2}.
\end{equation}
%If $x=c$ (degenerate case), sample a unit vector $\hat{u}$ uniformly on the $(d-1)$-sphere $S^{d-1}$ and set $\hat{d}\gets \hat{u}$.
Let $\kappa>0$ be the user-chosen split factor (denoted \texttt{split\_factor} in Algorithm~\ref{Algo-QuantumSMOTE}).
For each synthetic draw, sample a scalar step length
\begin{equation}
\label{eq:step-sampling}
r \;\sim\; \mathrm{Unif}\!\left(0,\, \frac{\alpha(x,c)}{\kappa}\right).
\end{equation}
This linear map couples the step scale to the quantum angle:
when $x$ is already well aligned with $c$ ($\alpha$ small), $r$ concentrates near $0$; when $x$ is poorly aligned ($\alpha$ large), $r$ admits larger values, encouraging stronger movement toward the cluster structure.
A single synthetic sample associated with a minority point $x$ and centroid $c=c(x)$ is constructed as
\begin{equation}
\label{eq:synthetic}
\widetilde{x}
%\;\coloneqq\;
=
x \;+\; r\,\hat{d}(x,c).
\qquad
%r\ \text{as in Eq. \eqref{eq:step-sampling}},\;\;\hat{d}\ \text{as in Eq. \eqref{eq:direction}}.
\end{equation}
The synthetic set for class $\ell$ is generated by repeating Eq. \eqref{eq:synthetic} independently $\mathrm{need}_\ell$ times for randomly chosen minority seeds $x$ with $y=\ell$.
Stacking all generated $\widetilde{x}$ across $\ell\in\mathcal{L}_{\min}$ produces the oversampled matrix $X_{\text{out}}$ and corresponding labels $y_{\text{out}}$. The method has the following consequences:
\begin{enumerate}
\item {Adaptivity to cluster geometry:} The step direction is \emph{cluster-informed} via $c=c(x)$, and the step length adapts to the \emph{quantum similarity} between $x$ and $c$ through $\alpha(x,c)$ in Eq. \eqref{eq:q-angle}.
\item {Scale compatibility:} Because $r$ is dimensionless while Eq. \eqref{eq:synthetic} outputs in feature units, standardization of $X$ is assumed so that angular scales translate meaningfully into step magnitudes.
\item {Convex movement:} If one wishes to constrain synthetic samples strictly within the ray segment from $x$ towards $c$, the optional projection
\[
\widetilde{x}\ \leftarrow\ x + \min\!\bigl\{r,\,\|c-x\|_2\bigr\}\,\hat{d}(x,c)
\]
can be applied; Algorithm~\ref{Algo-QuantumSMOTE} uses Eq. \eqref{eq:synthetic} without this clamp.
\end{enumerate}

\begin{algorithm}[H]
\caption{Quantum SMOTE}
\label{Algo-QuantumSMOTE}
\textbf{Input:} Feature matrix $X$, label vector $y$, minority labels $minority\_labels$, number of clusters $clusters$, split factor $split\_factor$ \\
\textbf{Output:} Oversampled feature matrix $X\_out$, label vector $y\_out$
\begin{algorithmic}[1]
\State Import KMeans and numpy
\State Initialize $X\_out \gets X$, $y\_out \gets y$
\State Fit KMeans with $clusters$ on $X$ to get centroids

\For{each label $lbl$ in $minority\_labels$}
    \State Extract minority samples $minority\_class \gets X[y == lbl]$
\EndFor

\State Compute majority class count $majority\_count \gets \max(\text{counts of each class})$
\For{each minority class array}
    \State Compute number of new samples needed $need \gets majority\_count - \text{len(minority class)}$
\EndFor

\For{each minority label $lbl$ and corresponding $need$}
    \If{$need = 0$ or minority class is empty} \textbf{continue} \EndIf
    \State Initialize empty list $synth\_list$
    
    \For{$i = 1$ to $need$}
        \State Pick random sample $x$ from minority class
        \State Predict cluster $cluster\_id$ of $x$ using KMeans
        \State Get cluster centroid $centroid \gets centroids[cluster\_id]$
        \State Compute quantum-inspired angle $angle \gets compact\_swap\_test(x, centroid)$
        \State Sample step fraction $frac \sim \text{Uniform}(0, angle / split\_factor)$
        \State Compute direction $direction \gets centroid - x$
        \If{norm(direction) $> 0$} 
            \State Normalize $direction \gets direction / \|direction\|$
        \Else
            \State Assign tiny random normalized direction
        \EndIf
        \State Generate synthetic sample $synthetic \gets x + direction * frac$
        \State Append $synthetic$ to $synth\_list$
    \EndFor

    \If{$synth\_list$ is not empty}
        \State Stack $synth\_list$ to array
        \State Append synthetic samples to $X\_out$, append labels to $y\_out$
    \EndIf
\EndFor

\State \Return $X\_out, y\_out$
\end{algorithmic}
\end{algorithm}

The QSMOTE method shown in Algorithm \ref{Algo-QuantumSMOTE} generates synthetic samples for minority classes in an imbalanced dataset using a quantum-inspired approach. It first fits a KMeans clustering model on the feature matrix $X$ to obtain cluster centroids. For each minority class, it calculates the number of new samples required to match the majority class size. For each required synthetic sample, it randomly selects a minority instance and predicts its cluster to obtain the corresponding centroid. It then computes a quantum-inspired angle between the sample and the centroid using the compact swap test method. A step fraction is sampled uniformly based on this angle and a split factor, and a direction vector from the sample to the centroid is computed and normalized. The synthetic sample is generated by moving along this direction scaled by the step fraction. All generated synthetic samples are concatenated to the original feature matrix and their labels to the label vector, producing a balanced dataset $X\_out$ and $y\_out$ for subsequent training of classical ML models. It is important to emphasize that the quantum-inspired noise channels considered in this work are applied to the compact-swap-test-based similarity estimation process rather than to the downstream classifiers. Since RF, SVM, DT, LR, and NB are classical machine-learning models, they are subsequently used only to evaluate the downstream impact of noisy-QSMOTE-generated synthetic samples on fault-classification performance. The noisy-QSMOTE framework follows a sequential propagation process. Noise first affects the compact swap test, which alters the overlap estimation between minority samples and cluster centroids. The modified overlap values influence the similarity score and angle computation used for synthetic sample generation. As a result, the generated balanced datasets differ from those obtained under noise-free conditions, allowing the downstream impact of noisy similarity estimation to be quantified through classification performance.

\begin{algorithm}[]
\caption{Noisy-QSMOTE}
\label{Algo-NoisyQSMOTE}
\textbf{Input:} Feature matrix $X$, label vector $y$, minority labels $minority_labels$, number of clusters $clusters$, split factor $split_factor$, noise channel $\mathcal{N}$, noise probability $p$ \\
\textbf{Output:} Noisy-QSMOTE oversampled feature matrix $X_{out}^{noise}$, label vector $y_{out}^{noise}$
\begin{algorithmic}[1]
\State Import KMeans, numpy, AerSimulator, NoiseModel, and the selected quantum-noise channel
\State Initialize $X_{out}^{noise} \gets X$, $y_{out}^{noise} \gets y$
\State Fit KMeans with $clusters$ on $X$ to obtain centroids

\For{each label $lbl$ in $minority\_labels$}
\State Extract minority samples $minority\_class \gets X[y == lbl]$
\EndFor

\State Compute majority class count $majority\_count \gets \max(\text{counts of each class})$

\For{each minority label $lbl$}
\State Compute number of new samples needed $need \gets majority\_count - \text{len}(minority\_class)$
\If{$need = 0$ or $minority\_class$ is empty} \textbf{continue} \EndIf
\State Initialize empty list $synth\_list^{noise}$

%```
\For{$i = 1$ to $need$}
    \State Pick random minority sample $x$ from $minority\_class$
    \State Predict cluster $cluster\_id$ of $x$ using KMeans
    \State Get cluster centroid $centroid \gets centroids[cluster\_id]$
    \State Construct compact swap-test circuit for $(x, centroid)$
    \State Inject selected noise channel $\mathcal{N}(p)$ into the compact swap-test circuit before ancilla measurement
    \State Run the noisy compact swap-test circuit and estimate noisy ancilla probability $p_0^{noise}$
    \State Compute noisy overlap estimate:
    \[
    \langle M|C\rangle_{noise} \gets \frac{(3-4p_0^{noise})Z}{2\|x\|\|centroid\|}
    \]
    \State Clip $\langle M|C\rangle_{noise}$ to $[-1,1]$
    \State Compute noisy quantum-inspired angle:
    \[
    angle^{noise} \gets \cos^{-1}\left(\langle M|C\rangle_{noise}\right)
    \]
    \State Sample step fraction $frac \sim \text{Uniform}(0, angle^{noise}/split\_factor)$
    \State Compute direction $direction \gets centroid - x$
    \If{$\|direction\| > 0$}
        \State Normalize $direction \gets direction/\|direction\|$
    \Else
        \State Assign tiny random normalized direction
    \EndIf
    \State Generate noisy-QSMOTE synthetic sample $synthetic^{noise} \gets x + direction * frac$
    \State Append $synthetic^{noise}$ to $synth\_list^{noise}$
\EndFor

\If{$synth\_list^{noise}$ is not empty}
    \State Stack $synth\_list^{noise}$ to array
    \State Append noisy synthetic samples to $X_{out}^{noise}$ and append labels to $y_{out}^{noise}$
\EndIf
%```

\EndFor

\State \Return $X_{out}^{noise}, y_{out}^{noise}$
\end{algorithmic}
\end{algorithm}

Algorithm~\ref{Algo-NoisyQSMOTE} presents the proposed Noisy-QSMOTE procedure, which extends the original QSMOTE framework by explicitly injecting quantum-inspired perturbations into the compact-swap-test-based similarity estimation process. Unlike standard QSMOTE, where the quantum-inspired angle is computed from a noise-free overlap estimate, Noisy-QSMOTE first applies a selected noise channel $\mathcal{N}(p)$ to the compact swap-test circuit before ancilla measurement. This modifies the measured probability $p_0$, which in turn affects the overlap estimate, the angle computation, and the final synthetic sample location. Therefore, the generated samples are derived from noise-corrupted similarity information rather than clean quantum-inspired similarity estimates. The resulting noisy-QSMOTE-generated dataset is then evaluated using classical ML models only as a downstream assessment tool. The classifiers are not directly exposed to the quantum-inspired noise channels; instead, their performance reflects the impact of noisy compact-swap-test-based similarity estimation on synthetic minority sample generation. This algorithm therefore represents the central technical contribution of the present work and enables systematic analysis of how quantum-inspired perturbations propagate through QSMOTE and influence industrial fault-classification performance.

\section{Experimental Results}\label{Sec4}

\subsection{Settings}
To comprehensively evaluate the effectiveness and robustness of QSMOTE, the experimental results are organized into three categories. First, baseline classification performance is reported on the original imbalanced datasets. Second, the impact of QSMOTE is assessed by comparing classification performance before and after oversampling. Third, noisy-QSMOTE results are presented, where quantum-inspired noise channels are injected into the compact-swap-test-based similarity estimation process used during synthetic sample generation. This structure allows the effects of imbalance mitigation, noisy-QSMOTE robustness, and the resulting changes in fault-classification performance to be examined separately.

\subsection{Datasets}
The SPID comprises six distinct classes: bird-drop (162 images), clean (159 images), dusty (147 images), snow-covered (103 images), electrical-damage (87 images), and physical-damage (50 images) \cite{afroz2023solar}. Each class represents a specific condition that can affect the efficiency and energy output of solar panels. While it offers diverse visual samples across categories, there is a noticeable imbalance in class distribution, with some categories (e.g., physical-damage) having fewer images compared to others (e.g., bird-drop). Then, the \emph{CWRUBD} hosted on Kaggle \cite{cwru_kaggle} is used, which repackage the Case Western Reserve University (CWRU) Bearing Data Center's motor condition monitoring signals for ML research in industrial settings. 
For supervised learning, nine time-domain features are computed per segment; maximum, minimum, mean, standard deviation, RMS, skewness, kurtosis, crest factor, and form factor-over fixed windows of 2300 samples (i.e., 0.04\,s at 48\,kHz). This segmentation yields labeled examples spanning multiple fault sizes, fault locations, and sensor positions, enabling robust classification and fault-identification studies on rotating machinery. The EFDD \cite{engine_failure_kaggle} is a synthetic collection simulating multi-sensor telemetry from automotive engines to support fault detection and severity classification. Records are time-stamped at 5-minute intervals beginning on December 24, 2024 at 10{:}00, and each row summarizes engine state via thermal, kinematic, vibration, and powertrain variables together with operational context and a discrete fault label. 
The dataset contains 1{,}000 samples and four target classes (\texttt{Fault\_Condition} $\in \{0,1,2,3\}$) ranging from normal to severe failure. The IFDD \cite{industrialfault2024} is designed for automatic fault detection in Industry 4.0 applications that use IoT-based smart sensors. It collects industrial data from several sensors, such as temperature (°C), vibration (m/s²), pressure (kPa), flow rate (L/min), current (A), and voltage (V). Each instance represents a snapshot of equipment operating conditions and sensor information, enabling real-time monitoring and intelligent diagnostics. The dataset includes a Fault\_Type variable that categorizes the system condition into four categories: 0 (Normal Operation), 1 (Overheating Fault), 2 (Leakage Fault), and 3 (Power Fluctuation Fault). 1000 samples are taken for the testing.

\subsection{Preprocessing}
In the preprocessing stage of SPID, first, the solar panel image dataset is downloaded from Kaggle and organized it into training and validation sets using TensorFlow's image\_dataset\_from\_directory, with an 80-20 split. Each image is resized to a fixed dimension of 244 $\times$ 244 pixels, batched, shuffled, and labeled according to its class. To visualize the data distribution, a few sample images are plotted along with their class names. For feature extraction, a ResNet50 model is pretrained on ImageNet, with the top classification layer removed and global average pooling applied, so that each image is mapped to a 2048-dimensional feature vector. These high-dimensional embeddings are then reduced to 32 principal components using PCA, ensuring compact yet informative features for downstream tasks. Finally, the reduced features and labels are stored in tabular form as Pandas DataFrames and exported as CSV files, making them suitable for classical ML pipelines for further experimentation. 
The CWRUBD is obtained from Kaggle, and the downloaded artifacts are counted to identify tabular files. The first available \texttt{.csv}/\texttt{.txt} file is ingested using \texttt{pandas.read\_csv} with \texttt{sep=None} and the Python engine, allowing for automated delimiter inference. To standardize supervision targets, the original fault indicator column was renamed from \texttt{fault} to \texttt{label}. The \texttt{label} column is transformed to a string, and a deterministic integer encoding is created by enumerating unique class names in order of appearance (\texttt{label\_map = \{name: i\}}). These identities are mapped back into the dataframe, so that \texttt{df["label"]} contains contiguous class IDs beginning with zero. For transparency and validation, the table's head, mapping dictionary, unique ID set, and number of separate errors are displayed. This technique creates a clean, model-ready dataframe with a canonical \texttt{label} field (integers) while keeping a traceable mapping to the original categorical labels, which serves as the foundation for subsequent dataset segmentation and training procedures. 
In case of EFDD, all available CSV files are automatically detected and loaded into a pandas \texttt{DataFrame}, where the target column is renamed to \texttt{label} for consistency and moved to the final position. The \texttt{Time\_Stamp} attribute, when present, is converted into a datetime format for potential temporal analysis but excluded from training to avoid leakage. The categorical feature \texttt{Operational\_Mode} is numerically encoded through a stable mapping $\mathrm{Operational\_Mode} \rightarrow {0, 1, 2, \dots, K-1}$ based on category appearance, with missing values replaced by a default ``Unknown" class. Non-essential columns, including \texttt{Time\_Stamp}, are dropped, and all numerical columns are memory-optimized by downcasting to lower-precision datatypes without compromising accuracy. The resulting dataset thus comprised normalized numerical and encoded categorical features, with the final structure denoted as $X \in \mathbb{R}^d$ and $y \in {0,1,2,3}$, representing four distinct operational fault categories; No Fault, Overheating, Leakage, and Power Fluctuation, ready for further processing. The IFDD is retrieved using the \texttt{kagglehub} API, after which all available CSV or text files are programmatically identified and loaded into a unified pandas \texttt{DataFrame}. The preprocessing pipeline standardized column names and structure by renaming the ground-truth column \texttt{Fault\_Type} to \texttt{label} for consistency across experiments and positioning it as the final column. When present, the \texttt{Time\_Stamp} attribute is parsed into a \texttt{datetime} format (errors coerced) to preserve temporal context but is excluded from the feature set to prevent time-based leakage. All numeric sensor readings, including temperature, vibration, pressure, flow rate, current, and voltage, are retained as predictors, while non-informative attributes are dropped. Categorical features are normalized to string type, missing entries are replaced with the placeholder ``Unknown" and a stable ordinal encoding map is applied based on the order of appearance to ensure deterministic label indices. Finally, numerical columns are downcast to \texttt{float32} or \texttt{int32} types using \texttt{pandas.to\_numeric()} to reduce memory usage without loss of precision. The resulting dataset, organized as $X \in \mathbb{R}^d$ and $y \in \{0,1,2,3\}$ representing No Fault, Overheating, Leakage, and Power Fluctuation classes, served as the clean and standardized input for the models.

\subsection{Comparative Models}
For comparison, several widely used classical ML models are evaluated. LR provides a linear baseline for assessing feature separability, while RF is used to capture non-linear patterns through an ensemble of DTs. SVM is included because it performs well in high-dimensional spaces and often provides strong class separation through margin-based learning. NB, a lightweight probabilistic classifier based on Bayes' theorem, is chosen for its efficiency and simplicity, despite the requirement of feature independence. Finally, the DT model is included because of its interpretability and ability to represent nonlinear decision limits with low complexity, making these models appropriate benchmarks against which to test our technique.

\subsection{Hyperparameters and Metrics}\label{Sec4.1}
Each baseline is trained with sensible, reproducible defaults and evaluated them using a consistent protocol. LR is run with max\_iter=500 on scaled features to ensure convergence under L2-regularized optimization. SVM uses an rbf kernel with probability=True (for downstream scoring/ROC needs) and is likewise fit on scaled inputs. GaussianNB is applied to scaled features without additional tuning, reflecting its parametric simplicity. Tree-based models operates on the unscaled tabular features: RF with n\_estimators=200, random\_state=42, and n\_jobs=-1 for robust, parallelized ensembling; and a DT with random\_state=42 as an interpretable non-linear baseline. For all models, stratified 5-fold cross-validation is performed on the training set to preserve class proportions and report mean $\pm$ standard deviation of accuracy, precision, recall, and F1-score; the latter three are computed as weighted averages (with zero\_division=0) to account for class imbalance. Once cross-validation is completed, the selected model is trained using the full training data and then evaluated on the test set to assess its performance on unseen samples. Random seeds are fixed (random\_state=42) for reproducibility.

\subsection{Noise Model}
The robustness of the proposed quantum algorithms is analyzed under six commonly used realistic noise models~\cite{satpathy2023analysis}. These include three unitary error models, bit flip, phase flip, and bit phase flip originating from imperfections in gate operations or interactions with the environment. In addition, the DP noise model captures random Pauli errors occurring with equal probability, while PD and AD represent non-unitary processes that model decoherence and energy relaxation, respectively. Together, these models provide a comprehensive framework for assessing algorithmic stability under both discrete and continuous noise channels.

%\subsubsection{Bit Flip Noise}
{1) Bit Flip Noise:} 
The bit flip channel represents the inversion of a qubit state $\ket{0} \leftrightarrow \ket{1}$ with probability $\eta_B$. Its Kraus operators are
\begin{equation}
    E_0 = \sqrt{1 - \eta_B} I, \qquad E_1 = \sqrt{\eta_B} X.
\end{equation}
The resulting density matrix after noise application is
\begin{equation}
    \zeta = (1 - \eta_B)\rho + \eta_B X\rho X.
\end{equation}

%\subsubsection{Phase Flip Noise}
{2) Phase Flip Noise:}
This model introduces a phase error, flipping $\ket{+}$ to $\ket{-}$ with probability $\eta_P$. The Kraus operators are
\begin{equation}
    E_0 = \sqrt{1 - \eta_P} I, \qquad E_1 = \sqrt{\eta_P} Z,
\end{equation}
and the noisy state evolves as
\begin{equation}
    \zeta = (1 - \eta_P)\rho + \eta_P Z\rho Z.
\end{equation}

%\subsubsection{Bit Phase Flip Noise}
{3) Bit Phase Flip Noise:}
The bit phase flip model combines both bit and phase inversions, represented by the Pauli-$Y$ operator, occurring with probability $\eta_{BP}$. The corresponding Kraus operators are
\begin{equation}
    E_0 = \sqrt{1 - \eta_{BP}} I, \qquad E_1 = \sqrt{\eta_{BP}} Y.
\end{equation}
The transformed state is
\begin{equation}
    \zeta = (1 - \eta_{BP})\rho + \eta_{BP} Y\rho Y.
\end{equation}

%\subsubsection{Depolarizing Noise}
{4) Depolarizing Noise:}
DP noise randomizes the state by applying any Pauli error ($X$, $Y$, or $Z$) with equal probability $\eta_D/3$. The Kraus operators are
\begin{align}
    E_0 &= \sqrt{1 - \eta_D}\, I, &
    E_1 &= \sqrt{\eta_D / 3}\, X, \nonumber\\
    E_2 &= \sqrt{\eta_D / 3}\, Y, &
    E_3 &= \sqrt{\eta_D / 3}\, Z.
\end{align}
The effect on $\rho$ is
\begin{equation}
    \zeta = (1 - \eta_D)\rho + \frac{\eta_D}{3}(X\rho X + Y\rho Y + Z\rho Z).
\end{equation}

%\subsubsection{Amplitude Damping Noise}
{5) Amplitude Damping Noise:}
AD models energy loss, such as spontaneous emission, where $\beta_A$ is the probability of relaxation. Its Kraus operators are
\begin{equation}
    E_0 = \ket{0}\bra{0} + \sqrt{1 - \beta_A}\ket{1}\bra{1}, \qquad
    E_1 = \sqrt{\beta_A}\ket{0}\bra{1},
\end{equation}
The final state is given by
\begin{equation}
    \zeta = E_0 \rho E_0^\dagger + E_1 \rho E_1^\dagger.
\end{equation}

%\subsubsection{Phase Damping Noise}
{6) Phase Damping Noise:}
PD captures the loss of quantum coherence without energy dissipation. The Kraus operators are given by
\begin{equation}
    E_0 = \ket{0}\bra{0} + \sqrt{1 - \lambda_P}\ket{1}\bra{1}, \qquad
    E_1 = \sqrt{\lambda_P}\ket{1}\bra{1},
\end{equation}
where $\lambda_P$ denotes the PD probability. The noisy evolution is
\begin{equation}
    \zeta = E_0 \rho E_0^\dagger + E_1 \rho E_1^\dagger.
\end{equation}

\subsection{Impact of QSMOTE-Based Imbalance Mitigation}\label{Sec4.3}
\begin{figure}
    \centering
    \begin{subfigure}{0.24\linewidth}
        \centering
        \includegraphics[width=\linewidth]{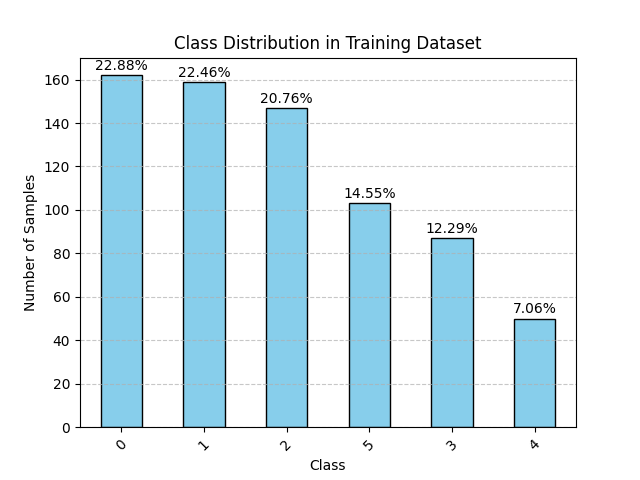} 
        \caption{}
        \label{fig:h_data_pca}
    \end{subfigure}
    \hfill
    \begin{subfigure}{0.24\linewidth}
        \centering
        \includegraphics[width=\linewidth]{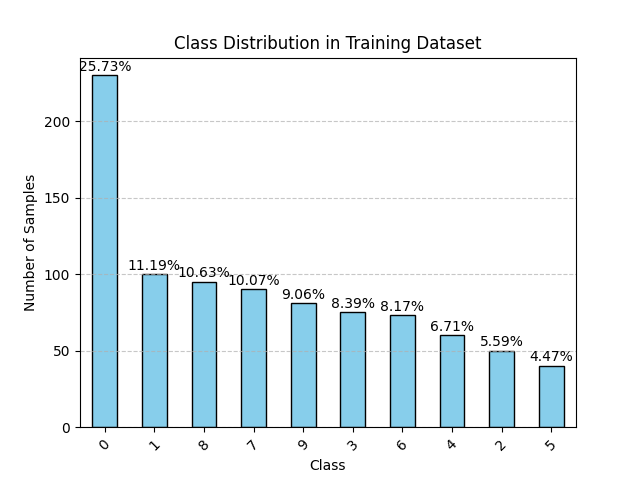}
        \caption{}
        \label{fig:m_data_pca}
    \end{subfigure}\hfill
    \begin{subfigure}{0.24\linewidth}
        \centering
        \includegraphics[width=\linewidth]{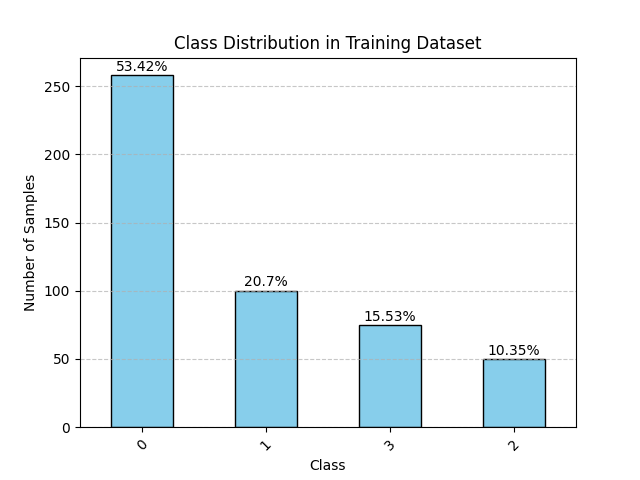} 
        \caption{}
        \label{fig:h_data_pca}
    \end{subfigure}
    \hfill
    \begin{subfigure}{0.24\linewidth}
        \centering
        \includegraphics[width=\linewidth]{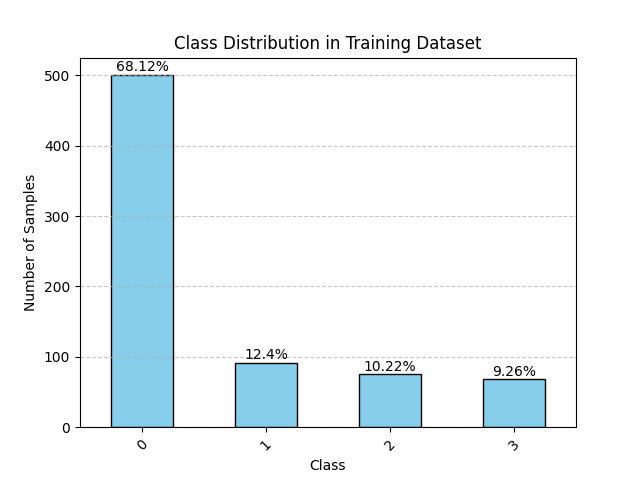} 
        \caption{}
        \label{fig:h_data_pca}
    \end{subfigure}
    \hfill
    \begin{subfigure}{0.24\linewidth}
        \centering
        \includegraphics[width=\linewidth]{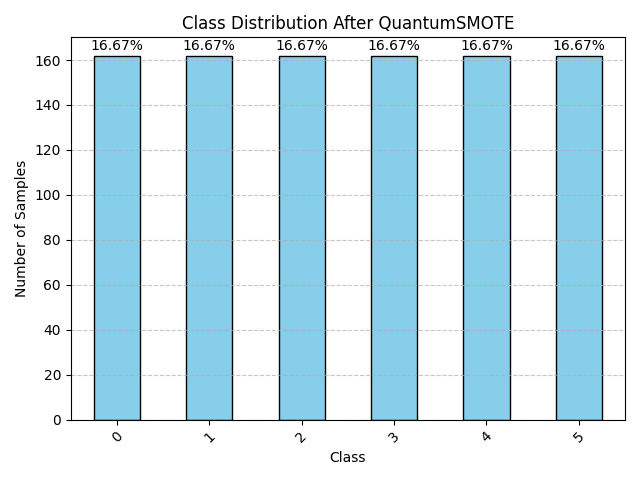} 
        \caption{}
        \label{fig:h_data_pca}
    \end{subfigure}
    \hfill
    \begin{subfigure}{0.24\linewidth}
        \centering
        \includegraphics[width=\linewidth]{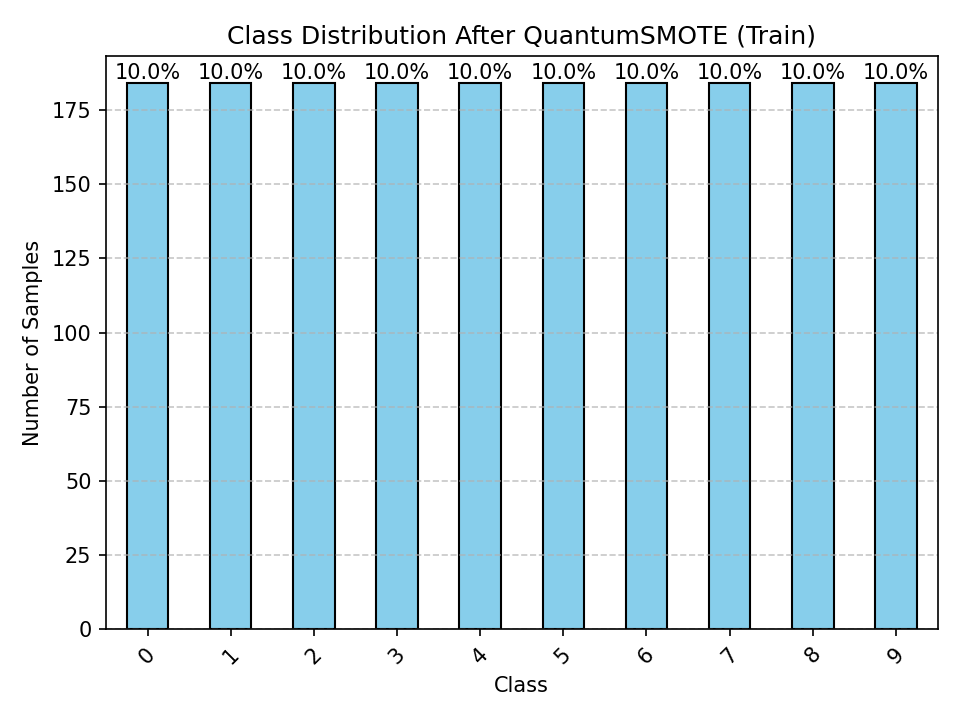} 
        \caption{}
        \label{fig:h_data_pca}
    \end{subfigure}
    \hfill
    \begin{subfigure}{0.24\linewidth}
        \centering
        \includegraphics[width=\linewidth]{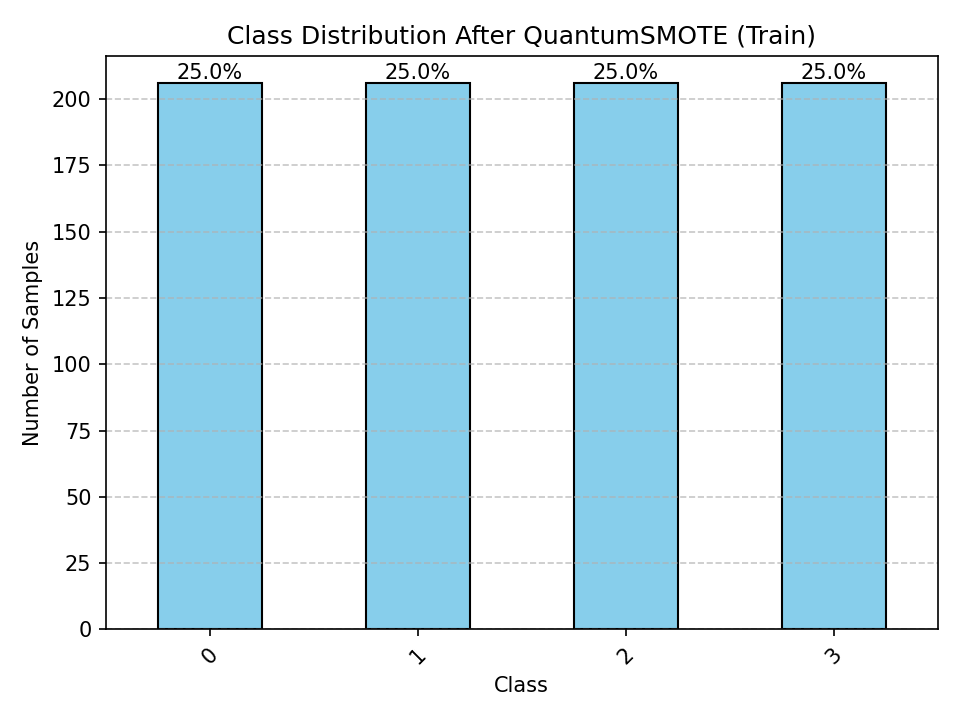} 
        \caption{}
        \label{fig:h_data_pca}
    \end{subfigure}
    \hfill
    \begin{subfigure}{0.24\linewidth}
        \centering
        \includegraphics[width=\linewidth]{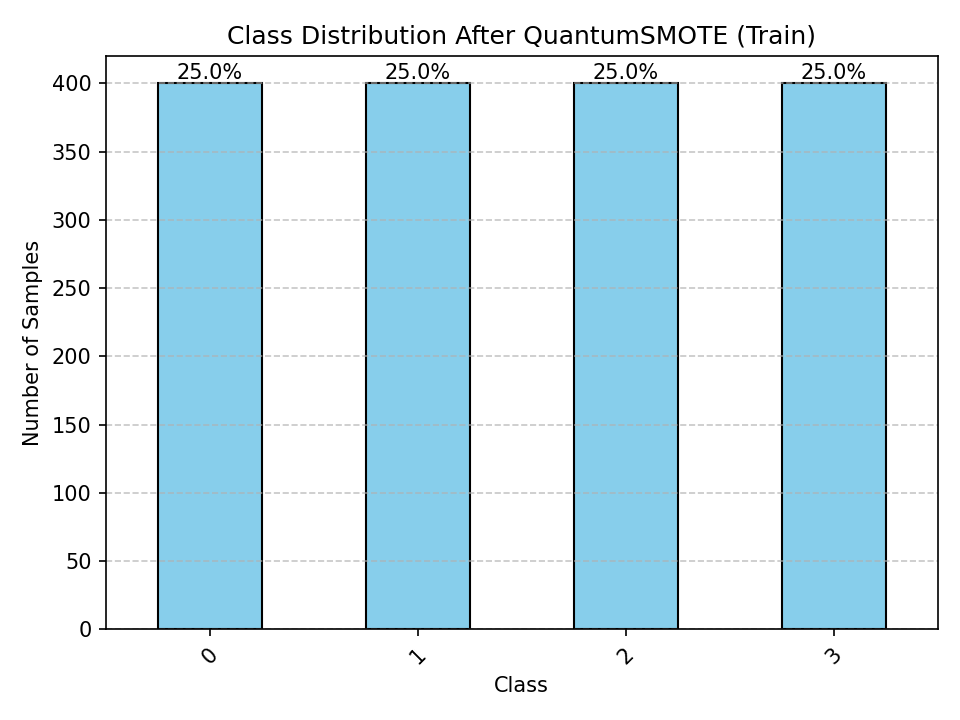} 
        \caption{}
        \label{fig:h_data_pca}
    \end{subfigure}\hfill
    \caption{Distribution of (a) SPID, (b) CWRUBD, (c) EFDD, (d) IFDD before QSMOTE. Distribution of (e) SPID, (f) CWRUBD, (g) EFDD, (h) IFDD after QSMOTE.}
    \label{fig:distribution_before_after_smote}
\end{figure}

\begin{table}[ht]
\caption{Performance Metrics of Classical Algorithms Before and After QSMOTE for Hexa Classification on SPID}
\label{table:performance_smote_hexa_1}
\centering
\begin{tabular*}{\textwidth}{@{\extracolsep\fill}lcccc}
\toprule
\textbf{Algorithm} 
& \textbf{Accuracy} 
& \textbf{Precision} 
& \textbf{Recall} 
& \textbf{F1 Measure} \\
\midrule
\multicolumn{5}{l}{\textbf{Before QSMOTE}} \\
\midrule
LR  
& 0.7880 $\pm$ 0.0342 
& 0.7976 $\pm$ 0.0293 
& 0.7880 $\pm$ 0.0342 
& 0.7881 $\pm$ 0.0313 \\

RF  
& 0.7756 $\pm$ 0.0259 
& 0.7847 $\pm$ 0.0259 
& 0.7756 $\pm$ 0.0259 
& 0.7726 $\pm$ 0.0257 \\

SVM 
& 0.8057 $\pm$ 0.0200 
& 0.8197 $\pm$ 0.0180 
& 0.8057 $\pm$ 0.0200 
& 0.8024 $\pm$ 0.0223 \\

NB  
& 0.7298 $\pm$ 0.0403 
& 0.7465 $\pm$ 0.0384 
& 0.7298 $\pm$ 0.0403 
& 0.7294 $\pm$ 0.0404 \\

DT  
& 0.5724 $\pm$ 0.0303 
& 0.5760 $\pm$ 0.0278 
& 0.5724 $\pm$ 0.0303 
& 0.5677 $\pm$ 0.0297 \\

\midrule
\multicolumn{5}{l}{\textbf{After QSMOTE}} \\
\midrule
LR  
& \textbf{0.8198 $\pm$ 0.0076} 
& \textbf{0.8255 $\pm$ 0.0101} 
& \textbf{0.8198 $\pm$ 0.0076} 
& \textbf{0.8201 $\pm$ 0.0095} \\

RF  
& \textbf{0.8533 $\pm$ 0.0147} 
& \textbf{0.8556 $\pm$ 0.0152} 
& \textbf{0.8533 $\pm$ 0.0147} 
& \textbf{0.8526 $\pm$ 0.0143} \\

SVM 
& \textbf{0.8893 $\pm$ 0.0149} 
& \textbf{0.8927 $\pm$ 0.0136} 
& \textbf{0.8893 $\pm$ 0.0149} 
& \textbf{0.8896 $\pm$ 0.0145} \\

NB  
& \textbf{0.7593 $\pm$ 0.0301} 
& \textbf{0.7659 $\pm$ 0.0303} 
& \textbf{0.7593 $\pm$ 0.0301} 
& \textbf{0.7592 $\pm$ 0.0298} \\

DT  
& \textbf{0.7439 $\pm$ 0.0184} 
& \textbf{0.7421 $\pm$ 0.0206} 
& \textbf{0.7439 $\pm$ 0.0184} 
& \textbf{0.7393 $\pm$ 0.0193} \\
\bottomrule
\end{tabular*}

%\footnotetext{Results are reported as mean $\pm$ standard deviation over multiple runs. QSMOTE denotes the proposed quantum-inspired synthetic minority oversampling technique.}
\end{table}

\begin{table}[ht]
\caption{Performance Metrics of Classical Algorithms Before and After QSMOTE for Deca-Class Classification on CWRUBD}
\label{table:performance_smote_ten_2}
\centering
\begin{tabular*}{\textwidth}{@{\extracolsep\fill}lcccc}
\toprule
\textbf{Algorithm} 
& \textbf{Accuracy} 
& \textbf{Precision} 
& \textbf{Recall} 
& \textbf{F1 Measure} \\
\midrule
\multicolumn{5}{l}{\textbf{Before QSMOTE}} \\
\midrule
LR  
& 0.8783 $\pm$ 0.0163 
& 0.8815 $\pm$ 0.0319 
& 0.8783 $\pm$ 0.0163 
& 0.8654 $\pm$ 0.0209 \\

RF  
& 0.9483 $\pm$ 0.0241 
& 0.9511 $\pm$ 0.0243 
& 0.9483 $\pm$ 0.0241 
& 0.9477 $\pm$ 0.0245 \\

SVM 
& 0.8797 $\pm$ 0.0174 
& 0.8745 $\pm$ 0.0247 
& 0.8797 $\pm$ 0.0174 
& 0.8664 $\pm$ 0.0169 \\

NB  
& 0.9245 $\pm$ 0.0168 
& 0.9303 $\pm$ 0.0171 
& 0.9245 $\pm$ 0.0168 
& 0.9229 $\pm$ 0.0170 \\

DT  
& 0.9077 $\pm$ 0.0174 
& 0.9152 $\pm$ 0.0176 
& 0.9077 $\pm$ 0.0174 
& 0.9072 $\pm$ 0.0193 \\

\midrule
\multicolumn{5}{l}{\textbf{After QSMOTE}} \\
\midrule
LR  
& \textbf{0.9332 $\pm$ 0.0066} 
& \textbf{0.9350 $\pm$ 0.0068} 
& \textbf{0.9332 $\pm$ 0.0066} 
& \textbf{0.9328 $\pm$ 0.0069} \\

RF  
& \textbf{0.9848 $\pm$ 0.0044} 
& \textbf{0.9856 $\pm$ 0.0037} 
& \textbf{0.9848 $\pm$ 0.0044} 
& \textbf{0.9848 $\pm$ 0.0044} \\

SVM 
& \textbf{0.9522 $\pm$ 0.0059} 
& \textbf{0.9528 $\pm$ 0.0063} 
& \textbf{0.9522 $\pm$ 0.0059} 
& \textbf{0.9518 $\pm$ 0.0062} \\

NB  
& 0.8875 $\pm$ 0.0182 
& 0.9002 $\pm$ 0.0162 
& 0.8875 $\pm$ 0.0182 
& 0.8856 $\pm$ 0.0196 \\

DT  
& \textbf{0.9750 $\pm$ 0.0080} 
& \textbf{0.9764 $\pm$ 0.0062} 
& \textbf{0.9750 $\pm$ 0.0080} 
& \textbf{0.9749 $\pm$ 0.0079} \\
\bottomrule
\end{tabular*}
\end{table}

\begin{table}[ht]
\caption{Performance Metrics of Classical Algorithms Before and After QSMOTE for Tetra-Class Classification on EFDD}
\label{table:performance_smote_four_third_3}
\centering
\begin{tabular*}{\textwidth}{@{\extracolsep\fill}lcccc}
\toprule
\textbf{Algorithm} 
& \textbf{Accuracy} 
& \textbf{Precision} 
& \textbf{Recall} 
& \textbf{F1 Measure} \\
\midrule
\multicolumn{5}{l}{\textbf{Before QSMOTE}} \\
\midrule
LR  
& 0.5337 $\pm$ 0.0024 
& 0.3003 $\pm$ 0.0285 
& 0.5337 $\pm$ 0.0024 
& 0.3765 $\pm$ 0.0083 \\

RF  
& 0.4923 $\pm$ 0.0201 
& 0.3403 $\pm$ 0.0788 
& 0.4923 $\pm$ 0.0201 
& 0.3664 $\pm$ 0.0147 \\

SVM 
& 0.5337 $\pm$ 0.0024 
& 0.2848 $\pm$ 0.0026 
& 0.5337 $\pm$ 0.0024 
& 0.3714 $\pm$ 0.0028 \\

NB  
& 0.5156 $\pm$ 0.0113 
& 0.2956 $\pm$ 0.0274 
& 0.5156 $\pm$ 0.0113 
& 0.3681 $\pm$ 0.0090 \\

DT  
& 0.2979 $\pm$ 0.0172 
& 0.3140 $\pm$ 0.0287 
& 0.2979 $\pm$ 0.0172 
& 0.3036 $\pm$ 0.0192 \\

\midrule
\multicolumn{5}{l}{\textbf{After QSMOTE}} \\
\midrule
LR  
& 0.2622 $\pm$ 0.0213 
& 0.2582 $\pm$ 0.0232 
& 0.2622 $\pm$ 0.0213 
& 0.2573 $\pm$ 0.0224 \\

RF  
& \textbf{0.9126 $\pm$ 0.0259} 
& \textbf{0.9142 $\pm$ 0.0254} 
& \textbf{0.9126 $\pm$ 0.0259} 
& \textbf{0.9116 $\pm$ 0.0263} \\

SVM 
& \textbf{0.6420 $\pm$ 0.0226} 
& \textbf{0.6400 $\pm$ 0.0202} 
& \textbf{0.6420 $\pm$ 0.0226} 
& \textbf{0.6358 $\pm$ 0.0207} \\

NB  
& 0.3362 $\pm$ 0.0221 
& 0.3373 $\pm$ 0.0233 
& 0.3362 $\pm$ 0.0221 
& 0.3353 $\pm$ 0.0224 \\

DT  
& \textbf{0.8228 $\pm$ 0.0115} 
& \textbf{0.8266 $\pm$ 0.0159} 
& \textbf{0.8228 $\pm$ 0.0115} 
& \textbf{0.8050 $\pm$ 0.0150} \\
\bottomrule
\end{tabular*}
\end{table}

\begin{table}[ht]
\caption{Performance Metrics of Classical Algorithms Before and After QSMOTE for Tetra-Class Classification on IFDD}
\label{table:performance_smote_hexa_fifth_5}
\centering
\begin{tabular*}{\textwidth}{@{\extracolsep\fill}lcccc}
\toprule
\textbf{Algorithm} 
& \textbf{Accuracy} 
& \textbf{Precision} 
& \textbf{Recall} 
& \textbf{F1 Measure} \\
\midrule
\multicolumn{5}{l}{\textbf{Before QSMOTE}} \\
\midrule
LR  
& 0.6712 $\pm$ 0.0118 
& 0.4728 $\pm$ 0.0222 
& 0.6712 $\pm$ 0.0118 
& 0.5505 $\pm$ 0.0096 \\

RF  
& 0.6763 $\pm$ 0.0089 
& 0.4825 $\pm$ 0.0416 
& 0.6763 $\pm$ 0.0089 
& 0.5527 $\pm$ 0.0108 \\

SVM 
& 0.6814 $\pm$ 0.0028 
& 0.4644 $\pm$ 0.0039 
& 0.6814 $\pm$ 0.0028 
& 0.5523 $\pm$ 0.0037 \\

NB  
& 0.3477 $\pm$ 0.0908 
& 0.4800 $\pm$ 0.0215 
& 0.3477 $\pm$ 0.0908 
& 0.3680 $\pm$ 0.0820 \\

DT  
& 0.4685 $\pm$ 0.0643 
& 0.4912 $\pm$ 0.0467 
& 0.4685 $\pm$ 0.0643 
& 0.4769 $\pm$ 0.0557 \\

\midrule
\multicolumn{5}{l}{\textbf{After QSMOTE}} \\
\midrule
LR  
& 0.4450 $\pm$ 0.0109 
& 0.4331 $\pm$ 0.0092 
& 0.4450 $\pm$ 0.0109 
& 0.4352 $\pm$ 0.0092 \\

RF  
& \textbf{0.9919 $\pm$ 0.0032} 
& \textbf{0.9920 $\pm$ 0.0031} 
& \textbf{0.9919 $\pm$ 0.0032} 
& \textbf{0.9918 $\pm$ 0.0032} \\

SVM 
& \textbf{0.8550 $\pm$ 0.0240} 
& \textbf{0.8540 $\pm$ 0.0262} 
& \textbf{0.8550 $\pm$ 0.0240} 
& \textbf{0.8498 $\pm$ 0.0237} \\

NB  
& \textbf{0.4475 $\pm$ 0.0306} 
& 0.4476 $\pm$ 0.0333 
& \textbf{0.4475 $\pm$ 0.0306} 
& \textbf{0.4417 $\pm$ 0.0305} \\

DT  
& \textbf{0.9006 $\pm$ 0.0111} 
& \textbf{0.9122 $\pm$ 0.0100} 
& \textbf{0.9006 $\pm$ 0.0111} 
& \textbf{0.8916 $\pm$ 0.0134} \\
\bottomrule
\end{tabular*}
\end{table}

This subsection evaluates the effectiveness of QSMOTE for class-imbalance mitigation. Classification performance is compared before and after QSMOTE-based oversampling in order to quantify the impact of quantum-inspired synthetic sample generation on fault-classification accuracy, precision, recall, and F1-score. Fig.~\ref{fig:distribution_before_after_smote} depicts the comparative class distributions of the datasets before and after applying the QSMOTE. The subfigures (a)-(d) illustrate the original imbalanced distributions for four datasets: (a) SPID, (b) CWRUBD, (c) EFDD, and (d) IFDD. In each case, certain classes contain substantially fewer samples than others, revealing a pronounced class imbalance that can hinder the performance of classical ML algorithms due to biased learning toward majority classes. 
Subfigures (e)–(h) show the relevant class distributions after applying QSMOTE. The minority classes are synthetically augmented to obtain a balanced representation in comparison to the majority classes. Comparing the two sets of subfigures demonstrates that QSMOTE effectively equalizes class frequencies across all datasets, resulting in a more uniform distribution. 
This balanced data structure is expected to improve model generalization, enhance metric stability, and mitigate the bias observed during training on the original imbalanced datasets.
The results in Table \ref{table:performance_smote_hexa_1} show that QSMOTE improves the performance of all classical algorithms for the hexa classification task of SPID. Prior to resampling, SVM achieved the highest accuracy (0.8057), followed by LR (0.7880) and RF (0.7756), whereas DT produced the lowest accuracy of 0.5724. After applying QSMOTE, every model shows clear improvement: LR rises to 0.8198, RF increases to 0.8533, and SVM achieves the highest accuracy of 0.8893. DT also improves notably from 0.5724 to 0.7439, showing that class balancing strongly helps weaker models. NB shows a smaller change, moving from 0.7298 to 0.7593, but still becomes more stable. Overall, QSMOTE balances the data and gives consistent gains in accuracy, precision, recall, and F1-score, with SVM and RF emerging as the best performers.
The results in Table~\ref{table:performance_smote_ten_2} show that QSMOTE consistently improves the performance of most classifiers on the deca-class dataset of CWRUBD. For LR, the accuracy rises from $0.8783$ to $0.9332 $, with precision increasing from $0.8815$ to $0.9350$, recall from $0.8783 $ to $0.9332 $, and F1-score from $0.8654 $ to $0.9328 $. RF achieves the highest performance, as accuracy improves from $0.9483 $ to $0.9848 $, and all other metrics exceed $0.984$. Similarly, SVM improves in accuracy from $0.8797 $ to $0.9522 $, and in F1-score from $0.8664 $ to $0.9518 $. In contrast, NB shows a decline after QSMOTE, with accuracy dropping from $0.9245 $ to $0.8875$ and F1-score decreasing from $0.9229$ to $0.8856 $, indicating possible oversensitivity to oversampling. DT benefits substantially, with accuracy increasing from $0.9077 $ to $0.9750 $, and F1-score improving from $0.9072 $ to $0.9749 $. QSMOTE improves categorization across most models, with RF and DT performing near-optimally.

Table~\ref{table:performance_smote_four_third_3} for EFDD demonstrates that class imbalance severely limited performance prior to resampling, with LR and SVM both at an accuracy of $0.5337 $ (precision $0.3003 $ and $0.2848 $, respectively), RF at $0.4923 $, NB at $0.5156 $, and DT markedly lower at $0.2979 $. After applying QSMOTE, there is a pronounced divergence across model families: RF achieves the best overall performance with accuracy $0.9126 $, precision $0.9142 $, recall $0.9126 $, and F1 $0.9116 $, corresponding to an absolute accuracy gain of $+0.4203$ (relative $\approx 85.4\%$) over its pre-QSMOTE baseline. DT rises to $0.8228 $ accuracy (F1 $0.8050 $), an absolute improvement of $+0.5249$ (relative $\approx 176.2\%$). SVM improves more moderately to $0.6420 $ accuracy (absolute $+0.1083$, relative $\approx 20.3\%$). In contrast, LR and NB deteriorate post-QSMOTE (LR to $0.2622 $, absolute $-0.2715$, relative $\approx -50.9\%$; NB to $0.3362 $, absolute $-0.1794$, relative $\approx -34.8\%$), indicating that oversampling benefited non-linear, tree-based methods substantially while offering limited or negative returns for linear/generative models under these features and class overlaps.
Table \ref{table:performance_smote_hexa_fifth_5} quantitatively compares the performance of classical classifiers before and after applying QSMOTE for IFDD. Before QSMOTE, the highest accuracy is achieved by the SVM classifier (0.6814), followed closely by RF (0.6763) and LR (0.6712). DT and NB perform relatively poorly with accuracies of 0.4685 and 0.3477, respectively. After applying QSMOTE, a significant improvement is observed across most classifiers, especially RF and DT, which attained accuracies of 0.9919  and 0.9006, respectively. The SVM also show a considerable gain to 0.8550, while LR and NB experience marginal changes, with accuracies of 0.4450  and 0.4475, respectively. In terms of overall enhancement, RF exhibits an impressive +46.7\% point increase in accuracy, and DT improves by +43.2 points, confirming their ability to leverage balanced samples effectively. The SVM improves by +17.4 points, but the LR and NB decrease by about -22.6 and -10.0 points, respectively. Oversampling is particularly effective for ensemble and tree-based models, resulting in near-perfect classification performance. Linear and probabilistic models are nonetheless susceptible to synthetic class distributions.

\subsection{Robustness of Noisy-QSMOTE}

\begin{figure}
    \centering
    \begin{subfigure}{0.32\linewidth}
        \centering
        \includegraphics[width=\linewidth]{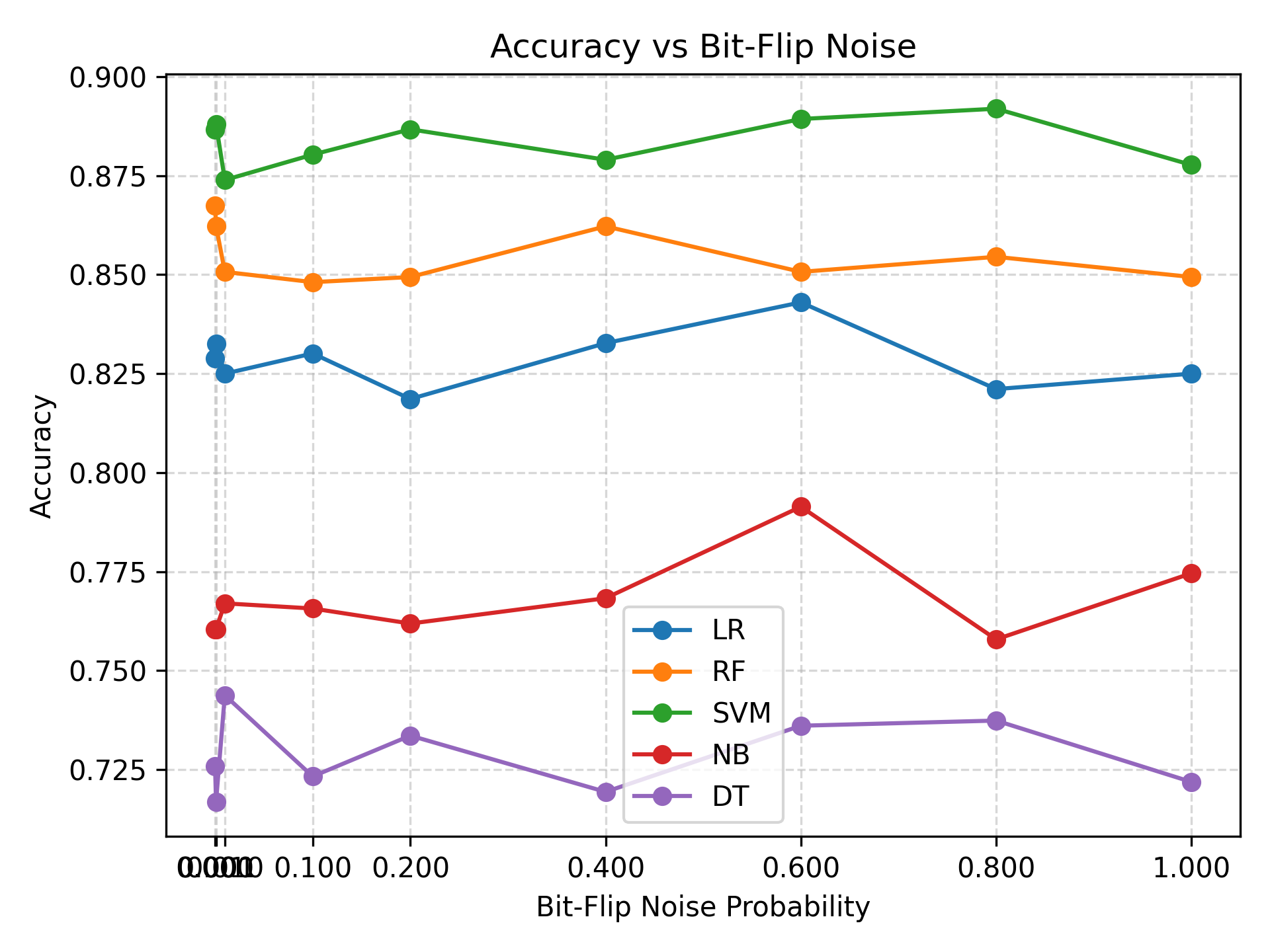} 
        \caption{}
        \label{fig:h_data_pca}
    \end{subfigure}
    \hfill
    \begin{subfigure}{0.32\linewidth}
        \centering
        \includegraphics[width=\linewidth]{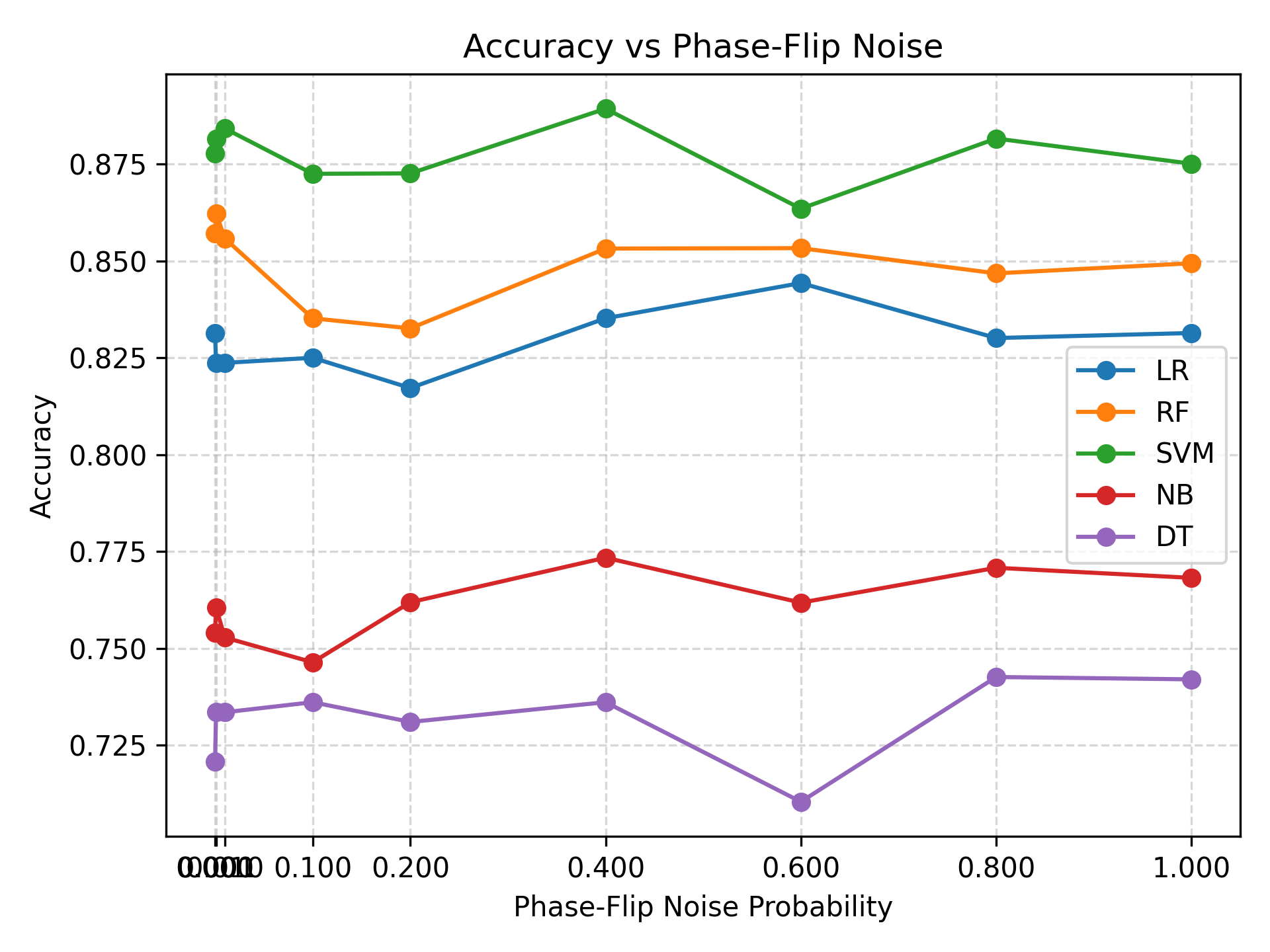}
        \caption{}
        \label{fig:m_data_pca}
    \end{subfigure}\hfill
    \begin{subfigure}{0.32\linewidth}
        \centering
        \includegraphics[width=\linewidth]{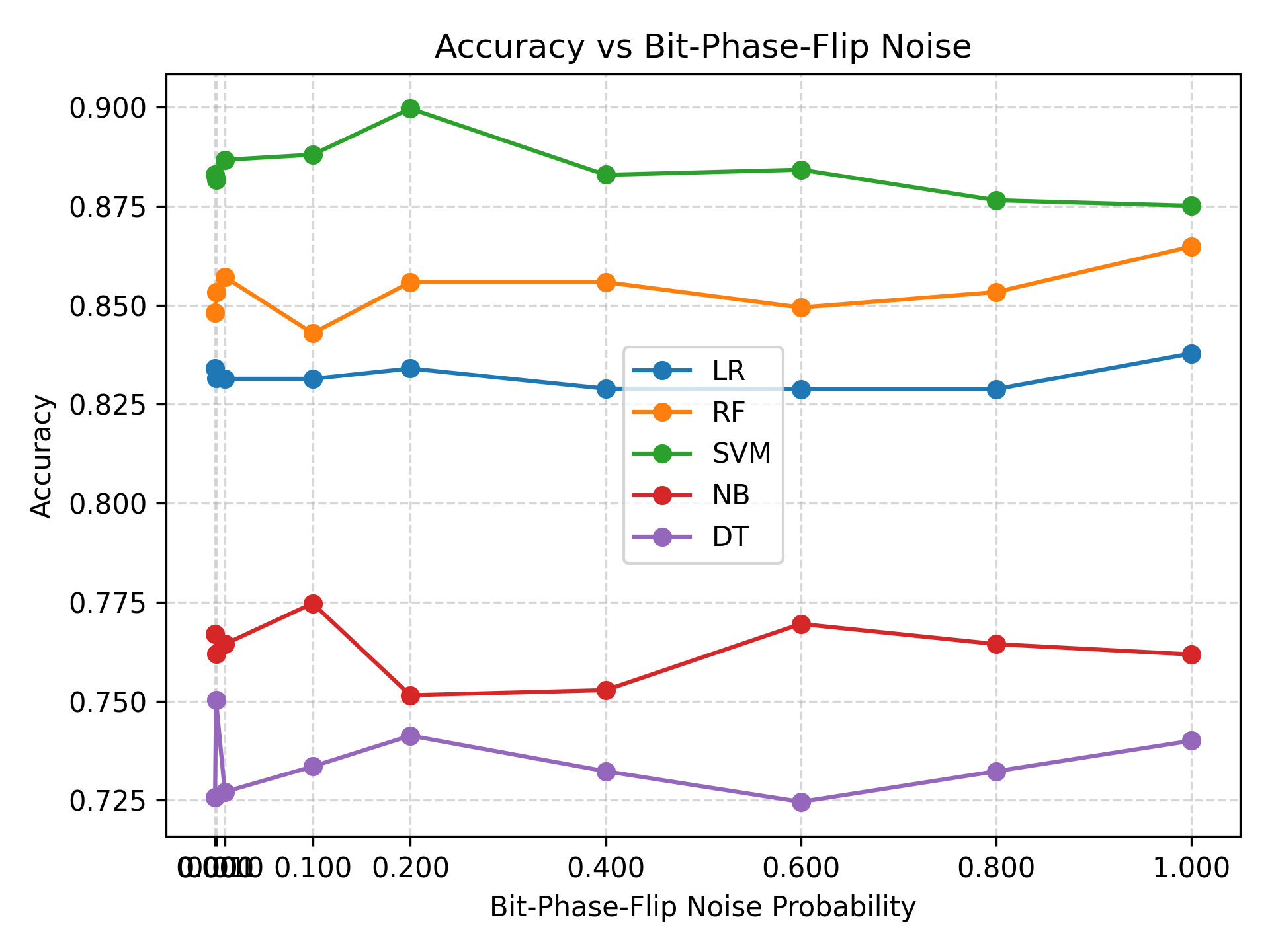} 
        \caption{}
        \label{fig:h_data_pca}
    \end{subfigure}
    \hfill
    \begin{subfigure}{0.32\linewidth}
        \centering
        \includegraphics[width=\linewidth]{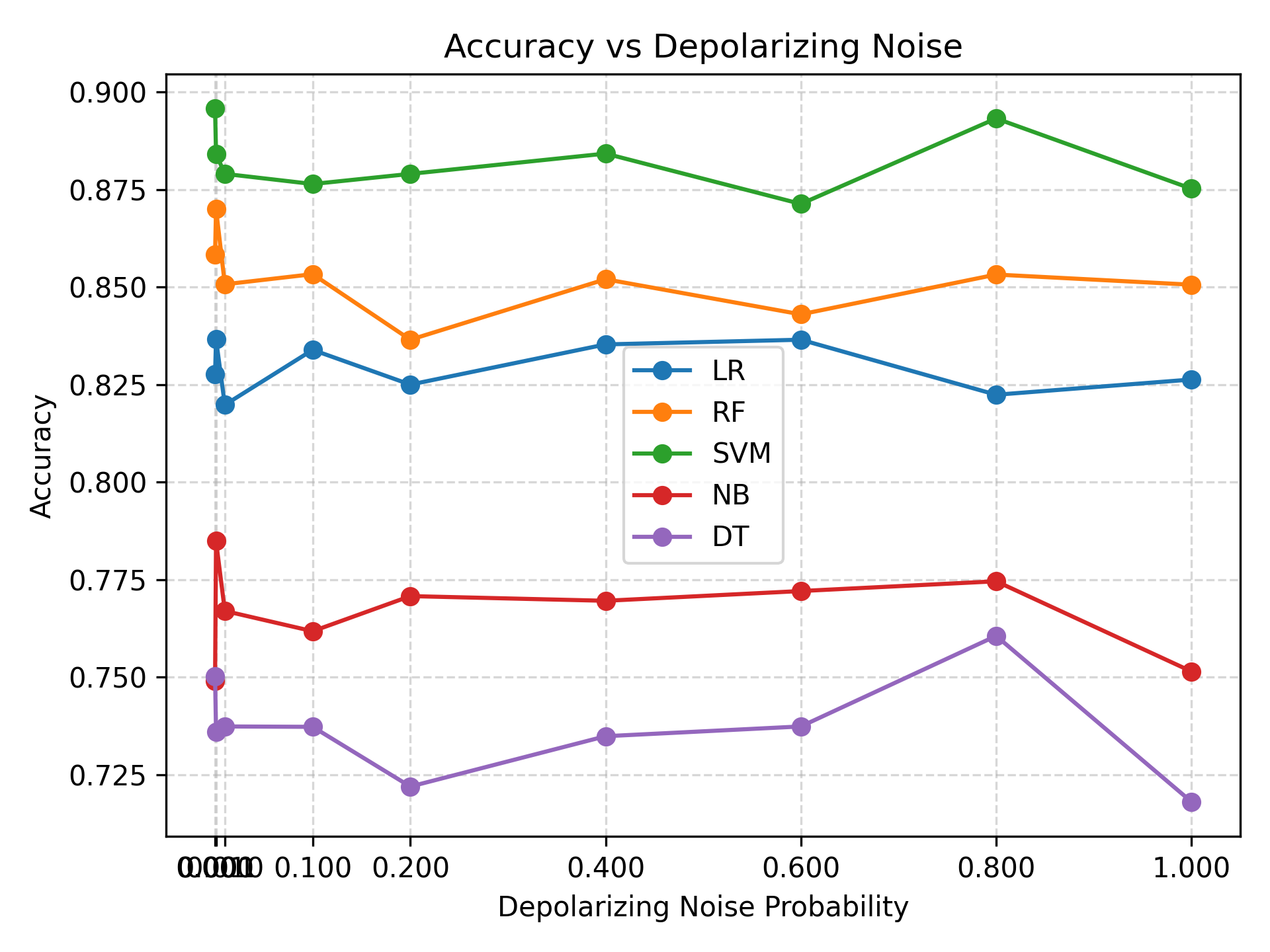} 
        \caption{}
        \label{fig:h_data_pca}
    \end{subfigure}
    \hfill
    \begin{subfigure}{0.32\linewidth}
        \centering
        \includegraphics[width=\linewidth]{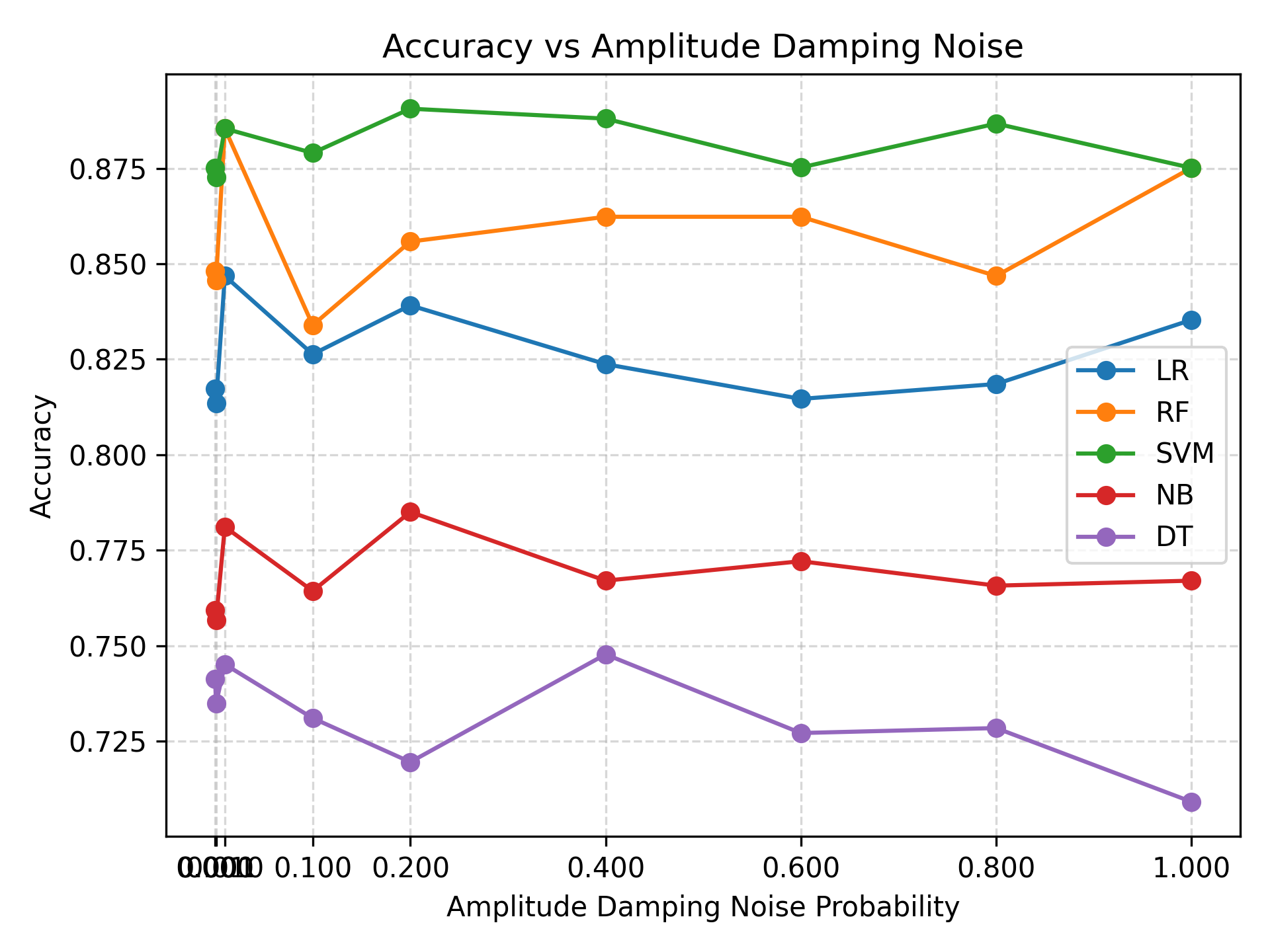} 
        \caption{}
        \label{fig:h_data_pca}
    \end{subfigure}
    \hfill
    \begin{subfigure}{0.32\linewidth}
        \centering
        \includegraphics[width=\linewidth]{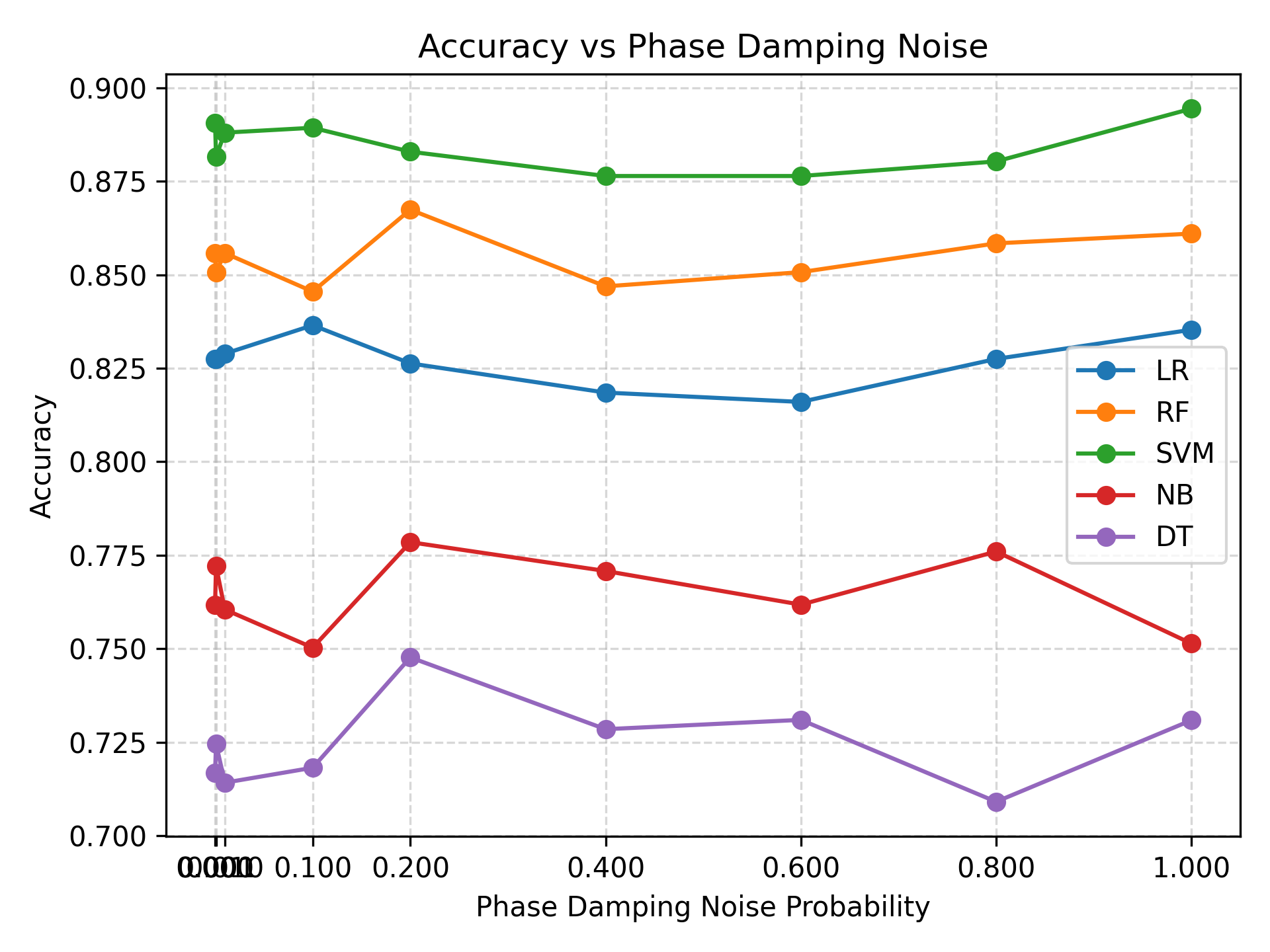} 
        \caption{}
        \label{fig:h_data_pca}
    \end{subfigure}
    \caption{Noisy-QSMOTE results for SPID. a) BF, b) PF, c) BPF, d) DP, e) AD, and f) PD.}
    \label{fig:noisy_d1}
\end{figure}

\begin{figure}
    \centering
    \begin{subfigure}{0.32\linewidth}
        \centering
        \includegraphics[width=\linewidth]{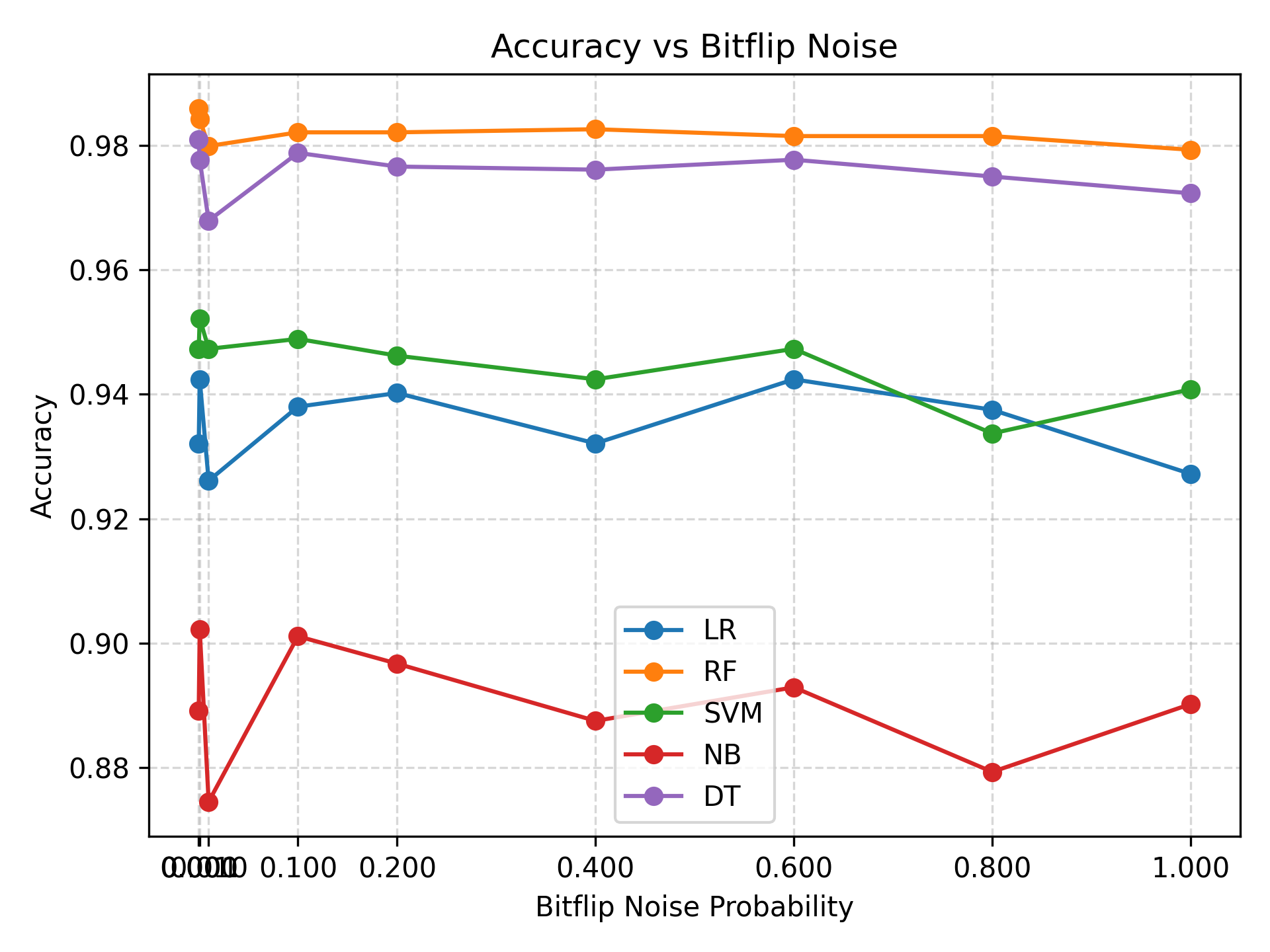} 
        \caption{}
        \label{fig:h_data_pca}
    \end{subfigure}
    \hfill
    \begin{subfigure}{0.32\linewidth}
        \centering
        \includegraphics[width=\linewidth]{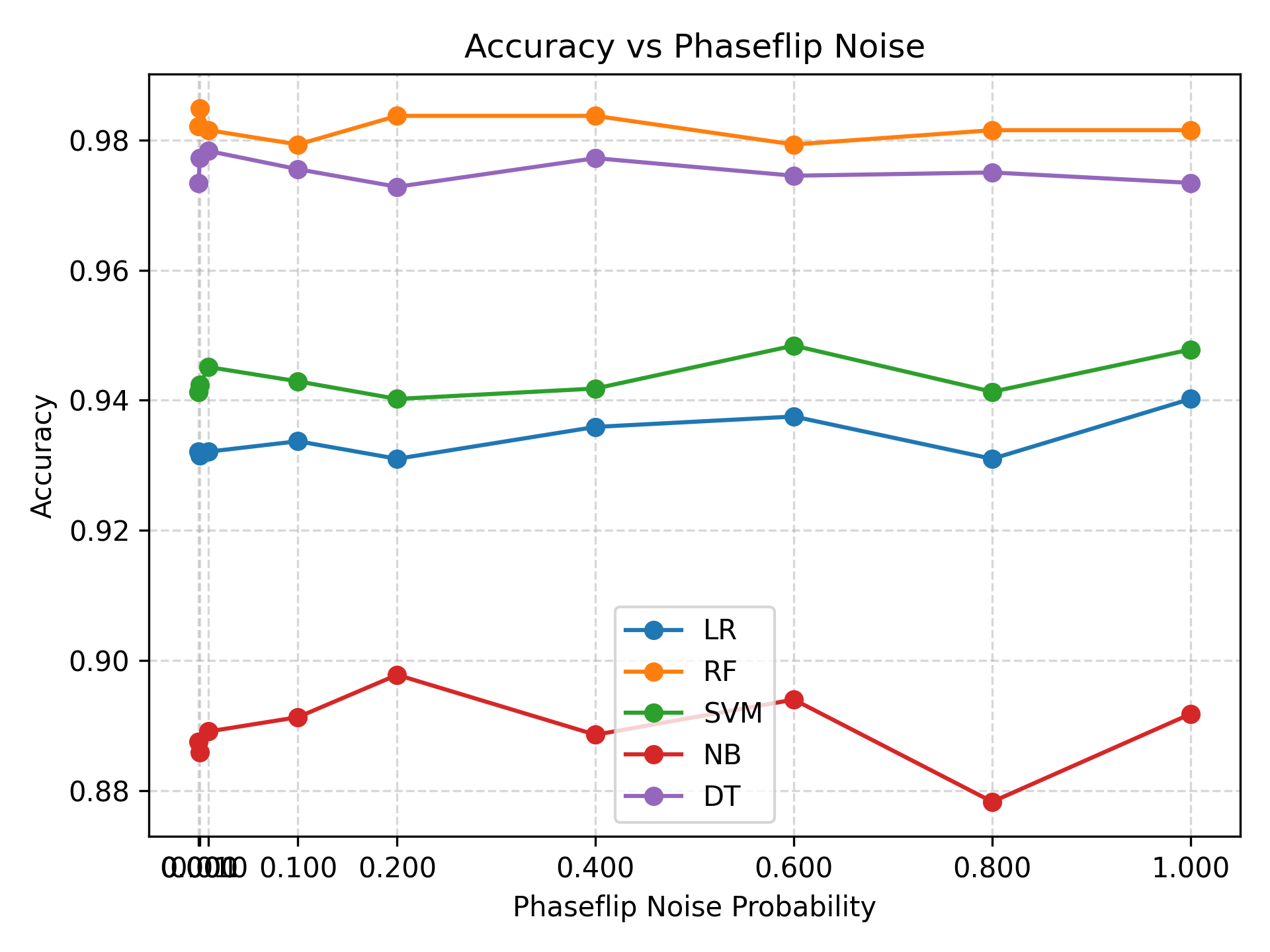}
        \caption{}
        \label{fig:m_data_pca}
    \end{subfigure}\hfill
    \begin{subfigure}{0.32\linewidth}
        \centering
        \includegraphics[width=\linewidth]{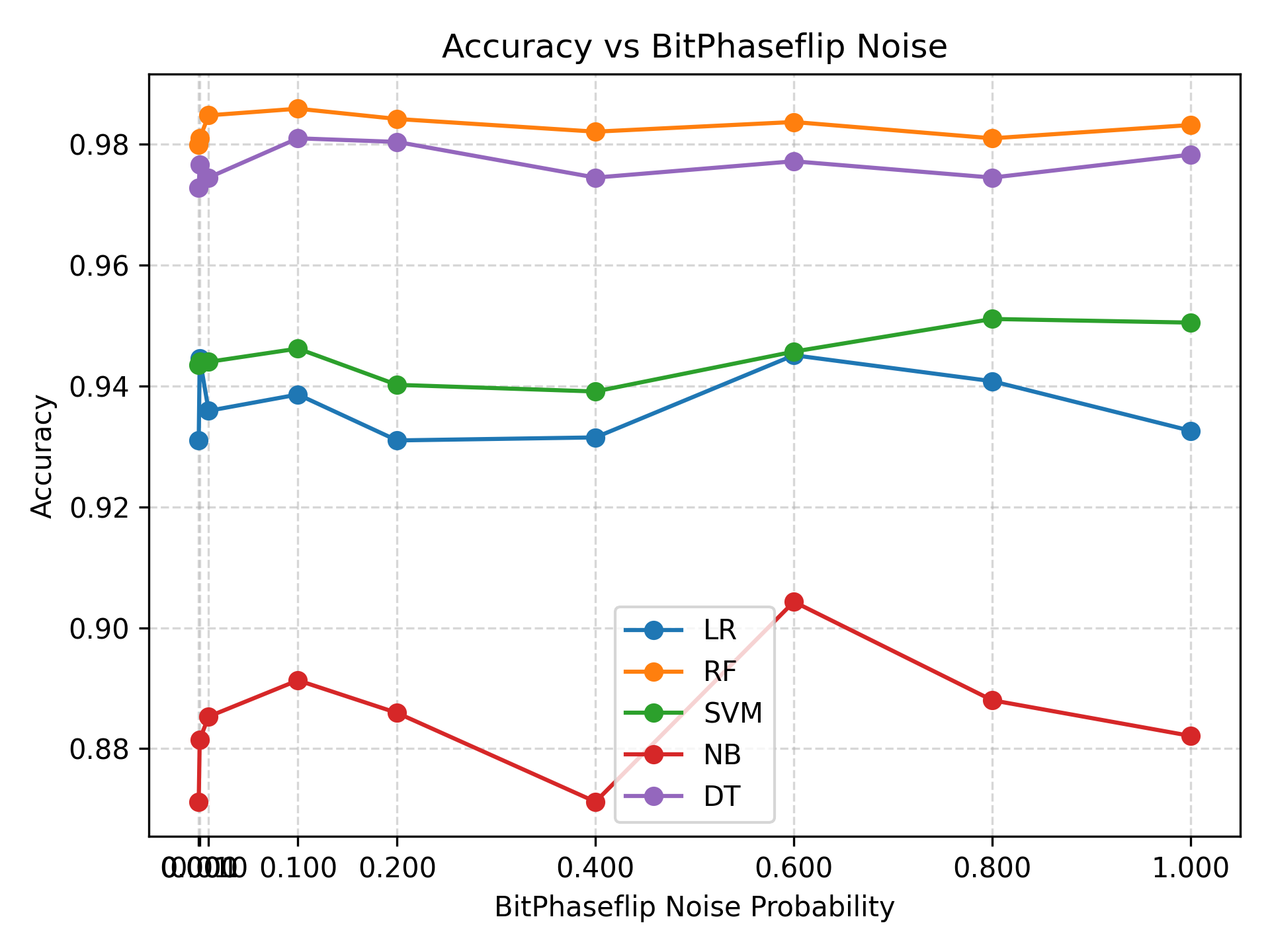} 
        \caption{}
        \label{fig:h_data_pca}
    \end{subfigure}
    \hfill
    \begin{subfigure}{0.32\linewidth}
        \centering
        \includegraphics[width=\linewidth]{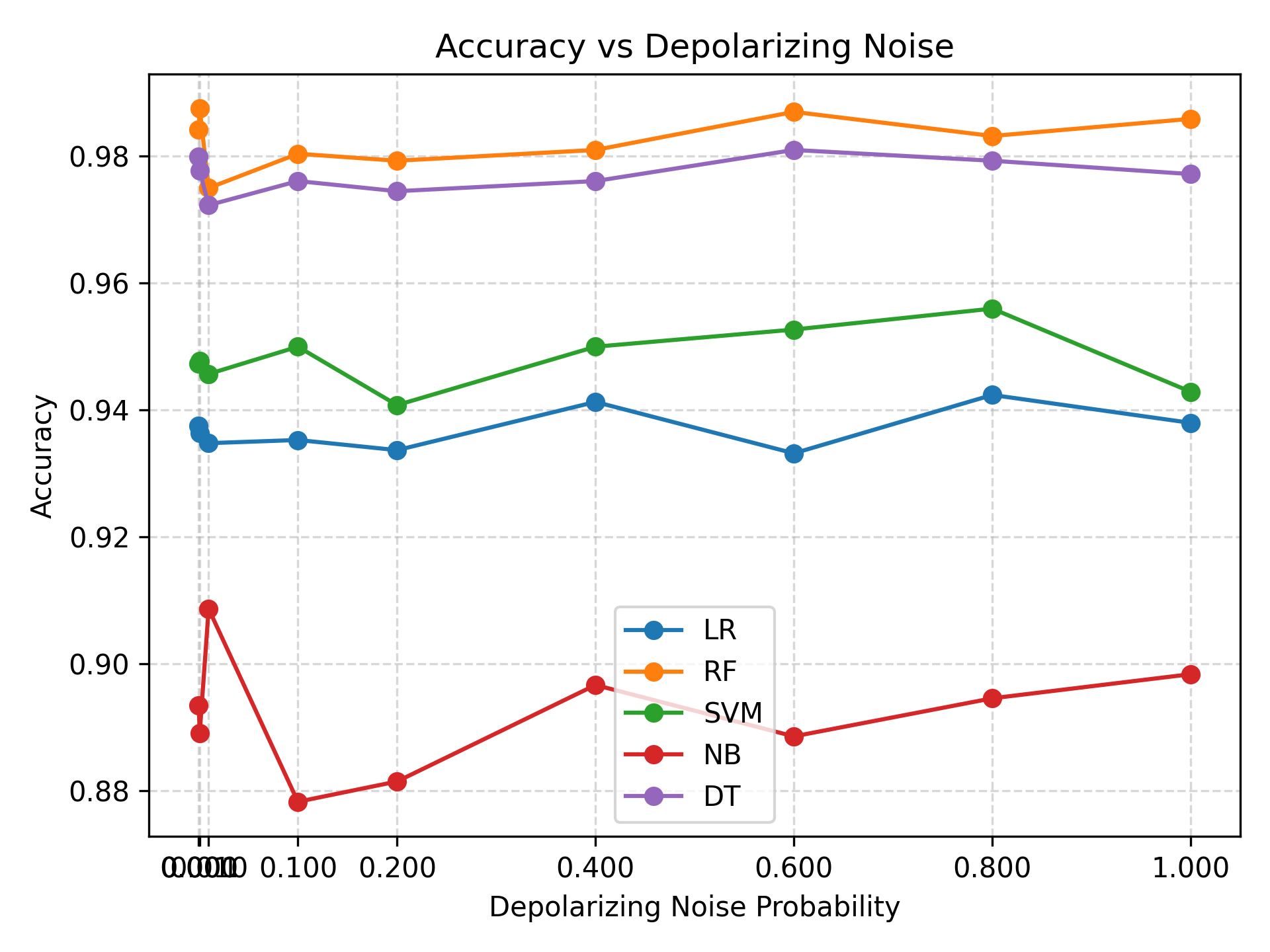} 
        \caption{}
        \label{fig:h_data_pca}
    \end{subfigure}
    \hfill
    \begin{subfigure}{0.32\linewidth}
        \centering
        \includegraphics[width=\linewidth]{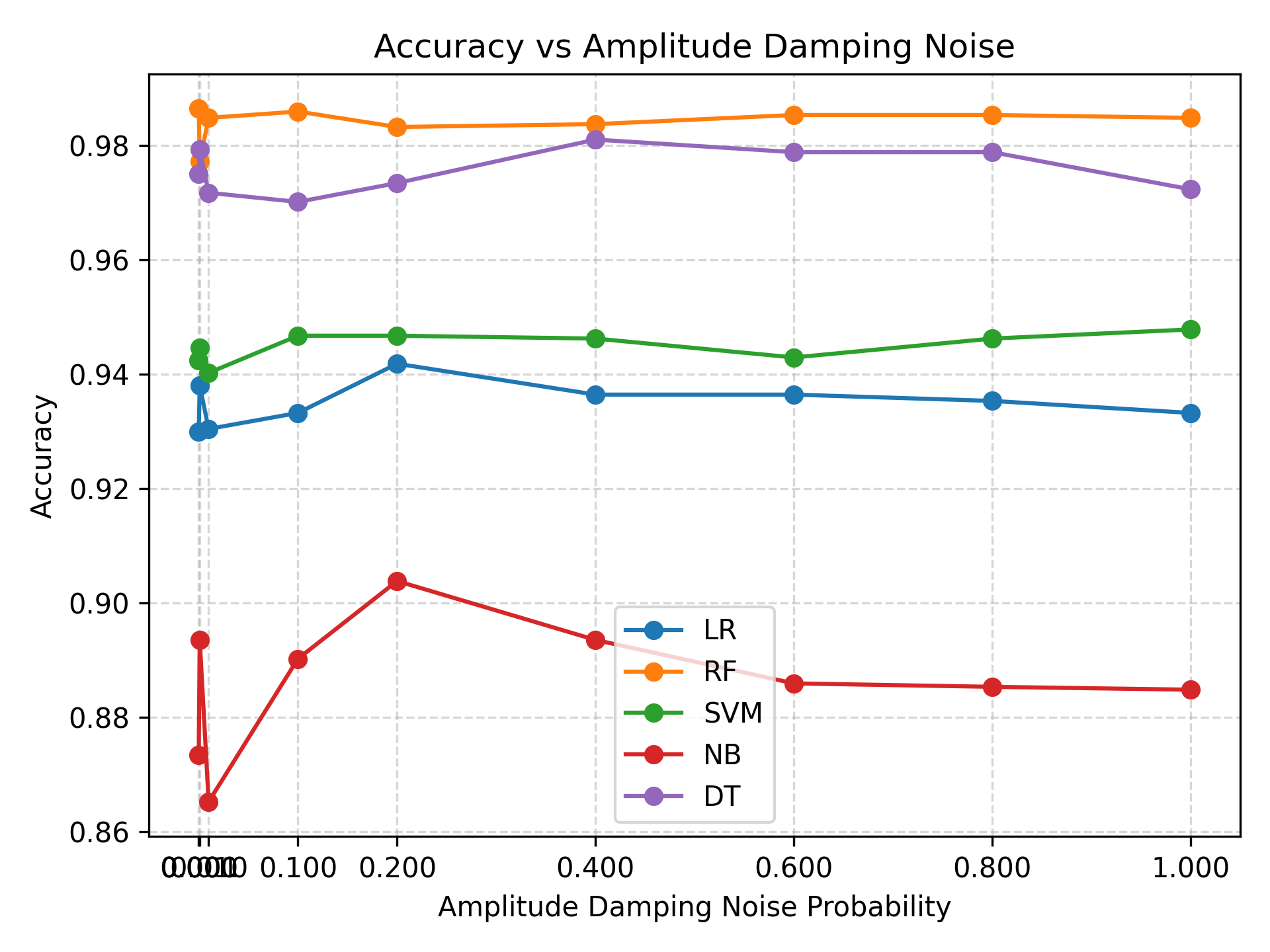} 
        \caption{}
        \label{fig:h_data_pca}
    \end{subfigure}
    \hfill
    \begin{subfigure}{0.32\linewidth}
        \centering
        \includegraphics[width=\linewidth]{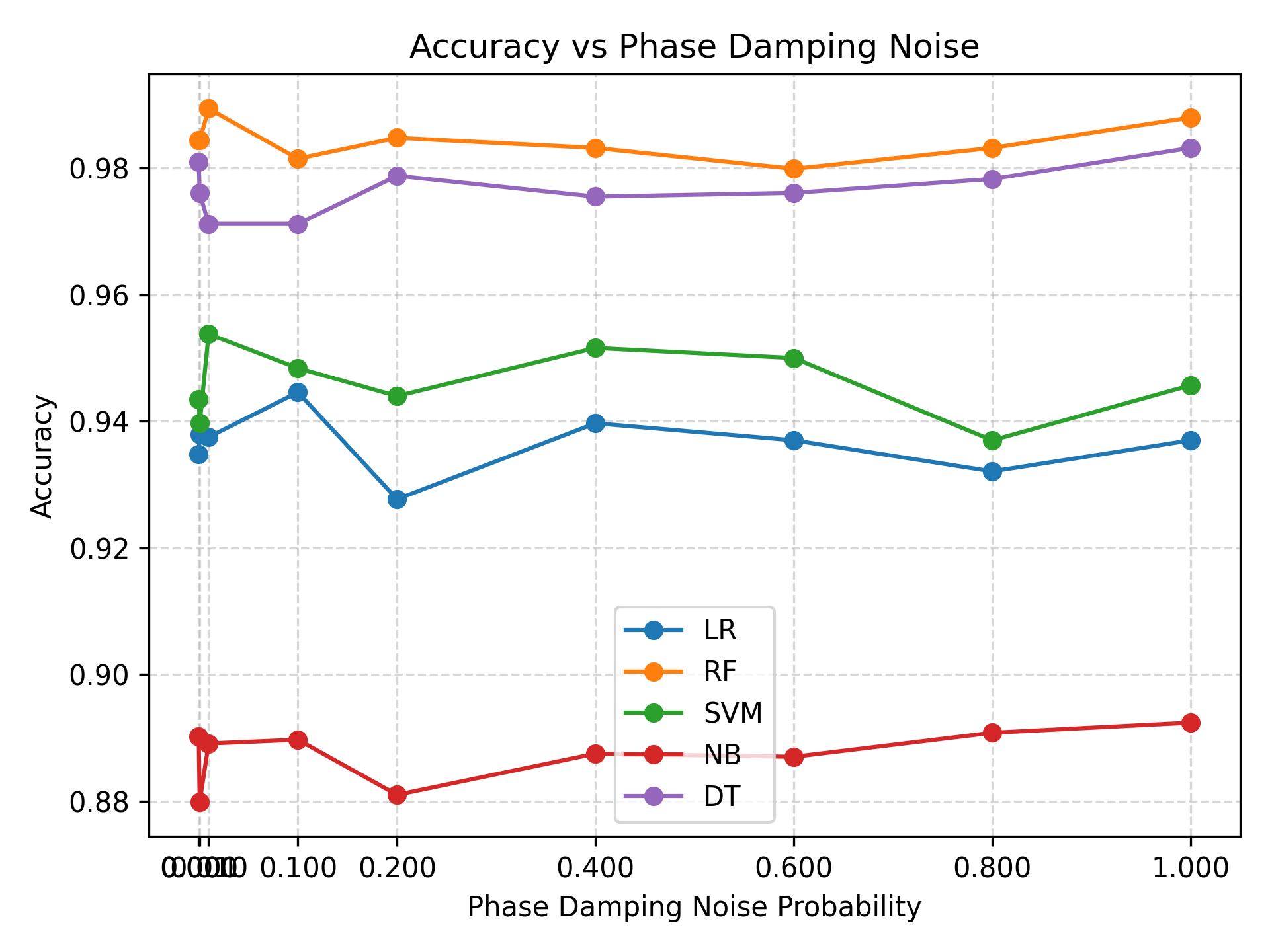} 
        \caption{}
        \label{fig:h_data_pca}
    \end{subfigure}
    \caption{Noisy-QSMOTE results for CWRUBD. a) BF, b) PF, c) BPF, d) DP, e) AD, and f) PD.}
    \label{fig:noisy_d2}
\end{figure}

\begin{figure}
    \centering
    \begin{subfigure}{0.32\linewidth}
        \centering
        \includegraphics[width=\linewidth]{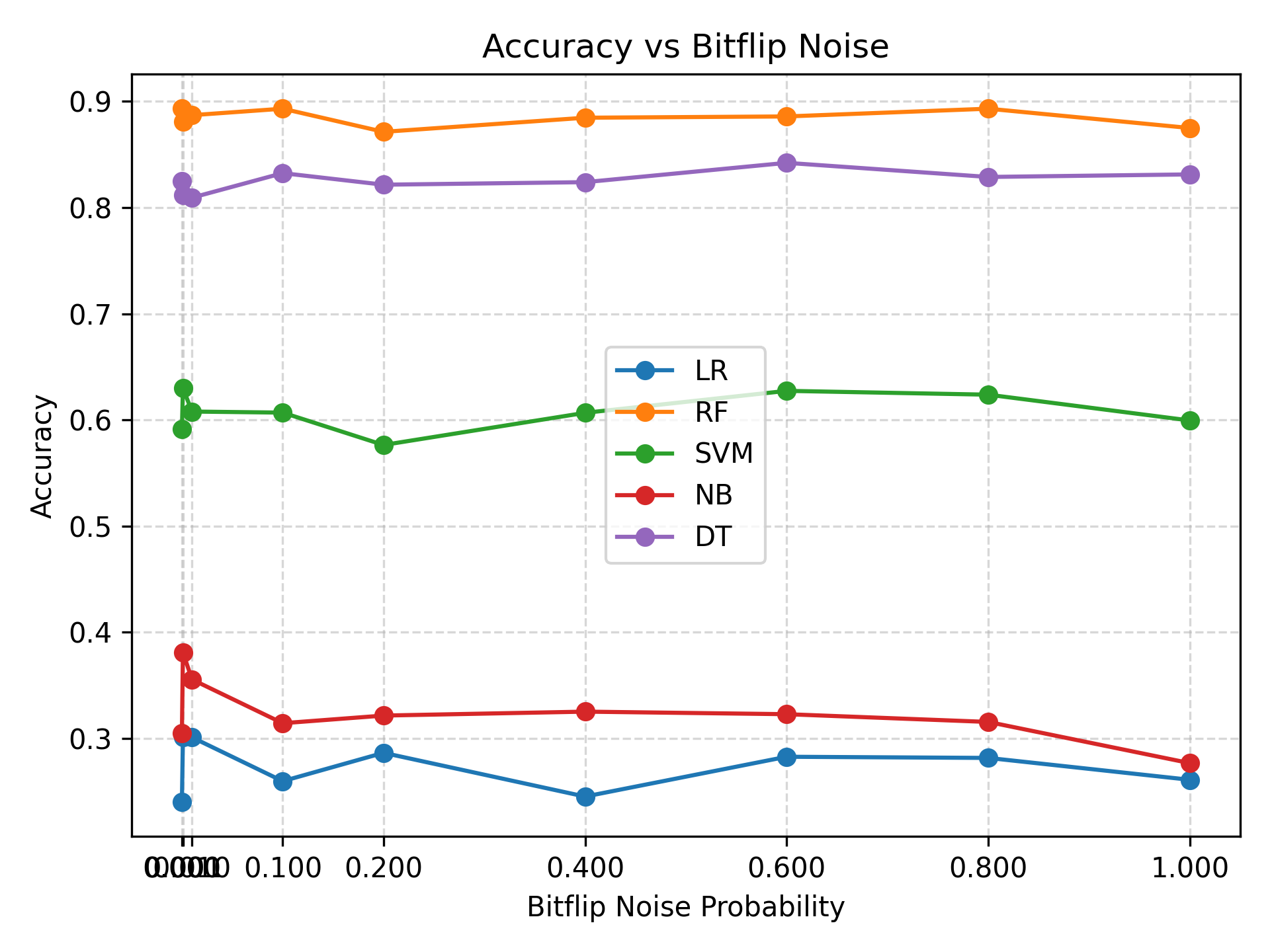} 
        \caption{}
        \label{fig:h_data_pca}
    \end{subfigure}
    \hfill
    \begin{subfigure}{0.32\linewidth}
        \centering
        \includegraphics[width=\linewidth]{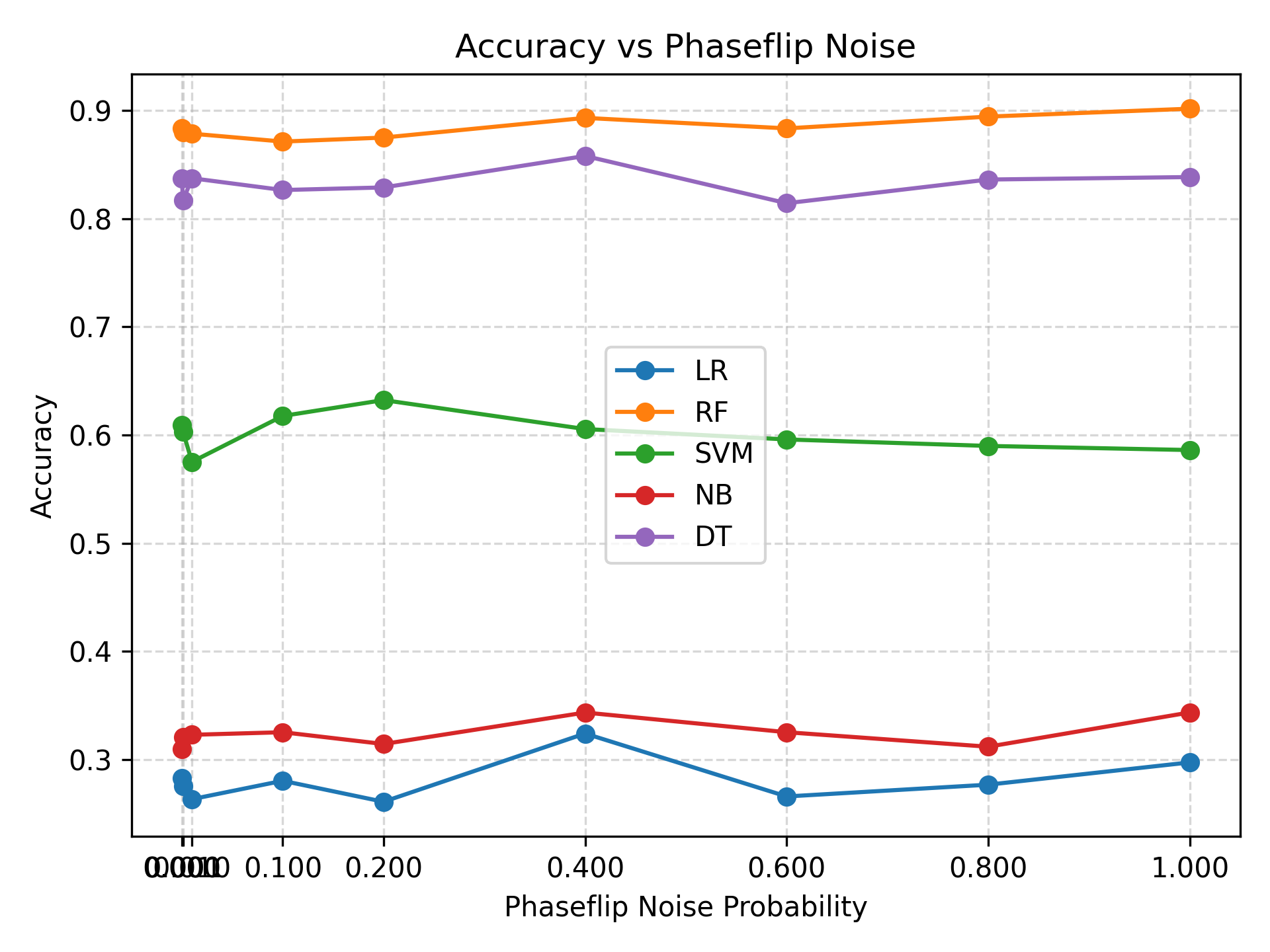}
        \caption{}
        \label{fig:m_data_pca}
    \end{subfigure}\hfill
    \begin{subfigure}{0.32\linewidth}
        \centering
        \includegraphics[width=\linewidth]{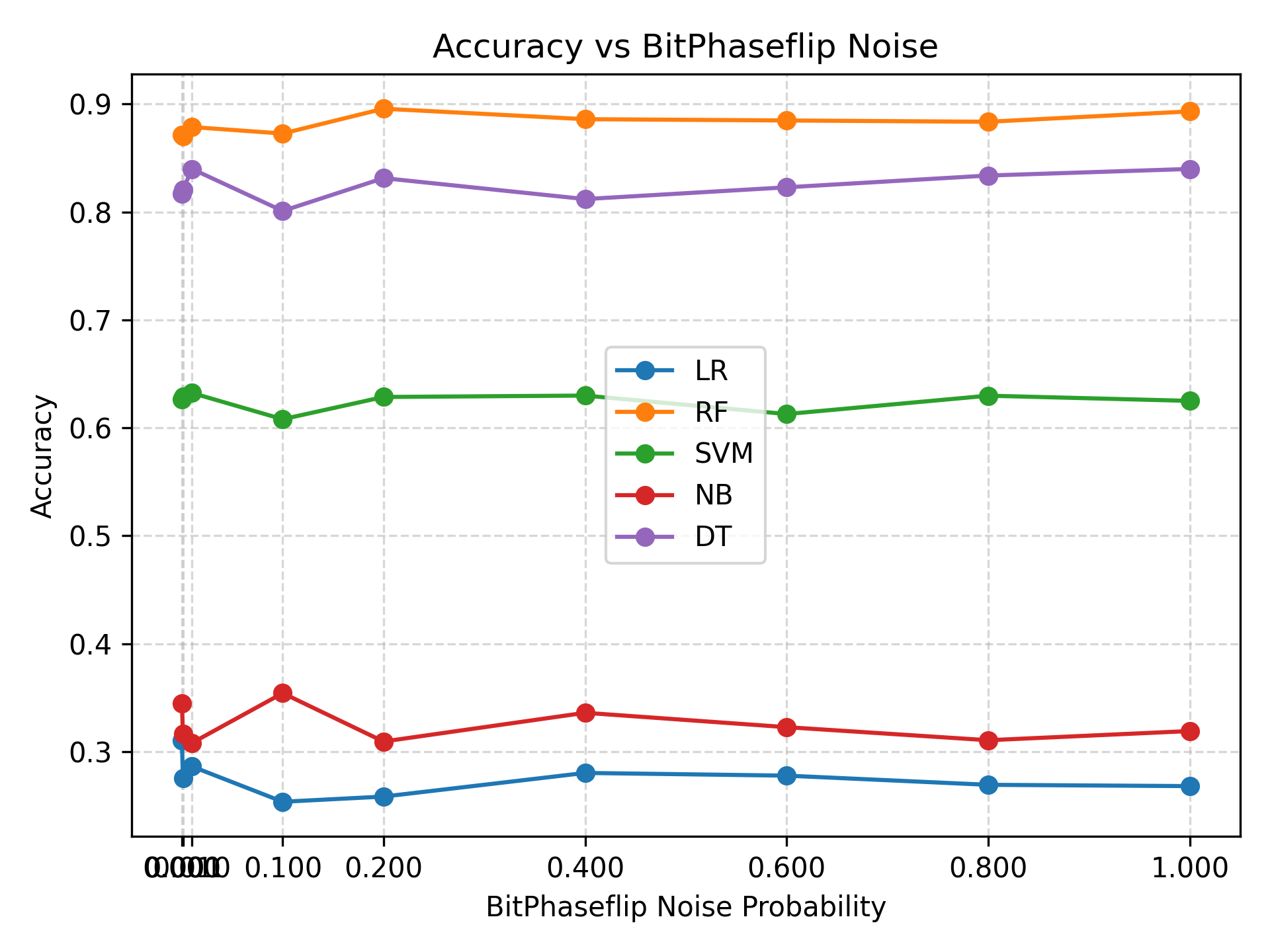} 
        \caption{}
        \label{fig:h_data_pca}
    \end{subfigure}
    \hfill
    \begin{subfigure}{0.32\linewidth}
        \centering
        \includegraphics[width=\linewidth]{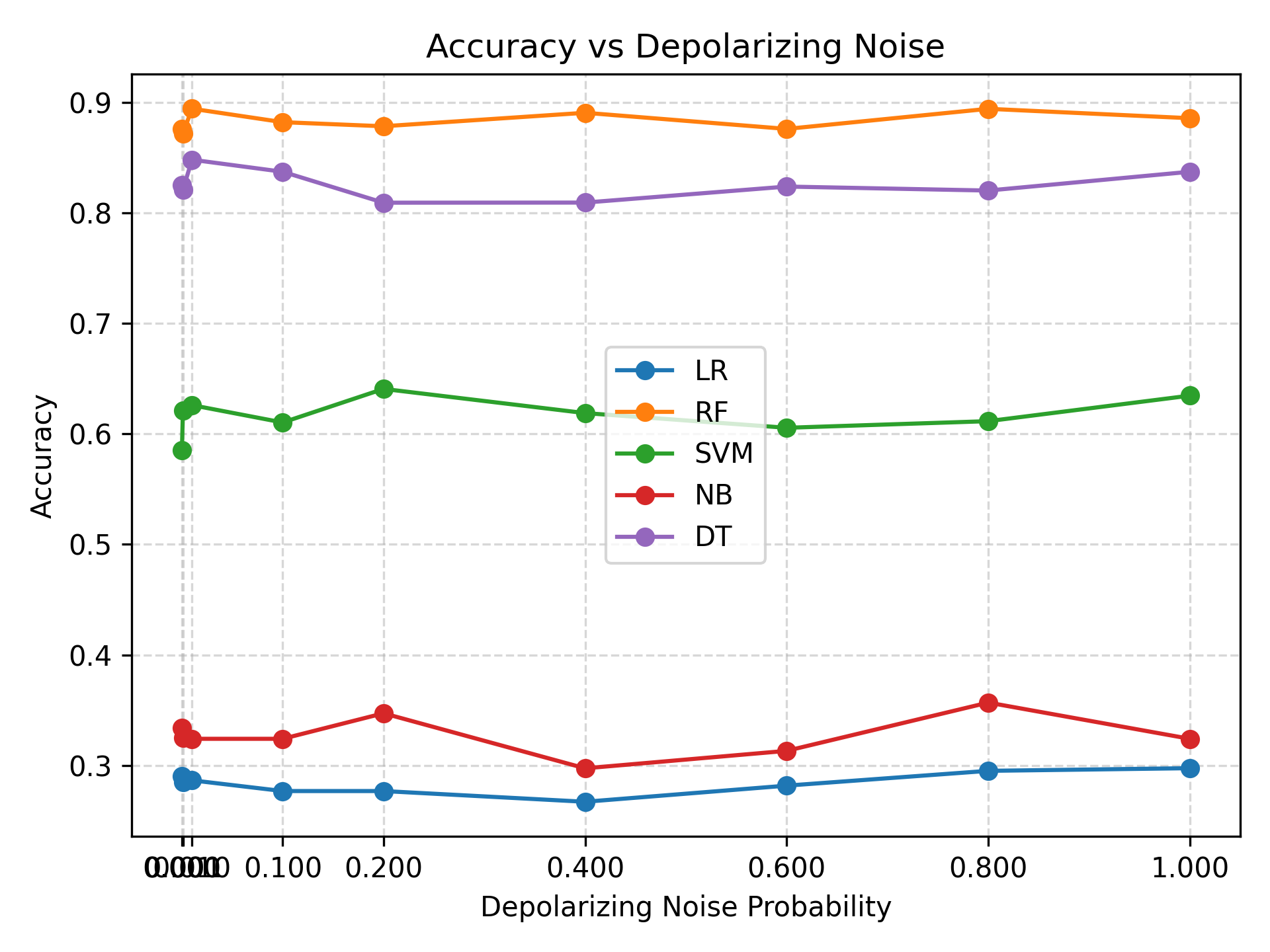} 
        \caption{}
        \label{fig:h_data_pca}
    \end{subfigure}
    \hfill
    \begin{subfigure}{0.32\linewidth}
        \centering
        \includegraphics[width=\linewidth]{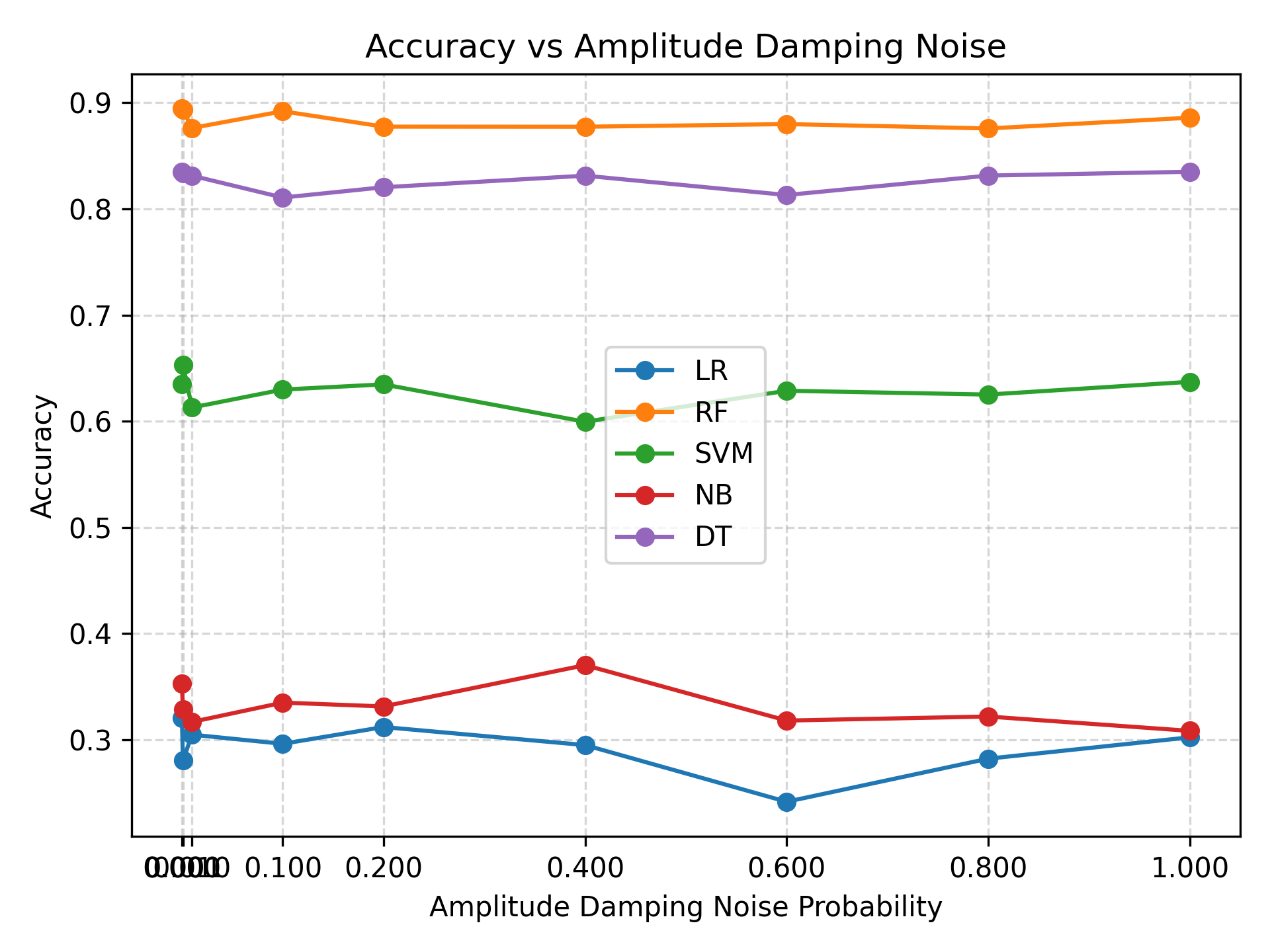} 
        \caption{}
        \label{fig:h_data_pca}
    \end{subfigure}
    \hfill
    \begin{subfigure}{0.32\linewidth}
        \centering
        \includegraphics[width=\linewidth]{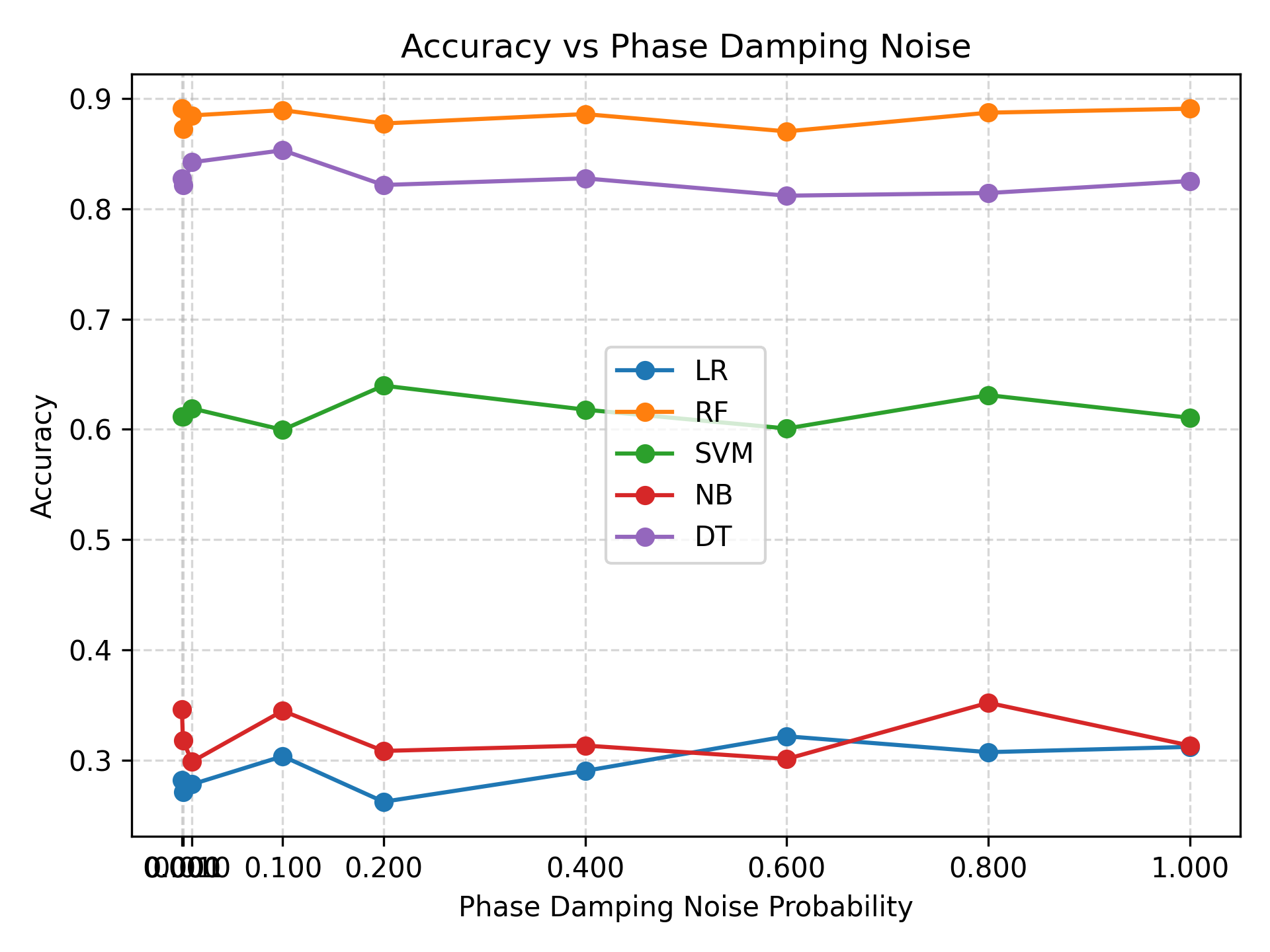} 
        \caption{}
        \label{fig:h_data_pca}
    \end{subfigure}
    \caption{Noisy-QSMOTE results for EFDD. a) BF, b) PF, c) BPF, d) DP, e) AD, and f) PD.}
    \label{fig:noisy_d3}
\end{figure}

\begin{figure}
    \centering
    \begin{subfigure}{0.32\linewidth}
        \centering
        \includegraphics[width=\linewidth]{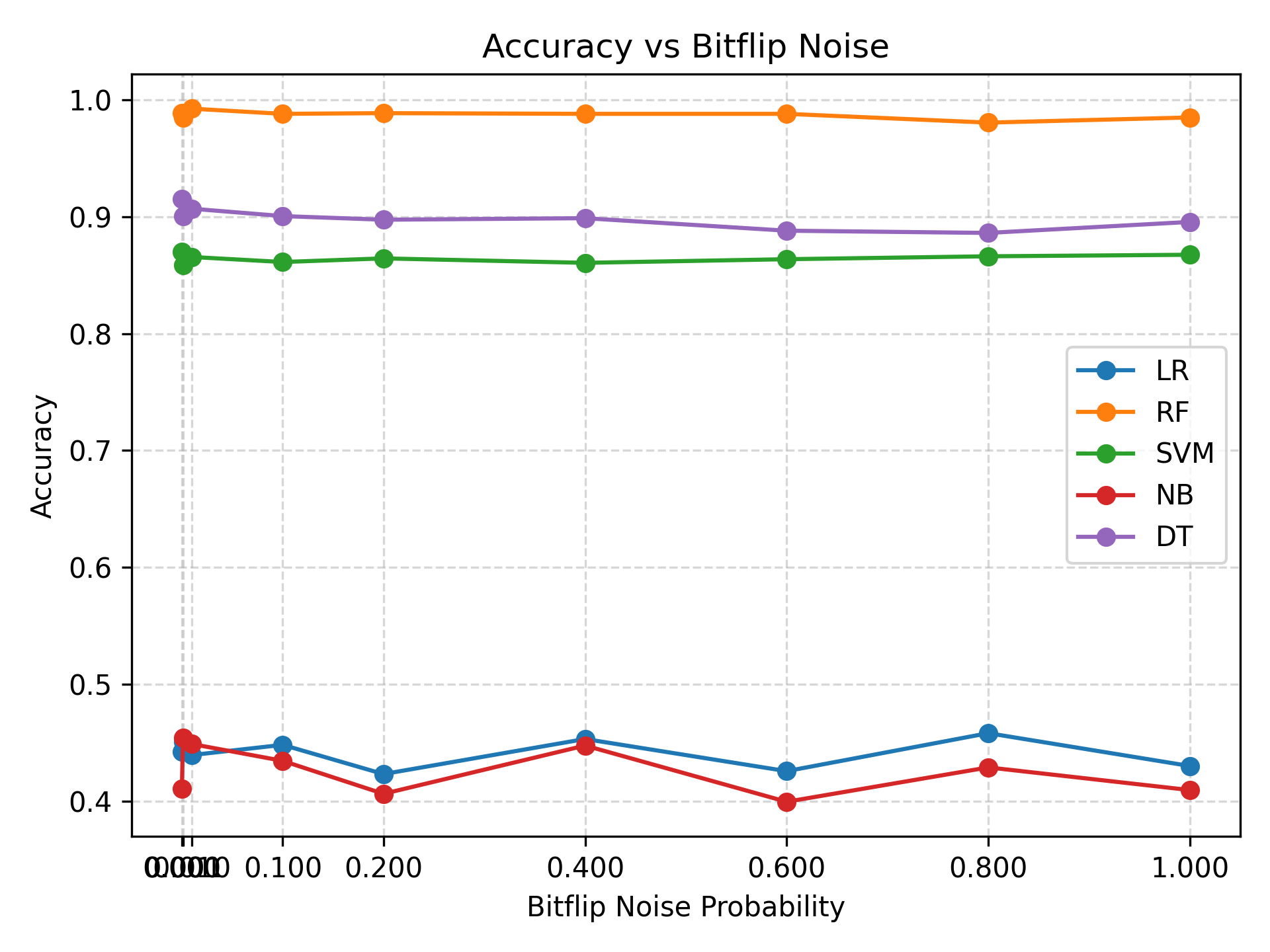} 
        \caption{}
        \label{fig:h_data_pca}
    \end{subfigure}
    \hfill
    \begin{subfigure}{0.32\linewidth}
        \centering
        \includegraphics[width=\linewidth]{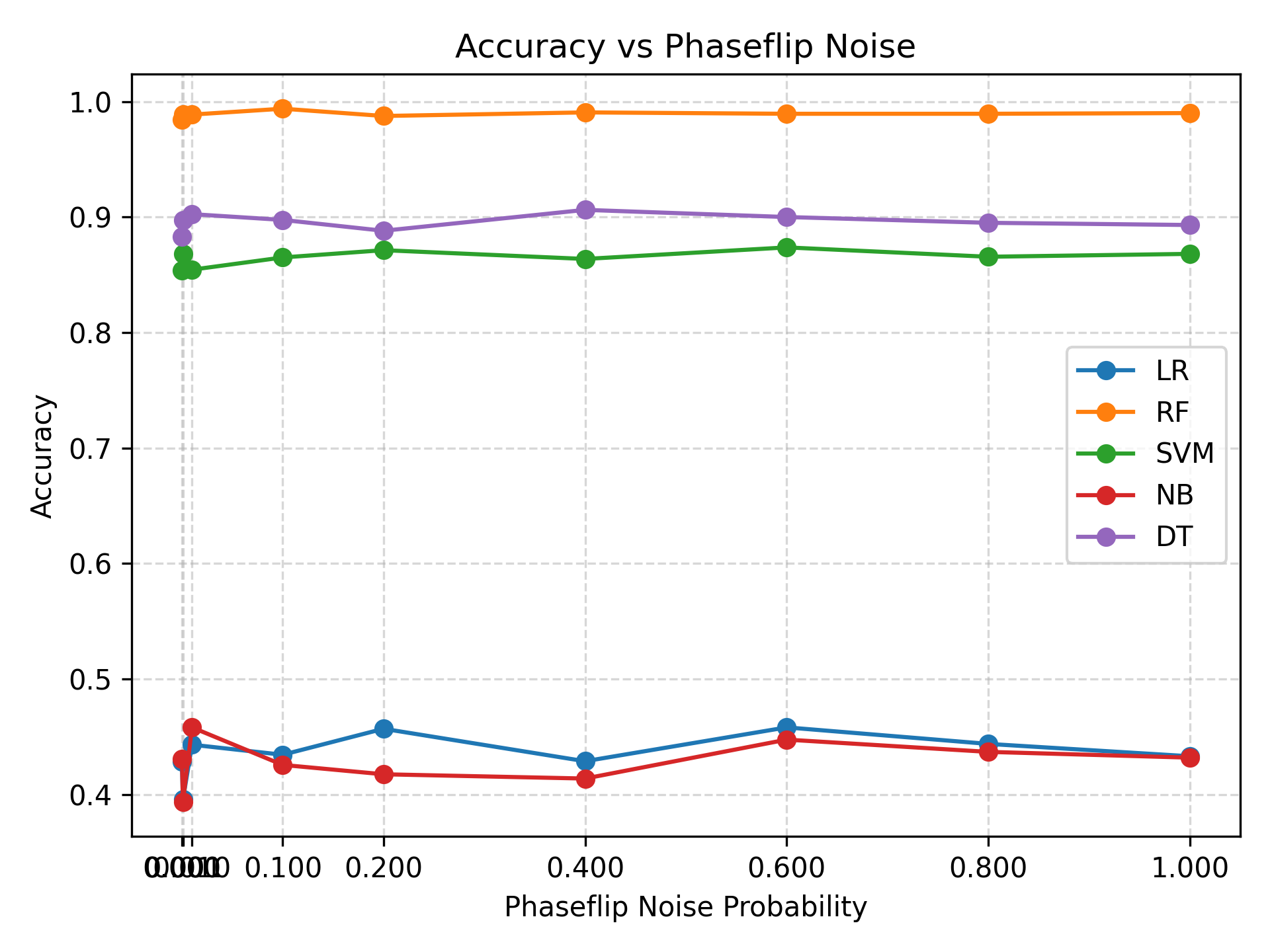}
        \caption{}
        \label{fig:m_data_pca}
    \end{subfigure}\hfill
    \begin{subfigure}{0.32\linewidth}
        \centering
        \includegraphics[width=\linewidth]{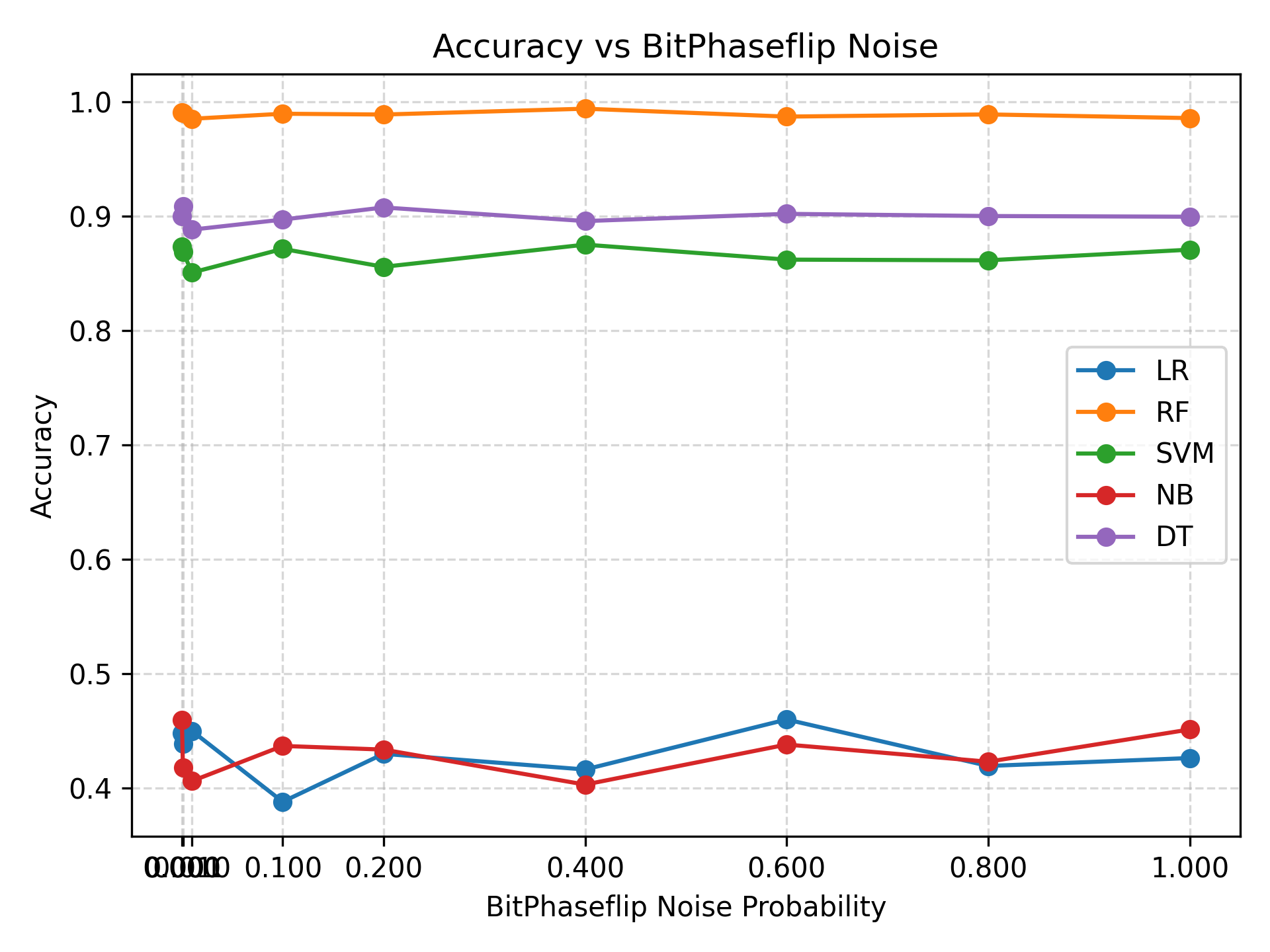} 
        \caption{}
        \label{fig:h_data_pca}
    \end{subfigure}
    \hfill
    \begin{subfigure}{0.32\linewidth}
        \centering
        \includegraphics[width=\linewidth]{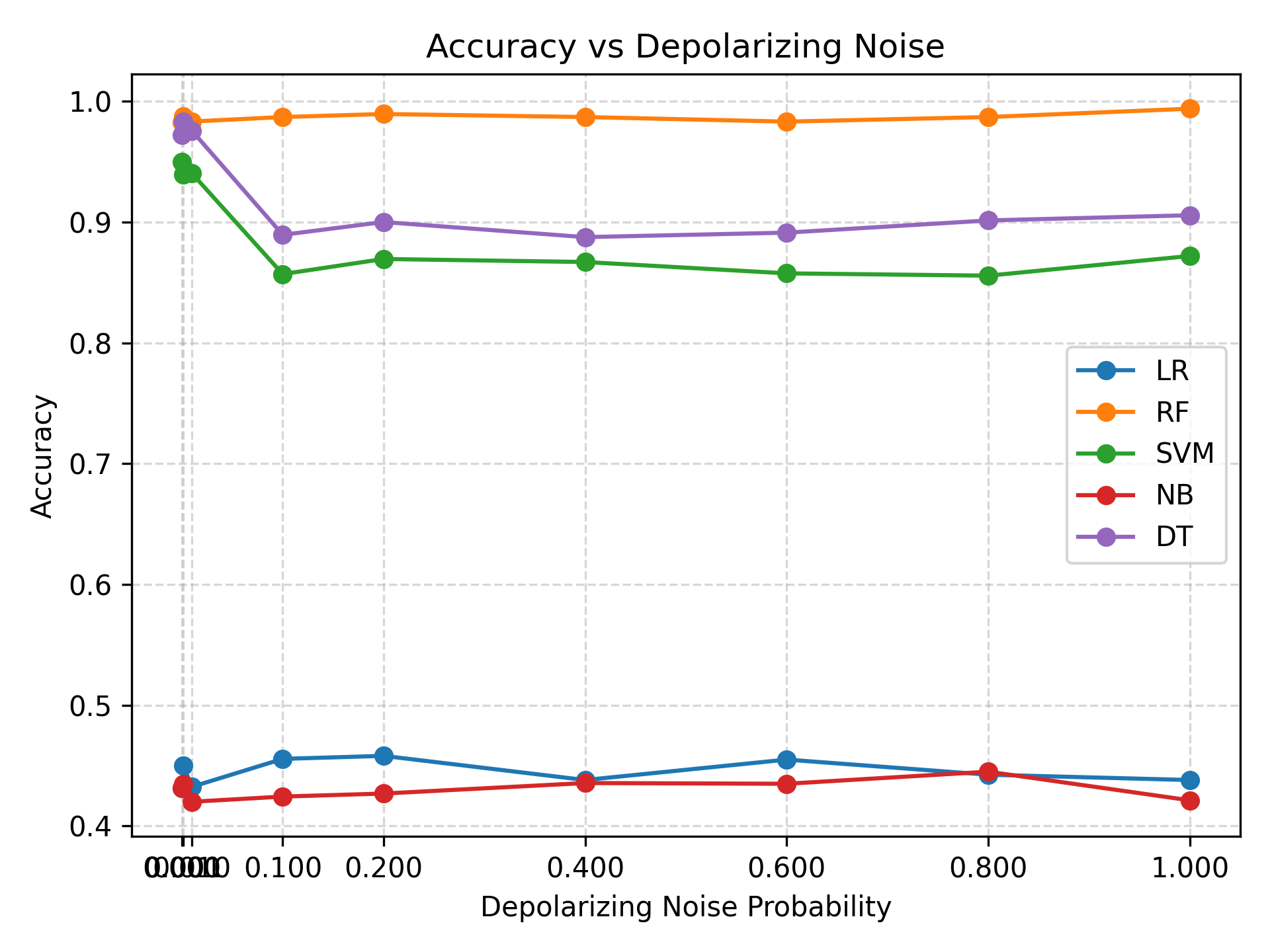} 
        \caption{}
        \label{fig:h_data_pca}
    \end{subfigure}
    \hfill
    \begin{subfigure}{0.32\linewidth}
        \centering
        \includegraphics[width=\linewidth]{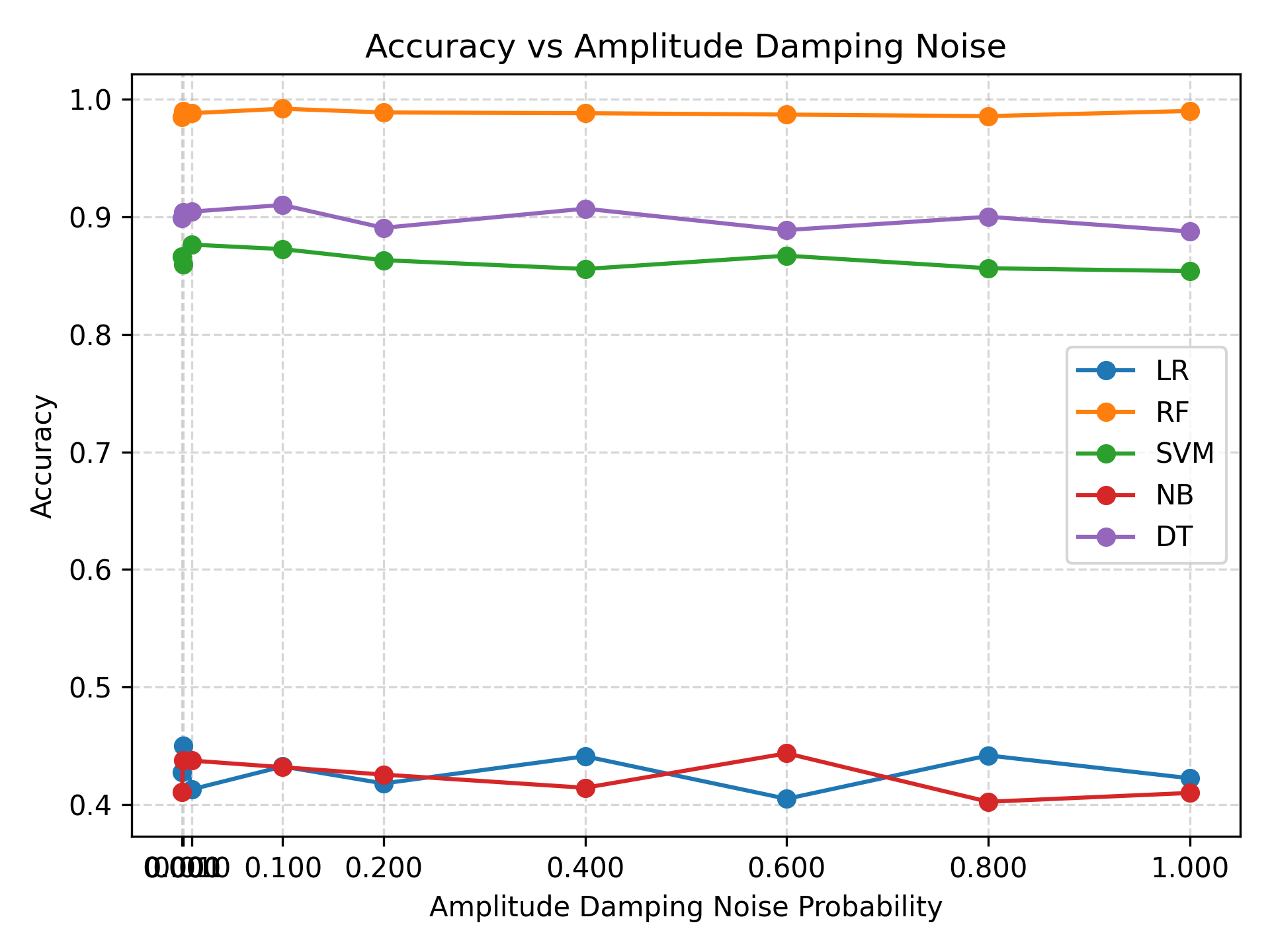} 
        \caption{}
        \label{fig:h_data_pca}
    \end{subfigure}
    \hfill
    \begin{subfigure}{0.32\linewidth}
        \centering
        \includegraphics[width=\linewidth]{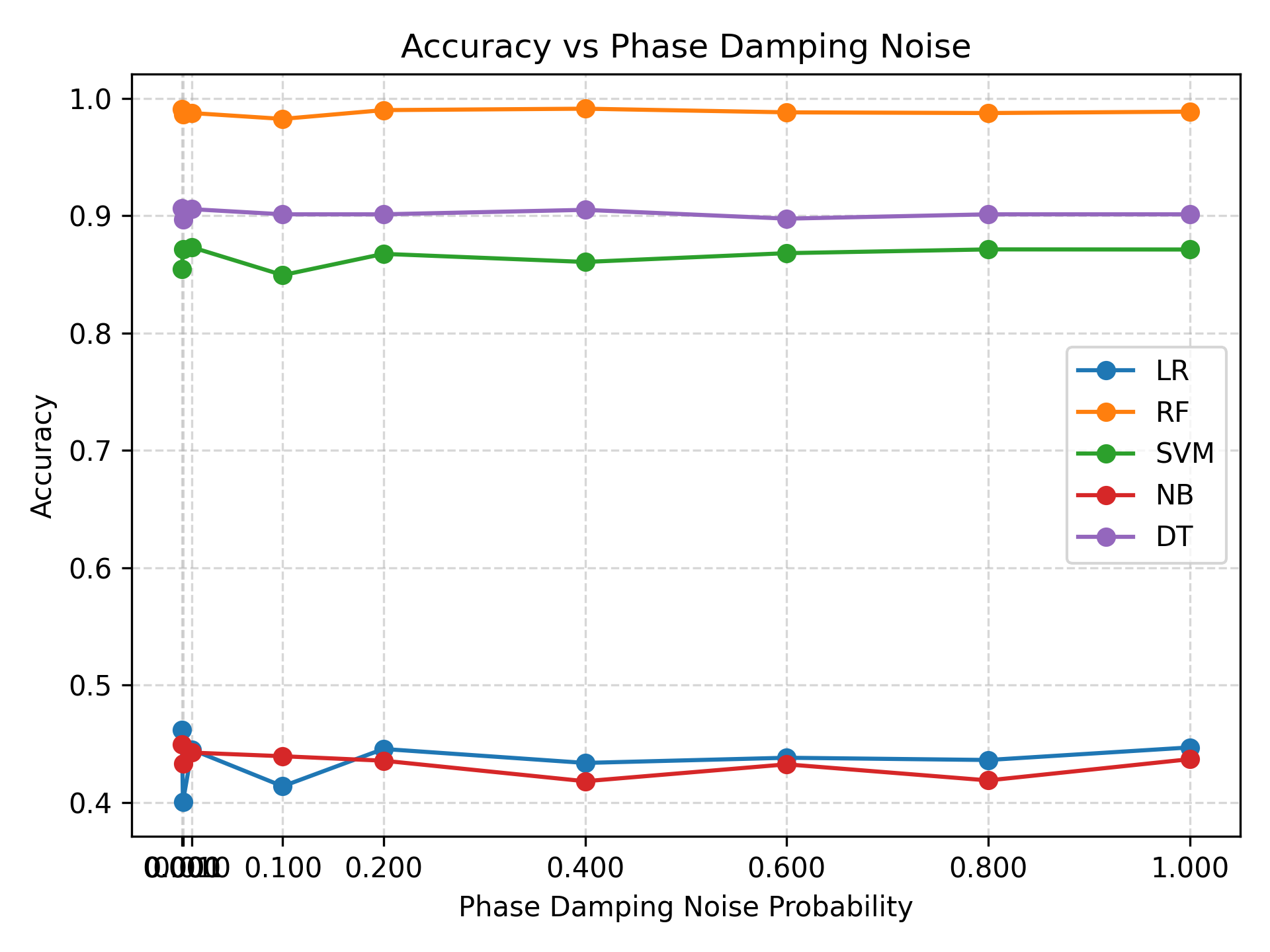} 
        \caption{}
        \label{fig:h_data_pca}
    \end{subfigure}
    \caption{Noisy-QSMOTE results for IFDD. a) BF, b) PF, c) BPF, d) DP, e) AD, and f) PD.}
    \label{fig:noisy_d4}
\end{figure}

In this section, the robustness of QSMOTE under quantum-inspired perturbations is investigated. The considered noise channels are injected directly into the compact swap test used during QSMOTE similarity estimation and synthetic sample generation. Consequently, the reported results quantify the downstream impact of noisy-QSMOTE-generated samples on classification performance. Fig. \ref{fig:noisy_d1} presents the classification accuracy of five ML models under varying strengths of six quantum noise channels. Across all noise types, BF, PF, BPF, DP, AD, and PD, SVM remains the most robust, consistently achieving 0.87–0.89 accuracy even at the highest noise probability, with fluctuations below 2-3\%. RF and LR show moderate stability, maintaining accuracy around 0.82-0.86, with only minor degradation ($<3\%$) as noise increases. In contrast, NB and DT exhibit strong sensitivity to noise, often dropping to 0.72–0.76 under severe conditions, particularly for DP and damping noise. Among all noise types, DP noise produces the highest overall degradation ($4\%$), whereas PD has the least impact, especially on SVM and RF. Overall, the results demonstrate that margin-based models (SVM) and ensemble methods (RF) maintain strong discriminative capability under noise, while simpler probabilistic or tree-based models degrade significantly with increasing quantum corruption. The noisy‐evaluation plots for the CWRUBD (Fig. \ref{fig:noisy_d2}) show that classical ML models exhibit high robustness across all six quantum-noise channels, BF, PF, BPF, DP, AD, and PD, although the degree of stability varies by algorithm. Across the evaluated noise settings, RF and DT maintained accuracies between 0.96 and 0.99. Even at the highest noise probability ($p=1.0$), the reduction in accuracy remained below 0.01, suggesting that the predictive performance of these models is only marginally affected by the perturbations introduced through the noisy-QSMOTE process. SVM and LR maintain moderate robustness, staying between 0.92–0.95, with slight oscillations as noise increases. NB shows noticeable sensitivity, fluctuating between 0.88–0.91, especially under BF and BPF noise, where sharp drops of ~0.02-0.03 occur at intermediate noise levels.

Fig. \ref{fig:noisy_d3} presents the robustness analysis of five classical classifiers, LR, RF, SVM, NB, and DT under increasing levels of six quantum noise models on the EFDD: BF, PF, BPF, DP, AD, and PD. Across all noise types, RF consistently achieves the highest accuracy ($\approx$ 0.88–0.90), showing minimal degradation even at maximum noise probability. Next, DT fluctuates around 0.8 accuracy. SVM forms a mid-tier cluster with accuracies around 0.68–0.70, remaining nearly noise-insensitive. NB and LR perform poorly, with NB ranging between 0.30–0.35 and DT sharply varying between 0.25–0.32 depending on noise type. Overall, RF is the most noise-robust model on EFDD, while NB and LR degrade significantly, confirming that EFDD is highly noise-sensitive for weak, variance-prone learners. Fig. \ref{fig:noisy_d4} illustrates how five classical ML models, LR, RF, SVM, NB, and DT, perform under increasing levels of quantum noise injected into the IFDD dataset. Across all six noise channels (BF, PF, BPF, DP, AD, and PD), RF consistently achieves the highest accuracy, remaining close to 0.95–0.99 even at maximum noise probability, demonstrating strong robustness. SVM and DT maintain stable performance ($\approx$0.80–0.90) across all noise types, showing only minor fluctuations ($<$3\%) as noise increases. NB and LR exhibit the lowest and most unstable accuracy, often dropping to $\approx$0.40–0.50 depending on the noise model. A general pattern across all subplots is that RF and SVM exhibit negligible degradation even when noise probability hits 1.0, demonstrating their robustness to quantum noise perturbations. In contrast, NB and LR decrease strongly across all noise types, demonstrating a high vulnerability.

\subsection{Discussion}

Fig.~\ref{fig:distribution_before_after_smote} and
Tables~\ref{table:performance_smote_hexa_1}-\ref{table:performance_smote_hexa_fifth_5}
collectively demonstrate that the effect of QSMOTE extends beyond restoring
class balance and depends strongly on the interaction between the dataset
geometry and the downstream classifier. As illustrated in
Fig.~\ref{fig:distribution_before_after_smote}, QSMOTE increases the
representation of minority classes and produces balanced training
distributions across the SPID, CWRUBD, EFDD, and IFDD datasets. However,
balancing the class distribution does not necessarily translate into uniform
performance gains across all learning algorithms. Instead, the results show
that the usefulness of the generated synthetic samples is
classifier-dependent.

For SPID under the hexa-class setting, SVM accuracy increases from 0.8057 to
0.8893 after QSMOTE, while RF improves from 0.7756 to 0.8533. More pronounced
gains are observed for CWRUBD, where RF and DT achieve accuracies of 0.9848
and 0.9750, respectively, after oversampling. The effect is particularly
substantial for EFDD: RF accuracy increases from 0.4923 to 0.9126 and DT
accuracy from 0.2979 to 0.8228, corresponding to large relative improvements
over the imbalanced-data baselines. Similarly, on IFDD, RF reaches an
accuracy of 0.9919 and DT reaches 0.9006 after QSMOTE. Together with the
corresponding precision, recall, and F1-score results, these observations
indicate that the additional minority-class representation generated by
QSMOTE can be particularly beneficial to classifiers capable of learning
non-linear and locally heterogeneous decision boundaries.

Importantly, the results also show that class balancing alone does not
guarantee improved classification. For example, LR accuracy on EFDD decreases
from 0.5337 before oversampling to 0.2622 after QSMOTE. This result is
important because it demonstrates that the effectiveness of QSMOTE depends on
how a classifier responds to the geometry introduced by synthetic
interpolation. One possible explanation is that interpolated minority samples
modify local feature distributions in ways that are advantageous to flexible
non-linear classifiers but less compatible with the decision structure
assumed by some linear or probabilistic models. The present experiments do
not directly establish the geometric mechanism responsible for these
differences; nevertheless, the contrasting behaviour across classifiers
shows that oversampling quality should not be assessed solely from the
resulting class counts. Classification performance after resampling remains
dependent on the interaction among the original feature distribution, the
synthetic samples, and the learning algorithm.

The noisy-QSMOTE experiments extend this analysis by examining the robustness
of the synthetic-sample generation mechanism itself. This distinction is
central to the present study. Noise is not applied directly to the input
features after oversampling or to the downstream classifiers. Instead, the
six quantum-inspired noise channels are introduced into the
compact-swap-test-based similarity-estimation process used by QSMOTE.
Perturbing this stage modifies the ancilla measurement probability and,
consequently, the estimated overlap between minority samples and their
associated centroids. The resulting change propagates to the angle used
during interpolation and therefore changes the synthetic samples supplied to
the classifiers. Variations in downstream performance should consequently be
interpreted as the end result of perturbing the similarity-estimation and
sample-generation process rather than as the direct noise tolerance of the
classifiers themselves.

The results in Figs.~\ref{fig:noisy_d1}-\ref{fig:noisy_d4} further show that this propagated
effect is strongly dataset- and classifier-dependent. For SPID
(Fig.~\ref{fig:noisy_d1}), SVM exhibits comparatively stable performance
across the investigated noise levels, while RF and LR also show relatively
limited variation under several noise conditions. DT and NB exhibit greater
sensitivity for particular channels and noise strengths. Among the evaluated
channels, depolarizing noise produces some of the more pronounced performance
changes, whereas phase damping generally produces smaller changes for several
classifier configurations. These observations indicate that the effect of
noisy similarity estimation cannot be characterized by noise strength alone;
the type of perturbation and the downstream decision mechanism are also
important.

A different robustness profile is observed for CWRUBD
(Fig.~\ref{fig:noisy_d2}). RF and DT maintain high accuracy over the
investigated noise conditions, while SVM and LR also retain comparatively
stable performance. NB shows greater variation, particularly under BF and BPF
noise. The comparatively limited degradation observed for several classifiers
suggests that the discriminative structure of CWRUBD is less affected by the
changes in synthetic samples produced by noisy similarity estimation.
However, this should not be interpreted as complete invariance to noise.
Rather, within the investigated noise range and experimental configuration,
the downstream classifiers preserve sufficient discriminative information
despite perturbations introduced during QSMOTE sample generation.

For EFDD (Fig.~\ref{fig:noisy_d3}), the separation between classifier
responses becomes more pronounced. RF maintains comparatively high and stable
performance across the investigated noise channels, whereas DT exhibits
greater variability. SVM remains comparatively stable at an intermediate
performance level, while NB and LR show substantially lower performance under
several noisy-QSMOTE configurations. The EFDD results are particularly
informative when considered together with the clean-QSMOTE experiment:
although oversampling produces a substantial improvement for RF and DT, the
quality of the perturbed synthetic samples does not affect all classifier
families in the same manner. Thus, robustness to noisy-QSMOTE depends not
only on the oversampling mechanism but also on the capacity of the downstream
model to accommodate changes in the generated training distribution.

A related pattern is observed for IFDD (Fig.~\ref{fig:noisy_d4}), where RF
maintains the highest performance among the evaluated models across much of
the investigated noise range. SVM and DT exhibit intermediate robustness,
whereas NB and LR generally show lower and more variable performance.
Considering SPID, CWRUBD, EFDD, and IFDD together, RF is the most consistently
stable classifier in the noisy-QSMOTE experiments, while SVM also demonstrates
strong robustness in several settings. In contrast, NB and, for the more
challenging datasets, LR exhibit greater sensitivity to changes in the
synthetic training distribution. These findings should be interpreted as
empirical robustness patterns for the datasets and noise configurations
considered here rather than as universal properties of the corresponding
classifier families.

The differences among the six noise channels provide an additional insight
into the QSMOTE mechanism. Because the channels perturb the
compact-swap-test-based similarity calculation in different ways, equal
nominal noise probabilities do not necessarily produce equivalent changes in
the resulting synthetic data. The experiments generally indicate stronger
performance variation under depolarizing noise in several dataset-classifier
combinations, whereas phase damping is comparatively less disruptive in a
number of cases. The practical implication is that robustness analysis of
quantum-inspired oversampling should consider the structure of the
perturbation rather than representing noise through a single generic
parameter. At the same time, the observed ordering is not identical for every
dataset and classifier, emphasizing that no single noise channel can be
assumed to be universally dominant or negligible.

From an industrial-AI perspective, these findings have two implications.
First, imbalance mitigation and robustness should be evaluated jointly.
A resampling method may substantially improve performance under its nominal
configuration while producing synthetic samples whose usefulness changes when
the underlying similarity estimates are perturbed. Second, selection of the
downstream classifier remains important even when the same balanced training
set is used. In the present experiments, ensemble and margin-based methods,
particularly RF and SVM, generally provide more stable downstream performance
than the more sensitive classifier configurations. Consequently, evaluating
only the balanced class distribution or only the nominal QSMOTE result may
provide an incomplete picture of the reliability of an imbalance-learning
pipeline intended for condition monitoring and fault diagnosis.

The three-stage experimental design used in this work therefore provides a
useful separation between two related questions: whether QSMOTE improves
learning from imbalanced industrial data and whether this improvement remains
stable when the quantum-inspired similarity-estimation mechanism is
perturbed. The baseline experiments quantify the original imbalance problem,
the QSMOTE experiments quantify the benefit of synthetic minority
oversampling, and the noisy-QSMOTE experiments characterize the downstream
consequences of perturbing the compact-swap-test-based similarity estimation.
This separation prevents changes caused by class balancing from being
conflated with those caused by noise and provides a more interpretable
assessment of the complete oversampling-classification pipeline.

Several limitations should nevertheless be considered when interpreting
these results. The robustness analysis is based on simulated
quantum-inspired noise applied to the compact-swap-test-based similarity
estimation mechanism rather than execution on physical quantum hardware.
Accordingly, the investigated channels provide controlled models of
perturbation but do not reproduce every source of device-specific error that
may arise in an actual quantum implementation. In addition, the empirical
conclusions are based on four industrial and energy-system datasets and five
classical classifiers; different feature distributions, imbalance ratios,
class overlaps, or learning architectures may produce different robustness
patterns. The present study also focuses on the propagation of noise through
similarity estimation and synthetic sample generation rather than isolating
the individual contribution of every intermediate quantity to the final
classification error. Future work can therefore extend the framework through
hardware-based experiments, broader industrial benchmarks, additional
learning architectures, and sensitivity analyses that explicitly relate
errors in estimated overlaps and interpolation angles to changes in synthetic
sample geometry and downstream predictive performance.

\section{Conclusion}\label{Sec5}

In this work, QSMOTE was evaluated for fault-diagnosis tasks in industrial and energy-system datasets under three different settings: the original imbalanced datasets, datasets balanced using QSMOTE, and noisy-QSMOTE-generated datasets obtained by introducing quantum-inspired perturbations into the similarity-estimation process used during synthetic sample generation.
Experimental investigations on four benchmark datasets, namely SPID, CWRUBD, EFDD, and IFDD, demonstrated that class imbalance and noisy similarity estimation have distinct effects on fault-classification performance. The results showed that QSMOTE effectively mitigates class imbalance by improving minority-class representation and generating more balanced training distributions. Consequently, significant improvements in accuracy, precision, recall, and F1-score were observed across most datasets and classifiers, with RF, DT, and SVM benefiting the most from QSMOTE-based oversampling. These findings confirm the effectiveness of compact-swap-test-based quantum-inspired similarity estimation for enhancing classification performance in imbalanced fault-diagnosis problems. At the same time, the observed degradation of certain linear and probabilistic models on selected datasets indicates that the effectiveness of oversampling remains dependent on the underlying data geometry and classifier characteristics.

Unlike conventional robustness studies, the noisy-QSMOTE analysis introduced BF, PF, BPF, DP, AD, and PD noise directly into the compact-swap-test-based similarity estimation process used during synthetic sample generation. Consequently, the perturbations affected overlap estimation, angle computation, and the resulting synthetic samples before propagating to downstream classification performance. The results showed that RF and SVM remained highly resilient to noisy-QSMOTE-generated samples, whereas LR and NB exhibited greater sensitivity, particularly on more complex datasets such as EFDD and IFDD. Furthermore, depolarizing noise consistently produced the largest performance degradation, while phase-damping noise generally had the least impact. The findings demonstrate that the effectiveness of quantum-inspired oversampling depends not only on imbalance correction but also on the robustness of the underlying similarity-estimation mechanism. The proposed framework establishes a systematic methodology for jointly evaluating baseline performance, QSMOTE effectiveness, and noisy-QSMOTE robustness within a unified setting. Future work will focus on developing noise-aware quantum-inspired oversampling techniques, more robust similarity-estimation strategies, and implementations on emerging quantum computing platforms for real-world industrial fault-diagnosis applications.

\section*{Supplementary information}
The supplementary information is provided in the Appendix section \ref{secA1}.

%\bmhead{Acknowledgements}

%Acknowledgements are not compulsory. Where included they should be brief. Grant or contribution numbers may be acknowledged.

%Please refer to Journal-level guidance for any specific requirements.

\section*{Declarations}

%Some journals require declarations to be submitted in a standardised format. Please check the Instructions for Authors of the journal to which you are submitting to see if you need to complete this section. If yes, your manuscript must contain the following sections under the heading `Declarations':

\begin{itemize}
\item Funding: This research did not receive any specific grant from funding agencies in the public, commercial, or not-for-profit sectors.
\item Conflict of interest/Competing interests: The authors have no financial or non-financial competing interests.
%\item Ethics approval and consent to participate
%\item Consent for publication
\item Data Availability: The datasets analyzed in this study are publicly available from their respective repositories. The dataset sources and access information are provided in Section 4.2 and the corresponding references. No new primary dataset was generated as part of this study.
%\item Materials availability
%\item Code availability 
\item CRediT authorship contribution statement:
A.S.P.: Conceptualization, Methodology, Data curation, Formal analysis,
Investigation, Writing - original draft, Writing - review \& editing.
H.R.P.: Conceptualization, Methodology, Formal analysis, Supervision,
Writing - original draft, Writing - review \& editing.
B.K.B.: Conceptualization, Methodology, Formal analysis,
Writing - original draft, Writing - review \& editing.
P.M.: Conceptualization, Methodology, Formal analysis,
Writing - original draft, Writing - review \& editing.
D.B.: Conceptualization, Methodology, Formal analysis,
Writing - original draft, Writing - review \& editing.
G.G.T.: Conceptualization, Methodology, Formal analysis,
Writing - original draft, Writing - review \& editing.
All authors reviewed and approved the final version of the manuscript.
\item Declaration of generative AI and AI-assisted technologies in the manuscript preparation process:
During the preparation of this work, the authors used ChatGPT (OpenAI) solely for language refinement and grammatical improvement. After using this tool, the authors reviewed and edited the content as needed and take full responsibility for the content of the published article.
\end{itemize}

\appendix

%\section{Section title of first appendix}\label{secA1}

\section{Supplementary Material}\label{secA1}

\begin{figure}
    \centering
    \begin{subfigure}{0.325\linewidth}
        \centering
        \includegraphics[width=\linewidth]{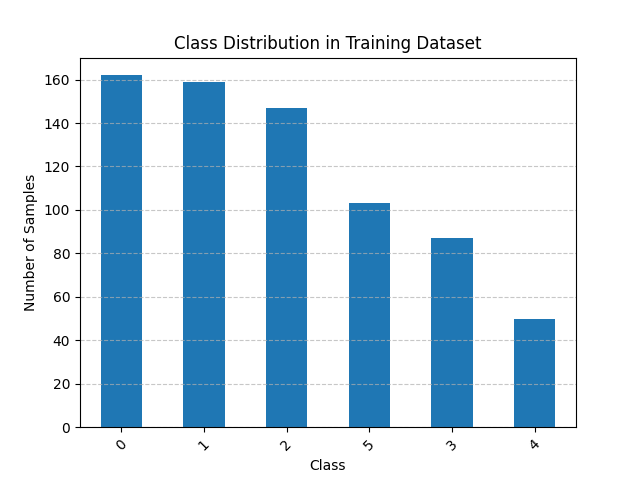} 
        \caption{}
        \label{fig:h_data_pca}
    \end{subfigure}
    \hfill
    \begin{subfigure}{0.325\linewidth}
        \centering
        \includegraphics[width=\linewidth]{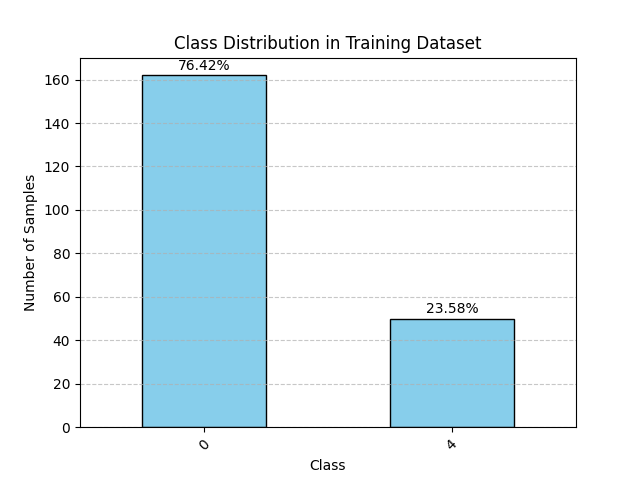}
        \caption{}
        \label{fig:m_data_pca}
    \end{subfigure}
    \begin{subfigure}{0.325\linewidth}
        \centering
        \includegraphics[width=\linewidth]{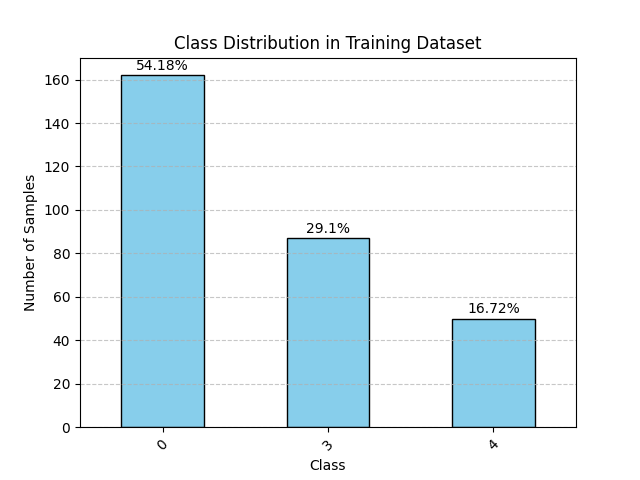} 
        \caption{}
        \label{fig:h_data_pca}
    \end{subfigure}
    \hfill
    \begin{subfigure}{0.325\linewidth}
        \centering
        \includegraphics[width=\linewidth]{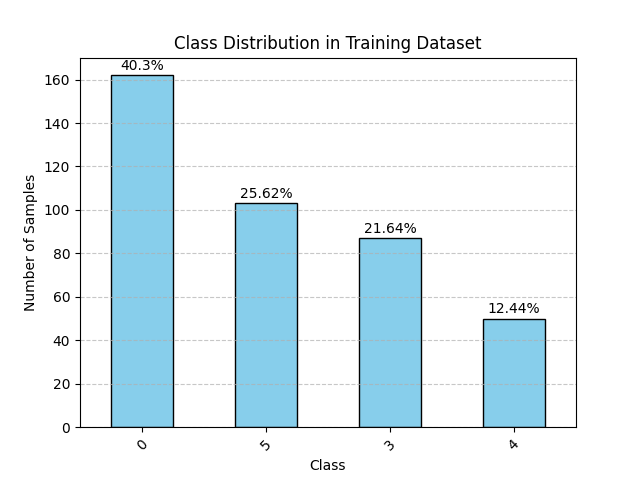} 
        \caption{}
        \label{fig:h_data_pca}
    \end{subfigure}
    \hfill
    \begin{subfigure}{0.325\linewidth}
        \centering
        \includegraphics[width=\linewidth]{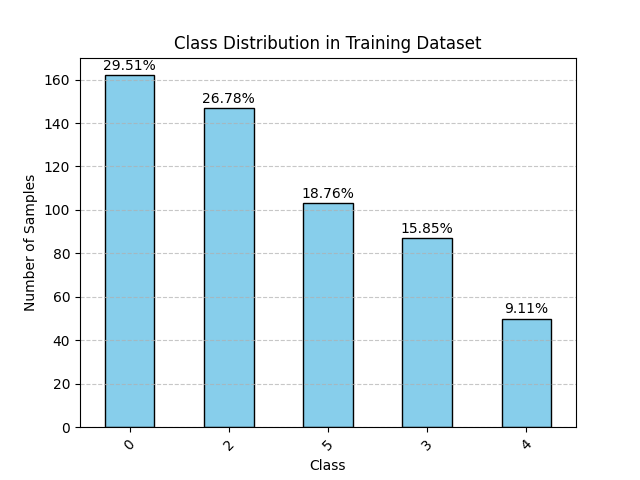} 
        \caption{}
        \label{fig:h_data_pca}
    \end{subfigure}
    \hfill
    \begin{subfigure}{0.325\linewidth}
        \centering
        \includegraphics[width=\linewidth]{figures/BeforeSmote-6.png} 
        \caption{}
        \label{fig:h_data_pca}
    \end{subfigure}
    \caption{(a) Initial Distribution of SPID. Class Percentages in (b) Binary, (c) Tertiary, (d) Tetra, (e) Penta, and (f) Hexa Classification Before QSMOTE.}
    \label{fig:distribution_before_smote}
\end{figure}

\begin{figure}
    \centering
%    \begin{subfigure}{0.325\linewidth}
%        \centering
%        \includegraphics[width=\linewidth]{figures/Before Smote.png} 
%        \caption{}
%        \label{fig:h_data_pca}
%    \end{subfigure}
%    \hfill
    \begin{subfigure}{0.325\linewidth}
        \centering
        \includegraphics[width=\linewidth]{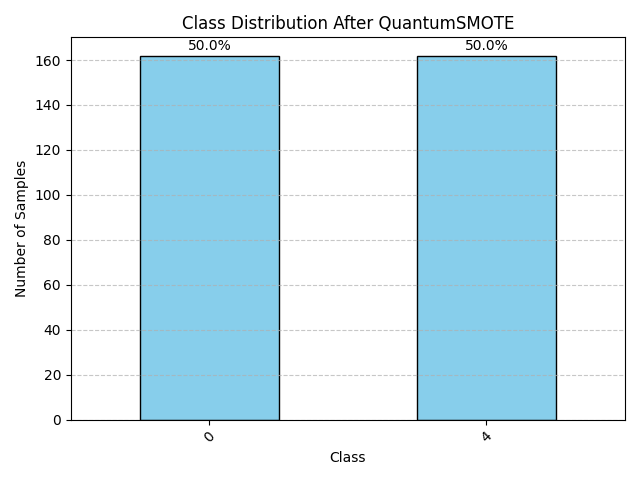}
        \caption{}
        \label{fig:m_data_pca}
    \end{subfigure}
    \begin{subfigure}{0.325\linewidth}
        \centering
        \includegraphics[width=\linewidth]{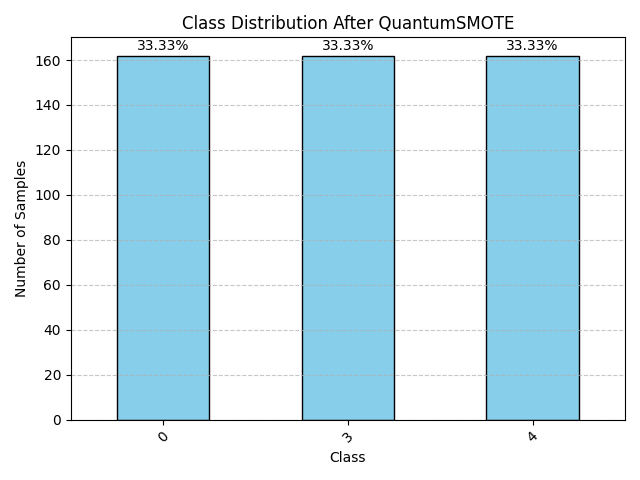} 
        \caption{}
        \label{fig:h_data_pca}
    \end{subfigure}
    \hfill
    \begin{subfigure}{0.325\linewidth}
        \centering
        \includegraphics[width=\linewidth]{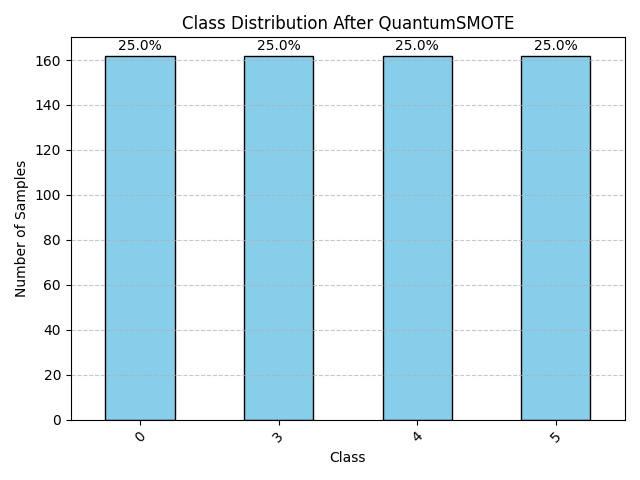} 
        \caption{}
        \label{fig:h_data_pca}
    \end{subfigure}
    \hfill
    \begin{subfigure}{0.325\linewidth}
        \centering
        \includegraphics[width=\linewidth]{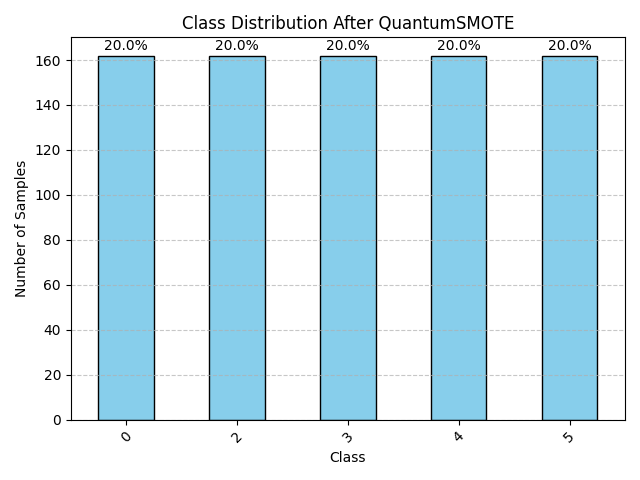} 
        \caption{}
        \label{fig:h_data_pca}
    \end{subfigure}
    \hfill
    \begin{subfigure}{0.325\linewidth}
        \centering
        \includegraphics[width=\linewidth]{figures/class_distribution_after_smote-6.png} 
        \caption{}
        \label{fig:h_data_pca}
    \end{subfigure}
    \caption{Final Distribution of SPID. Class Percentages in (b) Binary, (c) Tertiary, (d) Tetra, (e) Penta, and (f) Hexa Classification After QSMOTE.}
    \label{fig:distribution_after_smote}
\end{figure}

\begin{table}[ht]
\caption{Performance Metrics of Classical Algorithms Before and After QSMOTE for Binary Classification on SPID}
\label{table:performance_smote_binary_1}
\centering
\begin{tabular*}{\textwidth}{@{\extracolsep\fill}lcccc}
\toprule
\textbf{Algorithm} 
& \textbf{Accuracy} 
& \textbf{Precision} 
& \textbf{Recall} 
& \textbf{F1 Measure} \\
\midrule
\multicolumn{5}{l}{\textbf{Before QSMOTE}} \\
\midrule
LR  
& 0.9344 $\pm$ 0.0285 
& 0.9348 $\pm$ 0.0302 
& 0.9344 $\pm$ 0.0285 
& 0.9324 $\pm$ 0.0305 \\

RF  
& 0.8802 $\pm$ 0.0192 
& 0.8930 $\pm$ 0.0159 
& 0.8802 $\pm$ 0.0192 
& 0.8657 $\pm$ 0.0260 \\

SVM 
& 0.9103 $\pm$ 0.0263 
& 0.9204 $\pm$ 0.0211 
& 0.9103 $\pm$ 0.0263 
& 0.9013 $\pm$ 0.0319 \\

NB  
& 0.8813 $\pm$ 0.0760 
& 0.8815 $\pm$ 0.0787 
& 0.8813 $\pm$ 0.0760 
& 0.8749 $\pm$ 0.0792 \\

DT  
& 0.8205 $\pm$ 0.0496 
& 0.8213 $\pm$ 0.0585 
& 0.8205 $\pm$ 0.0496 
& 0.8168 $\pm$ 0.0524 \\

\midrule
\multicolumn{5}{l}{\textbf{After QSMOTE}} \\
\midrule
LR  
& \textbf{0.9685 $\pm$ 0.0200} 
& \textbf{0.9711 $\pm$ 0.0179} 
& \textbf{0.9685 $\pm$ 0.0200} 
& \textbf{0.9685 $\pm$ 0.0200} \\

RF  
& \textbf{0.9804 $\pm$ 0.0248} 
& \textbf{0.9822 $\pm$ 0.0224} 
& \textbf{0.9804 $\pm$ 0.0248} 
& \textbf{0.9803 $\pm$ 0.0249} \\

SVM 
& \textbf{0.9764 $\pm$ 0.0380} 
& \textbf{0.9798 $\pm$ 0.0316} 
& \textbf{0.9764 $\pm$ 0.0380} 
& \textbf{0.9762 $\pm$ 0.0383} \\

NB  
& \textbf{0.8900 $\pm$ 0.0338} 
& \textbf{0.9055 $\pm$ 0.0277} 
& \textbf{0.8900 $\pm$ 0.0338} 
& \textbf{0.8887 $\pm$ 0.0348} \\

DT  
& \textbf{0.9016 $\pm$ 0.0277} 
& \textbf{0.9082 $\pm$ 0.0285} 
& \textbf{0.9016 $\pm$ 0.0277} 
& \textbf{0.9012 $\pm$ 0.0278} \\
\bottomrule
\end{tabular*}
\end{table}

\begin{table}[ht]
\caption{Performance Metrics of Classical Algorithms Before and After QSMOTE for Tertiary Classification on SPID}
\label{table:performance_smote_tertiary_1}
\centering
\begin{tabular*}{\textwidth}{@{\extracolsep\fill}lcccc}
\toprule
\textbf{Algorithm} 
& \textbf{Accuracy} 
& \textbf{Precision} 
& \textbf{Recall} 
& \textbf{F1 Measure} \\
\midrule
\multicolumn{5}{l}{\textbf{Before QSMOTE}} \\
\midrule
LR  
& 0.8159 $\pm$ 0.0278 
& 0.8194 $\pm$ 0.0240 
& 0.8159 $\pm$ 0.0278 
& 0.8132 $\pm$ 0.0267 \\

RF  
& 0.8577 $\pm$ 0.0244 
& 0.8690 $\pm$ 0.0242 
& 0.8577 $\pm$ 0.0244 
& 0.8480 $\pm$ 0.0284 \\

SVM 
& 0.8618 $\pm$ 0.0290 
& 0.8734 $\pm$ 0.0352 
& 0.8618 $\pm$ 0.0290 
& 0.8469 $\pm$ 0.0333 \\

NB  
& 0.8325 $\pm$ 0.0377 
& 0.8333 $\pm$ 0.0399 
& 0.8325 $\pm$ 0.0377 
& 0.8248 $\pm$ 0.0437 \\

DT  
& 0.7746 $\pm$ 0.0669 
& 0.7956 $\pm$ 0.0652 
& 0.7746 $\pm$ 0.0669 
& 0.7778 $\pm$ 0.0627 \\

\midrule
\multicolumn{5}{l}{\textbf{After QSMOTE}} \\
\midrule
LR  
& \textbf{0.8531 $\pm$ 0.0191} 
& \textbf{0.8544 $\pm$ 0.0204} 
& \textbf{0.8531 $\pm$ 0.0191} 
& \textbf{0.8528 $\pm$ 0.0188} \\

RF  
& \textbf{0.9458 $\pm$ 0.0387} 
& \textbf{0.9474 $\pm$ 0.0384} 
& \textbf{0.9458 $\pm$ 0.0387} 
& \textbf{0.9455 $\pm$ 0.0391} \\

SVM 
& \textbf{0.9535 $\pm$ 0.0158} 
& \textbf{0.9561 $\pm$ 0.0142} 
& \textbf{0.9535 $\pm$ 0.0158} 
& \textbf{0.9536 $\pm$ 0.0157} \\

NB  
& \textbf{0.8427 $\pm$ 0.0502} 
& \textbf{0.8494 $\pm$ 0.0508} 
& \textbf{0.8427 $\pm$ 0.0502} 
& \textbf{0.8402 $\pm$ 0.0508} \\

DT  
& \textbf{0.8890 $\pm$ 0.0367} 
& \textbf{0.8946 $\pm$ 0.0298} 
& \textbf{0.8890 $\pm$ 0.0367} 
& \textbf{0.8887 $\pm$ 0.0361} \\
\bottomrule
\end{tabular*}
\end{table}

\begin{table}[ht]
\caption{Performance Metrics of Classical Algorithms Before and After QSMOTE for Tetra Classification on SPID}
\label{table:performance_smote_tetra_1}
\centering
\begin{tabular*}{\textwidth}{@{\extracolsep\fill}lcccc}
\toprule
\textbf{Algorithm} 
& \textbf{Accuracy} 
& \textbf{Precision} 
& \textbf{Recall} 
& \textbf{F1 Measure} \\
\midrule
\multicolumn{5}{l}{\textbf{Before QSMOTE}} \\
\midrule
LR  
& 0.8504 $\pm$ 0.0380 
& 0.8554 $\pm$ 0.0416 
& 0.8504 $\pm$ 0.0380 
& 0.8492 $\pm$ 0.0389 \\

RF  
& 0.8597 $\pm$ 0.0579 
& 0.8699 $\pm$ 0.0675 
& 0.8597 $\pm$ 0.0579 
& 0.8534 $\pm$ 0.0628 \\

SVM 
& 0.8785 $\pm$ 0.0348 
& 0.8912 $\pm$ 0.0425 
& 0.8785 $\pm$ 0.0348 
& 0.8692 $\pm$ 0.0375 \\

NB  
& 0.8287 $\pm$ 0.0310 
& 0.8352 $\pm$ 0.0345 
& 0.8287 $\pm$ 0.0310 
& 0.8243 $\pm$ 0.0340 \\

DT  
& 0.7632 $\pm$ 0.0391 
& 0.7714 $\pm$ 0.0388 
& 0.7632 $\pm$ 0.0391 
& 0.7625 $\pm$ 0.0384 \\

\midrule
\multicolumn{5}{l}{\textbf{After QSMOTE}} \\
\midrule
LR  
& \textbf{0.9016 $\pm$ 0.0421} 
& \textbf{0.9018 $\pm$ 0.0430} 
& \textbf{0.9016 $\pm$ 0.0421} 
& \textbf{0.9006 $\pm$ 0.0430} \\

RF  
& \textbf{0.9421 $\pm$ 0.0086} 
& \textbf{0.9438 $\pm$ 0.0075} 
& \textbf{0.9421 $\pm$ 0.0086} 
& \textbf{0.9421 $\pm$ 0.0086} \\

SVM 
& \textbf{0.9479 $\pm$ 0.0208} 
& \textbf{0.9494 $\pm$ 0.0203} 
& \textbf{0.9479 $\pm$ 0.0208} 
& \textbf{0.9478 $\pm$ 0.0210} \\

NB  
& \textbf{0.8456 $\pm$ 0.0120} 
& \textbf{0.8566 $\pm$ 0.0128} 
& \textbf{0.8456 $\pm$ 0.0120} 
& \textbf{0.8451 $\pm$ 0.0127} \\

DT  
& \textbf{0.8881 $\pm$ 0.0271} 
& \textbf{0.8904 $\pm$ 0.0271} 
& \textbf{0.8881 $\pm$ 0.0271} 
& \textbf{0.8868 $\pm$ 0.0268} \\
\bottomrule
\end{tabular*}
\end{table}

\begin{table}[ht]
\caption{Performance Metrics of Classical Algorithms Before and After QSMOTE for Penta Classification on SPID}
\label{table:performance_smote_penta_1}
\centering
\begin{tabular*}{\textwidth}{@{\extracolsep\fill}lcccc}
\toprule
\textbf{Algorithm} 
& \textbf{Accuracy} 
& \textbf{Precision} 
& \textbf{Recall} 
& \textbf{F1 Measure} \\
\midrule
\multicolumn{5}{l}{\textbf{Before QSMOTE}} \\
\midrule
LR  
& 0.7835 $\pm$ 0.0384 
& 0.7887 $\pm$ 0.0393 
& 0.7835 $\pm$ 0.0384 
& 0.7823 $\pm$ 0.0366 \\

RF  
& 0.7722 $\pm$ 0.0259 
& 0.7795 $\pm$ 0.0290 
& 0.7722 $\pm$ 0.0259 
& 0.7690 $\pm$ 0.0258 \\

SVM 
& 0.8176 $\pm$ 0.0331 
& 0.8305 $\pm$ 0.0277 
& 0.8176 $\pm$ 0.0331 
& 0.8143 $\pm$ 0.0330 \\

NB  
& 0.7835 $\pm$ 0.0343 
& 0.7908 $\pm$ 0.0338 
& 0.7835 $\pm$ 0.0343 
& 0.7810 $\pm$ 0.0363 \\

DT  
& 0.6263 $\pm$ 0.0448 
& 0.6355 $\pm$ 0.0425 
& 0.6263 $\pm$ 0.0448 
& 0.6250 $\pm$ 0.0410 \\

\midrule
\multicolumn{5}{l}{\textbf{After QSMOTE}} \\
\midrule
LR  
& \textbf{0.8364 $\pm$ 0.0286} 
& \textbf{0.8425 $\pm$ 0.0277} 
& \textbf{0.8364 $\pm$ 0.0286} 
& \textbf{0.8380 $\pm$ 0.0279} \\

RF  
& \textbf{0.8811 $\pm$ 0.0250} 
& \textbf{0.8831 $\pm$ 0.0254} 
& \textbf{0.8811 $\pm$ 0.0250} 
& \textbf{0.8801 $\pm$ 0.0259} \\

SVM 
& \textbf{0.8796 $\pm$ 0.0235} 
& \textbf{0.8837 $\pm$ 0.0223} 
& \textbf{0.8796 $\pm$ 0.0235} 
& \textbf{0.8800 $\pm$ 0.0235} \\

NB  
& \textbf{0.7870 $\pm$ 0.0132} 
& \textbf{0.7954 $\pm$ 0.0127} 
& \textbf{0.7870 $\pm$ 0.0132} 
& \textbf{0.7876 $\pm$ 0.0115} \\

DT  
& \textbf{0.7778 $\pm$ 0.0195} 
& \textbf{0.7784 $\pm$ 0.0194} 
& \textbf{0.7778 $\pm$ 0.0195} 
& \textbf{0.7708 $\pm$ 0.0198} \\
\bottomrule
\end{tabular*}
\end{table}

The sample distribution for Solar Panel Image Dataset (SPID) before and after Synthetic Minority Oversampling Technique (SMOTE) is shown in Figs. \ref{fig:distribution_before_smote} and \ref{fig:distribution_after_smote}. The results in Table \ref{table:performance_smote_binary_1} show that QSMOTE improves the performance of all classical algorithms in binary classification. Before resampling, logistic regression (LR) gives the best accuracy of 0.9344, followed by support vector machine (SVM) (0.9103) and random forest (RF) (0.8802). Decision tree (DT) shows the weakest result at 0.8205, while naive bayes (NB) stays moderate at 0.8813. After applying QSMOTE, every model performs better. RF gives the highest accuracy of 0.9804, precision of 0.9822, recall of 0.9804, and F1-score of 0.9803. SVM follows with 0.9764 accuracy and F1-score of 0.9762, showing strong balance across metrics. LR also improves to 0.9685 accuracy and F1-score. Weaker models also gain. DT increases from 0.8205 to 0.9016 accuracy (about 10\% gain) and F1-score of 0.9012, while NB improves from 0.8813 to 0.8900 accuracy with F1-score of 0.8887. Overall, QSMOTE boosts both strong models (RF, SVM, LR) and weaker models (DT, NB). RF shows the best performance after resampling, while SVM and LR also stay highly effective. The results in Table \ref{table:performance_smote_tertiary_1} show that QSMOTE improves all classical algorithms for tertiary classification. Before resampling, SVM gives the highest accuracy of 0.8618, followed by RF (0.8577) and NB (0.8325), while LR (0.8159) and DT (0.7746) perform lower. After QSMOTE, every model improves. SVM reaches 0.9535 accuracy and 0.9536 F1-score, and RF achieves 0.9458 accuracy with F1 0.9455, showing strong performance. LR rises to 0.8531 accuracy and F1 0.8528, and DT increases from 0.7746 to 0.8890 in accuracy, with F1 0.8887, showing the highest relative improvement. NB also improves modestly to 0.8427 accuracy and F1 0.8402. Overall, QSMOTE balances the classes and boosts both strong models (SVM, RF) and weaker models (LR, DT, NB), giving consistent gains across accuracy, precision, recall, and F1.

The results in Table \ref{table:performance_smote_tetra_1} highlight the quantitative impact of QSMOTE on tetra classification performance. Prior to resampling, SVM achieves the best accuracy at 0.8785, followed by RF at 0.8597 and LR at 0.8504, whereas DT performs the weakest with an accuracy of only 0.7632. After applying QSMOTE, all models record substantial improvements: SVM rises to 0.9479 (a gain of nearly +7\%), RF reaches 0.9421 (an increase of about +8\%), and LR improves to 0.9016 (a gain of around +5\%). DT also exhibits remarkable progress, moving from 0.7632 to 0.8881, representing the highest relative improvement of approximately +16\%. NB, while showing the smallest absolute change, still improves from 0.8287 to 0.8456. Similar upward trends are observed across precision, recall, and F1-measure, with SVM and RF consistently achieving the strongest overall metrics after resampling. These results quantitatively demonstrate that QSMOTE not only boosts average performance but also reduces variability, particularly benefiting weaker learners like DT while further enhancing strong performers such as SVM and RF. The results in Table \ref{table:performance_smote_penta_1} quantitatively demonstrate the effectiveness of QSMOTE in improving classification performance for the penta-class problem. Before resampling, SVM achieves the highest accuracy of 0.8176, followed by LR and NB at 0.7835, while RF slightly lags at 0.7722 and DT performs the weakest with only 0.6263. After applying QSMOTE, all models exhibit significant gains: RF improves from 0.7722 to 0.8811 (a relative increase of about +14\%), SVM rises from 0.8176 to 0.8796 (+7.6\%), and LR advances from 0.7835 to 0.8364 (+6.8\%). The most dramatic improvement is observed in DT, which jumps from 0.6263 to 0.7778, reflecting a relative gain of nearly +24\%, indicating that QSMOTE effectively mitigates its earlier limitations. NB shows the smallest change, moving from 0.7835 to 0.7870, but still gains stability with reduced variance. Consistently across precision, recall, and F1-measure, RF and SVM emerge as the best-performing models post-QSMOTE, underscoring that oversampling not only enhances accuracy but also ensures balanced generalization in complex multi-class settings.

\begin{table}[ht]
\caption{Performance Metrics of Classical Algorithms Before and After QSMOTE for Binary Classification on CWRUBD}
\label{table:performance_smote_binary_2}
\centering
\begin{tabular*}{\textwidth}{@{\extracolsep\fill}lcccc}
\toprule
\textbf{Algorithm} 
& \textbf{Accuracy} 
& \textbf{Precision} 
& \textbf{Recall} 
& \textbf{F1 Measure} \\
\midrule
\multicolumn{5}{l}{\textbf{Before QSMOTE}} \\
\midrule
LR  
& 0.9887 $\pm$ 0.0092 
& 0.9891 $\pm$ 0.0089 
& 0.9887 $\pm$ 0.0092 
& 0.9886 $\pm$ 0.0093 \\

RF  
& 0.9885 $\pm$ 0.0153 
& 0.9892 $\pm$ 0.0145 
& 0.9885 $\pm$ 0.0153 
& 0.9884 $\pm$ 0.0156 \\

SVM 
& 0.9962 $\pm$ 0.0075 
& 0.9964 $\pm$ 0.0071 
& 0.9962 $\pm$ 0.0075 
& 0.9963 $\pm$ 0.0075 \\

NB  
& 0.9962 $\pm$ 0.0075 
& 0.9964 $\pm$ 0.0071 
& 0.9962 $\pm$ 0.0075 
& 0.9963 $\pm$ 0.0075 \\

DT  
& 0.9847 $\pm$ 0.0224 
& 0.9858 $\pm$ 0.0207 
& 0.9847 $\pm$ 0.0224 
& 0.9844 $\pm$ 0.0231 \\

\midrule
\multicolumn{5}{l}{\textbf{After QSMOTE}} \\
\midrule
LR  
& \textbf{0.9946 $\pm$ 0.0067} 
& \textbf{0.9947 $\pm$ 0.0065} 
& \textbf{0.9946 $\pm$ 0.0067} 
& \textbf{0.9946 $\pm$ 0.0067} \\

RF  
& \textbf{0.9946 $\pm$ 0.0066} 
& \textbf{0.9947 $\pm$ 0.0064} 
& \textbf{0.9946 $\pm$ 0.0066} 
& \textbf{0.9946 $\pm$ 0.0066} \\

SVM 
& 0.9946 $\pm$ 0.0108 
& 0.9949 $\pm$ 0.0103 
& 0.9946 $\pm$ 0.0108 
& 0.9946 $\pm$ 0.0108 \\

NB  
& 0.9783 $\pm$ 0.0108 
& 0.9794 $\pm$ 0.0098 
& 0.9783 $\pm$ 0.0108 
& 0.9782 $\pm$ 0.0108 \\

DT  
& \textbf{0.9945 $\pm$ 0.0067} 
& \textbf{0.9947 $\pm$ 0.0065} 
& \textbf{0.9945 $\pm$ 0.0067} 
& \textbf{0.9945 $\pm$ 0.0067} \\
\bottomrule
\end{tabular*}
\end{table}

The results in Table~\ref{table:performance_smote_binary_2} for CWRU Bearing Dataset (CWRUBD) indicate that all classifiers perform very strongly in the binary classification setting, with minimal variation across models. Before QSMOTE, both SVM and NB achieve the best results, attaining accuracies of $0.9962 \pm 0.0075$, precisions of $0.9964 \pm 0.0071$, recalls of $0.9962 \pm 0.0075$, and F1 scores of $0.9963 \pm 0.0075$. LR and RF follow closely with accuracies around $0.9887 \pm 0.0092$ and $0.9885 \pm 0.0153$, respectively, while DT records the lowest accuracy of $0.9847 \pm 0.0224$. After applying QSMOTE, LR and RF improve to $0.9946 \pm 0.0067$ and $0.9946 \pm 0.0066$, respectively, while SVM slightly decreases to $0.9946 \pm 0.0108$, showing a marginal drop of about $0.16\%$ compared to its pre-QSMOTE performance. Interestingly, NB exhibits a notable decline after QSMOTE, dropping from $0.9962 \pm 0.0075$ to $0.9783 \pm 0.0108$, highlighting its sensitivity to synthetic oversampling. DT, in contrast, improves significantly, rising from $0.9847 \pm 0.0224$ to $0.9945 \pm 0.0067$, representing an absolute gain of nearly $1.0\%$. Overall, while most classifiers benefit from QSMOTE, SVM remained the most consistent high performer, whereas NB shows reduced robustness in this binary setting.

\begin{table}[ht]
\caption{Performance Metrics of Classical Algorithms Before and After QSMOTE for Tertiary Classification on CWRUBD}
\label{table:performance_smote_three_2}
\centering
\begin{tabular*}{\textwidth}{@{\extracolsep\fill}lcccc}
\toprule
\textbf{Algorithm} 
& \textbf{Accuracy} 
& \textbf{Precision} 
& \textbf{Recall} 
& \textbf{F1 Measure} \\
\midrule
\multicolumn{5}{l}{\textbf{Before QSMOTE}} \\
\midrule
LR  
& 0.9637 $\pm$ 0.0165 
& 0.9651 $\pm$ 0.0178 
& 0.9637 $\pm$ 0.0165 
& 0.9636 $\pm$ 0.0172 \\

RF  
& 0.9639 $\pm$ 0.0191 
& 0.9657 $\pm$ 0.0188 
& 0.9639 $\pm$ 0.0191 
& 0.9633 $\pm$ 0.0194 \\

SVM 
& 0.9639 $\pm$ 0.0191 
& 0.9671 $\pm$ 0.0174 
& 0.9639 $\pm$ 0.0191 
& 0.9623 $\pm$ 0.0202 \\

NB  
& 0.9507 $\pm$ 0.0180 
& 0.9558 $\pm$ 0.0160 
& 0.9507 $\pm$ 0.0180 
& 0.9506 $\pm$ 0.0174 \\

DT  
& 0.9442 $\pm$ 0.0382 
& 0.9484 $\pm$ 0.0361 
& 0.9442 $\pm$ 0.0382 
& 0.9439 $\pm$ 0.0373 \\

\midrule
\multicolumn{5}{l}{\textbf{After QSMOTE}} \\
\midrule
LR  
& \textbf{0.9765 $\pm$ 0.0072} 
& \textbf{0.9768 $\pm$ 0.0072} 
& \textbf{0.9765 $\pm$ 0.0072} 
& \textbf{0.9764 $\pm$ 0.0072} \\

RF  
& \textbf{0.9855 $\pm$ 0.0122} 
& \textbf{0.9861 $\pm$ 0.0116} 
& \textbf{0.9855 $\pm$ 0.0122} 
& \textbf{0.9855 $\pm$ 0.0122} \\

SVM 
& \textbf{0.9765 $\pm$ 0.0157} 
& \textbf{0.9771 $\pm$ 0.0156} 
& \textbf{0.9765 $\pm$ 0.0157} 
& \textbf{0.9765 $\pm$ 0.0157} \\

NB  
& \textbf{0.9656 $\pm$ 0.0155} 
& \textbf{0.9665 $\pm$ 0.0151} 
& \textbf{0.9656 $\pm$ 0.0155} 
& \textbf{0.9657 $\pm$ 0.0154} \\

DT  
& \textbf{0.9729 $\pm$ 0.0213} 
& \textbf{0.9738 $\pm$ 0.0203} 
& \textbf{0.9729 $\pm$ 0.0213} 
& \textbf{0.9728 $\pm$ 0.0214} \\
\bottomrule
\end{tabular*}
\end{table}

The results in Table~\ref{table:performance_smote_three_2} demonstrate the effectiveness of QSMOTE in improving tertiary classification performance across most classifiers. Before QSMOTE, LR, RF, and SVM achieve comparable accuracies of $0.9637 \pm 0.0165$, $0.9639 \pm 0.0191$, and $0.9639 \pm 0.0191$, respectively, with SVM attaining the highest precision of $0.9671 \pm 0.0174$. NB and DT underperform with accuracies of $0.9507 \pm 0.0180$ and $0.9442 \pm 0.0382$, respectively. After applying QSMOTE, all classifiers exhibit performance gains, with RF again emerging as the best performer, achieving an accuracy of $0.9855 \pm 0.0122$, precision of $0.9861 \pm 0.0116$, recall of $0.9855 \pm 0.0122$, and F1 score of $0.9855 \pm 0.0122$. This corresponds to a relative improvement of approximately $2.2\%$ in accuracy compared to its pre-QSMOTE counterpart. LR and SVM also improve to accuracies of $0.9765 \pm 0.0072$ and $0.9765 \pm 0.0157$, respectively, while DT rises markedly from $0.9442 \pm 0.0382$ to $0.9729 \pm 0.0213$, showing a relative gain of nearly $3.0\%$. NB, although showing a modest increase, improves to $0.9656 \pm 0.0155$. Overall, the application of QSMOTE enhances the robustness and balance of classifiers, with RF consistently outperforming all other methods across all evaluation metrics.

\begin{table}[ht]
\caption{Performance Metrics of Classical Algorithms Before and After QSMOTE for Tetra-Class Classification on CWRUBD}
\label{table:performance_smote_four_2}
\centering
\begin{tabular*}{\textwidth}{@{\extracolsep\fill}lcccc}
\toprule
\textbf{Algorithm} 
& \textbf{Accuracy} 
& \textbf{Precision} 
& \textbf{Recall} 
& \textbf{F1 Measure} \\
\midrule
\multicolumn{5}{l}{\textbf{Before QSMOTE}} \\
\midrule
LR  
& 0.9698 $\pm$ 0.0201 
& 0.9706 $\pm$ 0.0207 
& 0.9698 $\pm$ 0.0201 
& 0.9696 $\pm$ 0.0204 \\

RF  
& 0.9781 $\pm$ 0.0164 
& 0.9794 $\pm$ 0.0153 
& 0.9781 $\pm$ 0.0164 
& 0.9773 $\pm$ 0.0172 \\

SVM 
& 0.9753 $\pm$ 0.0159 
& 0.9781 $\pm$ 0.0123 
& 0.9753 $\pm$ 0.0159 
& 0.9740 $\pm$ 0.0183 \\

NB  
& 0.9588 $\pm$ 0.0149 
& 0.9627 $\pm$ 0.0121 
& 0.9588 $\pm$ 0.0149 
& 0.9589 $\pm$ 0.0141 \\

DT  
& 0.9616 $\pm$ 0.0201 
& 0.9626 $\pm$ 0.0203 
& 0.9616 $\pm$ 0.0201 
& 0.9615 $\pm$ 0.0202 \\

\midrule
\multicolumn{5}{l}{\textbf{After QSMOTE}} \\
\midrule
LR  
& \textbf{0.9755 $\pm$ 0.0133} 
& \textbf{0.9762 $\pm$ 0.0131} 
& \textbf{0.9755 $\pm$ 0.0133} 
& \textbf{0.9756 $\pm$ 0.0133} \\

RF  
& \textbf{0.9918 $\pm$ 0.0027} 
& \textbf{0.9921 $\pm$ 0.0027} 
& \textbf{0.9918 $\pm$ 0.0027} 
& \textbf{0.9918 $\pm$ 0.0028} \\

SVM 
& \textbf{0.9878 $\pm$ 0.0100} 
& \textbf{0.9882 $\pm$ 0.0097} 
& \textbf{0.9878 $\pm$ 0.0100} 
& \textbf{0.9878 $\pm$ 0.0100} \\

NB  
& 0.9497 $\pm$ 0.0092 
& 0.9507 $\pm$ 0.0093 
& 0.9497 $\pm$ 0.0092 
& 0.9497 $\pm$ 0.0092 \\

DT  
& \textbf{0.9891 $\pm$ 0.0033} 
& \textbf{0.9895 $\pm$ 0.0032} 
& \textbf{0.9891 $\pm$ 0.0033} 
& \textbf{0.9891 $\pm$ 0.0034} \\
\bottomrule
\end{tabular*}
\end{table}

The results in Table~\ref{table:performance_smote_four_2} highlight the quantitative improvements achieved through QSMOTE in the tetra-class classification task. Before resampling, RF attains the best performance with an accuracy of $0.9781 \pm 0.0164$, precision of $0.9794 \pm 0.0153$, recall of $0.9781 \pm 0.0164$, and F1 score of $0.9773 \pm 0.0172$. LR and SVM follow closely with accuracies of $0.9698 \pm 0.0201$ and $0.9753 \pm 0.0159$, respectively, while NB perform the weakest with accuracy $0.9588 \pm 0.0149$. After applying QSMOTE, all classifiers except NB demonstrate consistent improvements. RF achieves the highest overall metrics, reaching an accuracy of $0.9918 \pm 0.0027$, precision of $0.9921 \pm 0.0027$, recall of $0.9918 \pm 0.0027$, and F1 score of $0.9918 \pm 0.0028$, representing an approximate $1.4\%$ accuracy gain compared to its performance on the imbalanced dataset. SVM and DT also show significant improvements, achieving accuracies of $0.9878 \pm 0.0100$ and $0.9891 \pm 0.0033$, respectively. In contrast, NB experiences a slight decrease in performance after QSMOTE, dropping from $0.9588 \pm 0.0149$ to $0.9497 \pm 0.0092$ in accuracy. Overall, QSMOTE proves effective in improving the robustness of most classifiers, particularly RF, which outperformed all other methods across all evaluation metrics.

\begin{table}[ht]
\caption{Performance Metrics of Classical Algorithms Before and After QSMOTE for Penta-Class Classification on CWRUBD}
\label{table:performance_smote_five_2}
\centering
\begin{tabular*}{\textwidth}{@{\extracolsep\fill}lcccc}
\toprule
\textbf{Algorithm} 
& \textbf{Accuracy} 
& \textbf{Precision} 
& \textbf{Recall} 
& \textbf{F1 Measure} \\
\midrule
\multicolumn{5}{l}{\textbf{Before QSMOTE}} \\
\midrule
LR  
& 0.9757 $\pm$ 0.0204 
& 0.9767 $\pm$ 0.0198 
& 0.9757 $\pm$ 0.0204 
& 0.9753 $\pm$ 0.0208 \\

RF  
& 0.9685 $\pm$ 0.0122 
& 0.9697 $\pm$ 0.0123 
& 0.9685 $\pm$ 0.0122 
& 0.9682 $\pm$ 0.0121 \\

SVM 
& 0.9758 $\pm$ 0.0107 
& 0.9767 $\pm$ 0.0106 
& 0.9758 $\pm$ 0.0107 
& 0.9751 $\pm$ 0.0111 \\

NB  
& 0.9733 $\pm$ 0.0142 
& 0.9736 $\pm$ 0.0156 
& 0.9733 $\pm$ 0.0142 
& 0.9727 $\pm$ 0.0153 \\

DT  
& 0.9612 $\pm$ 0.0140 
& 0.9648 $\pm$ 0.0126 
& 0.9612 $\pm$ 0.0140 
& 0.9609 $\pm$ 0.0141 \\

\midrule
\multicolumn{5}{l}{\textbf{After QSMOTE}} \\
\midrule
LR  
& \textbf{0.9880 $\pm$ 0.0063} 
& \textbf{0.9882 $\pm$ 0.0062} 
& \textbf{0.9880 $\pm$ 0.0063} 
& \textbf{0.9880 $\pm$ 0.0063} \\

RF  
& \textbf{0.9957 $\pm$ 0.0041} 
& \textbf{0.9958 $\pm$ 0.0039} 
& \textbf{0.9957 $\pm$ 0.0041} 
& \textbf{0.9956 $\pm$ 0.0041} \\

SVM 
& \textbf{0.9880 $\pm$ 0.0087} 
& \textbf{0.9882 $\pm$ 0.0086} 
& \textbf{0.9880 $\pm$ 0.0087} 
& \textbf{0.9880 $\pm$ 0.0087} \\

NB  
& 0.9446 $\pm$ 0.0121 
& 0.9471 $\pm$ 0.0112 
& 0.9446 $\pm$ 0.0121 
& 0.9439 $\pm$ 0.0124 \\

DT  
& \textbf{0.9815 $\pm$ 0.0027} 
& \textbf{0.9823 $\pm$ 0.0024} 
& \textbf{0.9815 $\pm$ 0.0027} 
& \textbf{0.9814 $\pm$ 0.0026} \\
\bottomrule
\end{tabular*}
\end{table}

The results in Table~\ref{table:performance_smote_five_2} show that QSMOTE consistently improves most classifiers for the penta-class dataset, with the exception of NB. LR improves from an accuracy of $0.9757 \pm 0.0204$ to $0.9880 \pm 0.0063$, with corresponding increases in precision ($0.9767 \pm 0.0198$ to $0.9882 \pm 0.0062$), recall ($0.9757 \pm 0.0204$ to $0.9880 \pm 0.0063$), and F1-score ($0.9753 \pm 0.0208$ to $0.9880 \pm 0.0063$). RF achieves the highest performance, with accuracy rising from $0.9685 \pm 0.0122$ to $0.9957 \pm 0.0041$, and all other metrics exceeding $0.995$. Similarly, SVM improves in accuracy from $0.9758 \pm 0.0107$ to $0.9880 \pm 0.0087$, and in F1-score from $0.9751 \pm 0.0111$ to $0.9880 \pm 0.0087$. DT also benefits, with accuracy increasing from $0.9612 \pm 0.0140$ to $0.9815 \pm 0.0027$, and F1-score from $0.9609 \pm 0.0141$ to $0.9814 \pm 0.0026$. In contrast, NB shows a decline, as accuracy drops from $0.9733 \pm 0.0142$ to $0.9446 \pm 0.0121$, and F1-score decreases from $0.9727 \pm 0.0153$ to $0.9439 \pm 0.0124$, indicating oversensitivity to synthetic samples. Overall, QSMOTE strongly boosts LR, RF, SVM, and DT, with RF achieving near-perfect classification, while NB remains adversely affected.

\begin{table}[ht]
\caption{Performance Metrics of Classical Algorithms Before and After QSMOTE for Hexa-Class Classification on CWRUBD}
\label{table:performance_smote_six_2}
\centering
\begin{tabular*}{\textwidth}{@{\extracolsep\fill}lcccc}
\toprule
\textbf{Algorithm} 
& \textbf{Accuracy} 
& \textbf{Precision} 
& \textbf{Recall} 
& \textbf{F1 Measure} \\
\midrule
\multicolumn{5}{l}{\textbf{Before QSMOTE}} \\
\midrule
LR  
& 0.9707 $\pm$ 0.0090 
& 0.9718 $\pm$ 0.0099 
& 0.9707 $\pm$ 0.0090 
& 0.9702 $\pm$ 0.0091 \\

RF  
& 0.9752 $\pm$ 0.0166 
& 0.9771 $\pm$ 0.0157 
& 0.9752 $\pm$ 0.0166 
& 0.9748 $\pm$ 0.0169 \\

SVM 
& 0.9775 $\pm$ 0.0071 
& 0.9790 $\pm$ 0.0069 
& 0.9775 $\pm$ 0.0071 
& 0.9771 $\pm$ 0.0071 \\

NB  
& 0.9662 $\pm$ 0.0144 
& 0.9677 $\pm$ 0.0153 
& 0.9662 $\pm$ 0.0144 
& 0.9655 $\pm$ 0.0145 \\

DT  
& 0.9370 $\pm$ 0.0088 
& 0.9428 $\pm$ 0.0109 
& 0.9370 $\pm$ 0.0088 
& 0.9358 $\pm$ 0.0063 \\

\midrule
\multicolumn{5}{l}{\textbf{After QSMOTE}} \\
\midrule
LR  
& \textbf{0.9882 $\pm$ 0.0068} 
& \textbf{0.9884 $\pm$ 0.0067} 
& \textbf{0.9882 $\pm$ 0.0068} 
& \textbf{0.9882 $\pm$ 0.0068} \\

RF  
& \textbf{0.9909 $\pm$ 0.0050} 
& \textbf{0.9911 $\pm$ 0.0050} 
& \textbf{0.9909 $\pm$ 0.0050} 
& \textbf{0.9909 $\pm$ 0.0050} \\

SVM 
& \textbf{0.9855 $\pm$ 0.0060} 
& \textbf{0.9858 $\pm$ 0.0059} 
& \textbf{0.9855 $\pm$ 0.0060} 
& \textbf{0.9855 $\pm$ 0.0060} \\

NB  
& 0.9339 $\pm$ 0.0142 
& 0.9378 $\pm$ 0.0149 
& 0.9339 $\pm$ 0.0142 
& 0.9325 $\pm$ 0.0152 \\

DT  
& \textbf{0.9783 $\pm$ 0.0101} 
& \textbf{0.9791 $\pm$ 0.0098} 
& \textbf{0.9783 $\pm$ 0.0101} 
& \textbf{0.9780 $\pm$ 0.0104} \\
\bottomrule
\end{tabular*}
\end{table}

The results in Table~\ref{table:performance_smote_six_2} show that QSMOTE considerably improves classifier performance for the hexa-class dataset, with the exception of NB. LR exhibits accuracy gains from $0.9707 \pm 0.0090$ to $0.9882 \pm 0.0068$, along with improvements in precision ($0.9718 \pm 0.0099$ to $0.9884 \pm 0.0067$), recall ($0.9707 \pm 0.0090$ to $0.9882 \pm 0.0068$), and F1-score ($0.9702 \pm 0.0091$ to $0.9882 \pm 0.0068$). RF achieves the highest overall performance, with accuracy increasing from $0.9752 \pm 0.0166$ to $0.9909 \pm 0.0050$, and all other metrics converging above $0.990$. SVM also demonstrates notable improvement, as accuracy rises from $0.9775 \pm 0.0071$ to $0.9855 \pm 0.0060$, and F1-score from $0.9771 \pm 0.0071$ to $0.9855 \pm 0.0060$. DT benefits substantially, with accuracy increasing from $0.9370 \pm 0.0088$ to $0.9783 \pm 0.0101$, and F1-score from $0.9358 \pm 0.0063$ to $0.9780 \pm 0.0104$. In contrast, NB experiences a significant drop, with accuracy declining from $0.9662 \pm 0.0144$ to $0.9339 \pm 0.0142$, and F1-score decreasing from $0.9655 \pm 0.0145$ to $0.9325 \pm 0.0152$. Overall, QSMOTE yields strong improvements for LR, RF, SVM, and DT, with RF achieving near-perfect classification results, while NB appears oversensitive to the synthetic data.

\begin{table}[ht]
\caption{Performance Metrics of Classical Algorithms Before and After QSMOTE for Hepta-Class Classification on CWRUBD}
\label{table:performance_smote_seven_2}
\centering
\begin{tabular*}{\textwidth}{@{\extracolsep\fill}lcccc}
\toprule
\textbf{Algorithm} 
& \textbf{Accuracy} 
& \textbf{Precision} 
& \textbf{Recall} 
& \textbf{F1 Measure} \\
\midrule
\multicolumn{5}{l}{\textbf{Before QSMOTE}} \\
\midrule
LR  
& 0.9701 $\pm$ 0.0062 
& 0.9717 $\pm$ 0.0057 
& 0.9701 $\pm$ 0.0062 
& 0.9688 $\pm$ 0.0070 \\

RF  
& 0.9782 $\pm$ 0.0170 
& 0.9793 $\pm$ 0.0168 
& 0.9782 $\pm$ 0.0170 
& 0.9777 $\pm$ 0.0177 \\

SVM 
& 0.9661 $\pm$ 0.0173 
& 0.9671 $\pm$ 0.0178 
& 0.9661 $\pm$ 0.0173 
& 0.9654 $\pm$ 0.0175 \\

NB  
& 0.9661 $\pm$ 0.0101 
& 0.9683 $\pm$ 0.0103 
& 0.9661 $\pm$ 0.0101 
& 0.9655 $\pm$ 0.0106 \\

DT  
& 0.9562 $\pm$ 0.0203 
& 0.9588 $\pm$ 0.0226 
& 0.9562 $\pm$ 0.0203 
& 0.9555 $\pm$ 0.0220 \\

\midrule
\multicolumn{5}{l}{\textbf{After QSMOTE}} \\
\midrule
LR  
& \textbf{0.9767 $\pm$ 0.0073} 
& \textbf{0.9771 $\pm$ 0.0071} 
& \textbf{0.9767 $\pm$ 0.0073} 
& \textbf{0.9766 $\pm$ 0.0075} \\

RF  
& \textbf{0.9984 $\pm$ 0.0019} 
& \textbf{0.9985 $\pm$ 0.0018} 
& \textbf{0.9984 $\pm$ 0.0019} 
& \textbf{0.9984 $\pm$ 0.0019} \\

SVM 
& \textbf{0.9860 $\pm$ 0.0031} 
& \textbf{0.9865 $\pm$ 0.0032} 
& \textbf{0.9860 $\pm$ 0.0031} 
& \textbf{0.9860 $\pm$ 0.0031} \\

NB  
& 0.9293 $\pm$ 0.0045 
& 0.9322 $\pm$ 0.0052 
& 0.9293 $\pm$ 0.0045 
& 0.9282 $\pm$ 0.0050 \\

DT  
& \textbf{0.9876 $\pm$ 0.0062} 
& \textbf{0.9881 $\pm$ 0.0060} 
& \textbf{0.9876 $\pm$ 0.0062} 
& \textbf{0.9875 $\pm$ 0.0063} \\
\bottomrule
\end{tabular*}
\end{table}

The results in Table~\ref{table:performance_smote_seven_2} demonstrate that QSMOTE consistently enhances classifier performance for the hepta-class dataset, with the exception of NB. For LR, the accuracy improves from $0.9701 \pm 0.0062$ to $0.9767 \pm 0.0073$, accompanied by corresponding increases in precision ($0.9717 \pm 0.0057$ to $0.9771 \pm 0.0071$), recall ($0.9701 \pm 0.0062$ to $0.9767 \pm 0.0073$), and F1-score ($0.9688 \pm 0.0070$ to $0.9766 \pm 0.0075$). RF achieves the highest overall performance, with accuracy rising from $0.9782 \pm 0.0170$ to $0.9984 \pm 0.0019$, while all other metrics converge above $0.998$. Similarly, SVM improves from $0.9661 \pm 0.0173$ to $0.9860 \pm 0.0031$ in accuracy, and from $0.9654 \pm 0.0175$ to $0.9860 \pm 0.0031$ in F1-score. DT also benefits considerably, with accuracy increasing from $0.9562 \pm 0.0203$ to $0.9876 \pm 0.0062$ and F1-score from $0.9555 \pm 0.0220$ to $0.9875 \pm 0.0063$. In contrast, NB experiences a significant drop after QSMOTE, with accuracy decreasing from $0.9661 \pm 0.0101$ to $0.9293 \pm 0.0045$ and F1-score from $0.9655 \pm 0.0106$ to $0.9282 \pm 0.0050$, indicating sensitivity to synthetic sample generation. Overall, QSMOTE boosts the performance of LR, RF, SVM, and DT, with RF achieving near-perfect classification results.

\begin{table}[ht]
\caption{Performance Metrics of Classical Algorithms Before and After QSMOTE for Octa-Class Classification on CWRUBD}
\label{table:performance_smote_eight_2}
\centering
\begin{tabular*}{\textwidth}{@{\extracolsep\fill}lcccc}
\toprule
\textbf{Algorithm} 
& \textbf{Accuracy} 
& \textbf{Precision} 
& \textbf{Recall} 
& \textbf{F1 Measure} \\
\midrule
\multicolumn{5}{l}{\textbf{Before QSMOTE}} \\
\midrule
LR  
& 0.8745 $\pm$ 0.0163 
& 0.8945 $\pm$ 0.0166 
& 0.8745 $\pm$ 0.0163 
& 0.8665 $\pm$ 0.0159 \\

RF  
& 0.9495 $\pm$ 0.0140 
& 0.9519 $\pm$ 0.0135 
& 0.9495 $\pm$ 0.0140 
& 0.9487 $\pm$ 0.0148 \\

SVM 
& 0.8780 $\pm$ 0.0269 
& 0.8865 $\pm$ 0.0385 
& 0.8780 $\pm$ 0.0269 
& 0.8613 $\pm$ 0.0255 \\

NB  
& 0.9163 $\pm$ 0.0311 
& 0.9220 $\pm$ 0.0291 
& 0.9163 $\pm$ 0.0311 
& 0.9147 $\pm$ 0.0320 \\

DT  
& 0.9129 $\pm$ 0.0207 
& 0.9239 $\pm$ 0.0134 
& 0.9129 $\pm$ 0.0207 
& 0.9148 $\pm$ 0.0181 \\

\midrule
\multicolumn{5}{l}{\textbf{After QSMOTE}} \\
\midrule
LR  
& \textbf{0.9409 $\pm$ 0.0153} 
& \textbf{0.9448 $\pm$ 0.0142} 
& \textbf{0.9409 $\pm$ 0.0153} 
& \textbf{0.9415 $\pm$ 0.0149} \\

RF  
& \textbf{0.9844 $\pm$ 0.0046} 
& \textbf{0.9850 $\pm$ 0.0047} 
& \textbf{0.9844 $\pm$ 0.0046} 
& \textbf{0.9844 $\pm$ 0.0046} \\

SVM 
& \textbf{0.9341 $\pm$ 0.0090} 
& \textbf{0.9437 $\pm$ 0.0041} 
& \textbf{0.9341 $\pm$ 0.0090} 
& \textbf{0.9353 $\pm$ 0.0088} \\

NB  
& 0.8818 $\pm$ 0.0140 
& 0.9056 $\pm$ 0.0087 
& 0.8818 $\pm$ 0.0140 
& 0.8827 $\pm$ 0.0148 \\

DT  
& \textbf{0.9790 $\pm$ 0.0084} 
& \textbf{0.9796 $\pm$ 0.0081} 
& \textbf{0.9790 $\pm$ 0.0084} 
& \textbf{0.9789 $\pm$ 0.0085} \\
\bottomrule
\end{tabular*}
\end{table}

The results in Table~\ref{table:performance_smote_eight_2} show that QSMOTE substantially improves performance across most classifiers for the octa-class dataset. For LR, the accuracy increases from $0.8745 \pm 0.0163$ to $0.9409 \pm 0.0153$, with corresponding gains in precision ($0.8945 \pm 0.0166$ to $0.9448 \pm 0.0142$), recall ($0.8745 \pm 0.0163$ to $0.9409 \pm 0.0153$), and F1-score ($0.8665 \pm 0.0159$ to $0.9415 \pm 0.0149$). RF achieves the strongest results, with accuracy improving from $0.9495 \pm 0.0140$ to $0.9844 \pm 0.0046$, while all other metrics remain above $0.984$. Similarly, SVM improves from $0.8780 \pm 0.0269$ to $0.9341 \pm 0.0090$ in accuracy, and from $0.8613 \pm 0.0255$ to $0.9353 \pm 0.0088$ in F1-score. In contrast, NB shows a decline after QSMOTE, with accuracy dropping from $0.9163 \pm 0.0311$ to $0.8818 \pm 0.0140$ and F1-score decreasing from $0.9147 \pm 0.0320$ to $0.8827 \pm 0.0148$, suggesting oversensitivity to synthetic samples. DT demonstrates notable improvement, with accuracy rising from $0.9129 \pm 0.0207$ to $0.9790 \pm 0.0084$, and F1-score increasing from $0.9148 \pm 0.0181$ to $0.9789 \pm 0.0085$. Overall, QSMOTE enhances minority class representation and leads to strong gains in LR, RF, SVM, and DT, with RF and DT achieving near-optimal classification performance.

\begin{table}[ht]
\caption{Performance Metrics of Classical Algorithms Before and After QSMOTE for Nona-Class Classification on CWRUBD}
\label{table:performance_smote_nine_2}
\centering
\begin{tabular*}{\textwidth}{@{\extracolsep\fill}lcccc}
\toprule
\textbf{Algorithm} 
& \textbf{Accuracy} 
& \textbf{Precision} 
& \textbf{Recall} 
& \textbf{F1 Measure} \\
\midrule
\multicolumn{5}{l}{\textbf{Before QSMOTE}} \\
\midrule
LR  
& 0.9118 $\pm$ 0.0153 
& 0.9231 $\pm$ 0.0135 
& 0.9118 $\pm$ 0.0153 
& 0.9026 $\pm$ 0.0182 \\

RF  
& 0.9638 $\pm$ 0.0107 
& 0.9669 $\pm$ 0.0099 
& 0.9638 $\pm$ 0.0107 
& 0.9636 $\pm$ 0.0107 \\

SVM 
& 0.9102 $\pm$ 0.0080 
& 0.8963 $\pm$ 0.0324 
& 0.9102 $\pm$ 0.0080 
& 0.8905 $\pm$ 0.0103 \\

NB  
& 0.9638 $\pm$ 0.0080 
& 0.9669 $\pm$ 0.0081 
& 0.9638 $\pm$ 0.0080 
& 0.9635 $\pm$ 0.0080 \\

DT  
& 0.9276 $\pm$ 0.0195 
& 0.9376 $\pm$ 0.0159 
& 0.9276 $\pm$ 0.0195 
& 0.9284 $\pm$ 0.0198 \\

\midrule
\multicolumn{5}{l}{\textbf{After QSMOTE}} \\
\midrule
LR  
& \textbf{0.9571 $\pm$ 0.0049} 
& \textbf{0.9582 $\pm$ 0.0041} 
& \textbf{0.9571 $\pm$ 0.0049} 
& \textbf{0.9567 $\pm$ 0.0049} \\

RF  
& \textbf{0.9934 $\pm$ 0.0044} 
& \textbf{0.9936 $\pm$ 0.0043} 
& \textbf{0.9934 $\pm$ 0.0044} 
& \textbf{0.9933 $\pm$ 0.0045} \\

SVM 
& \textbf{0.9656 $\pm$ 0.0123} 
& \textbf{0.9675 $\pm$ 0.0120} 
& \textbf{0.9656 $\pm$ 0.0123} 
& \textbf{0.9653 $\pm$ 0.0126} \\

NB  
& 0.9293 $\pm$ 0.0078 
& 0.9312 $\pm$ 0.0093 
& 0.9293 $\pm$ 0.0078 
& 0.9281 $\pm$ 0.0085 \\

DT  
& \textbf{0.9885 $\pm$ 0.0029} 
& \textbf{0.9890 $\pm$ 0.0028} 
& \textbf{0.9885 $\pm$ 0.0029} 
& \textbf{0.9885 $\pm$ 0.0030} \\
\bottomrule
\end{tabular*}
\end{table}

The results in Table~\ref{table:performance_smote_nine_2} show a clear improvement in performance metrics after applying QSMOTE to the nona-class dataset. For LR, the accuracy rises from $0.9118 \pm 0.0153$ to $0.9571 \pm 0.0049$, with corresponding gains in precision ($0.9231 \pm 0.0135$ to $0.9582 \pm 0.0041$), recall ($0.9118 \pm 0.0153$ to $0.9571 \pm 0.0049$), and F1-score ($0.9026 \pm 0.0182$ to $0.9567 \pm 0.0049$). RF shows the strongest performance, with accuracy improving from $0.9638 \pm 0.0107$ to $0.9934 \pm 0.0044$, and precision, recall, and F1-score all staying above $0.993$. Similarly, SVM improves in accuracy from $0.9102 \pm 0.0080$ to $0.9656 \pm 0.0123$, and in F1-score from $0.8905 \pm 0.0103$ to $0.9653 \pm 0.0126$. On the other hand, NB, while initially competitive with accuracy $0.9638 \pm 0.0080$, drops to $0.9293 \pm 0.0078$ after QSMOTE, along with a slight decrease in F1-score from $0.9635 \pm 0.0080$ to $0.9281 \pm 0.0085$, indicating possible oversampling sensitivity. DT benefits significantly, as accuracy rises from $0.9276 \pm 0.0195$ to $0.9885 \pm 0.0029$, and F1-score increases from $0.9284 \pm 0.0198$ to $0.9885 \pm 0.0030$. Overall, QSMOTE improves the balance of class distributions, leading to performance boosts across most classifiers, with RF and DT showing near-perfect classification performance.

\begin{table}[ht]
\caption{Performance Metrics of Classical Algorithms Before and After QSMOTE for Binary-Class Classification on EFDD}
\label{table:performance_smote_binary_third_3}
\centering
\begin{tabular*}{\textwidth}{@{\extracolsep\fill}lcccc}
\toprule
\textbf{Algorithm}
& \textbf{Accuracy}
& \textbf{Precision}
& \textbf{Recall}
& \textbf{F1 Measure} \\
\midrule
\multicolumn{5}{l}{\textbf{Before QSMOTE}} \\
\midrule
LR
& 0.7132 $\pm$ 0.0146
& 0.5465 $\pm$ 0.0586
& 0.7132 $\pm$ 0.0146
& 0.6054 $\pm$ 0.0149 \\

RF
& 0.7098 $\pm$ 0.0393
& 0.6518 $\pm$ 0.1384
& 0.7098 $\pm$ 0.0393
& 0.6293 $\pm$ 0.0410 \\

SVM
& 0.7203 $\pm$ 0.0019
& 0.5188 $\pm$ 0.0028
& 0.7203 $\pm$ 0.0019
& 0.6031 $\pm$ 0.0026 \\

NB
& 0.7097 $\pm$ 0.0149
& 0.5458 $\pm$ 0.0590
& 0.7097 $\pm$ 0.0149
& 0.6037 $\pm$ 0.0157 \\

DT
& 0.5910 $\pm$ 0.0490
& 0.6088 $\pm$ 0.0324
& 0.5910 $\pm$ 0.0490
& 0.5978 $\pm$ 0.0407 \\

\midrule
\multicolumn{5}{l}{\textbf{After QSMOTE}} \\
\midrule
LR
& 0.5464 $\pm$ 0.0653
& 0.5470 $\pm$ 0.0660
& 0.5464 $\pm$ 0.0653
& 0.5455 $\pm$ 0.0657 \\

RF
& \textbf{0.8544 $\pm$ 0.0148}
& \textbf{0.8618 $\pm$ 0.0198}
& \textbf{0.8544 $\pm$ 0.0148}
& \textbf{0.8538 $\pm$ 0.0145} \\

SVM
& \textbf{0.7283 $\pm$ 0.0380}
& \textbf{0.7361 $\pm$ 0.0365}
& \textbf{0.7283 $\pm$ 0.0380}
& \textbf{0.7260 $\pm$ 0.0391} \\

NB
& 0.5825 $\pm$ 0.0519
& 0.5837 $\pm$ 0.0525
& 0.5825 $\pm$ 0.0519
& 0.5800 $\pm$ 0.0535 \\

DT
& \textbf{0.8132 $\pm$ 0.0188}
& \textbf{0.8278 $\pm$ 0.0306}
& \textbf{0.8132 $\pm$ 0.0188}
& \textbf{0.8114 $\pm$ 0.0177} \\
\bottomrule
\end{tabular*}
\end{table}

The performance of the classical algorithms for binary-class classification on the Engine Failure Detection Dataset (EFDD) is summarized in Table~\ref{table:performance_smote_binary_third_3}. Prior to applying QSMOTE, SVM achieves the highest accuracy (0.7203 $\pm$ 0.0019), while RF exhibits strong precision (0.6518 $\pm$ 0.1384) but with considerable variance, reflecting instability under class imbalance. LR and NB produce comparable outcomes, with F1-scores around 0.60, whereas DT lags behind with lower accuracy (0.5910 $\pm$ 0.0490). After resampling with QSMOTE, substantial improvements are observed for RF and DT, with RF achieving the best overall performance (accuracy: 0.8544 $\pm$ 0.0148, F1: 0.8538 $\pm$ 0.0145) and DT following closely (accuracy: 0.8132 $\pm$ 0.0188, F1: 0.8114 $\pm$ 0.0177). SVM also benefits moderately from QSMOTE, with improvements across all metrics, whereas LR and NB show a decline in performance, suggesting oversensitivity to the resampled data distribution. These results highlight that ensemble- and tree-based models are more effective and robust for binary classification under imbalanced conditions when augmented with QSMOTE.

\begin{table}[ht]
\caption{Performance Metrics of Classical Algorithms Before and After QSMOTE for Tertiary-Class Classification on EFDD}
\label{table:performance_smote_three_third_3}
\centering
\begin{tabular*}{\textwidth}{@{\extracolsep\fill}lcccc}
\toprule
\textbf{Algorithm}
& \textbf{Accuracy}
& \textbf{Precision}
& \textbf{Recall}
& \textbf{F1 Measure} \\
\midrule
\multicolumn{5}{l}{\textbf{Before QSMOTE}} \\
\midrule
LR
& 0.6227 $\pm$ 0.0133
& 0.3983 $\pm$ 0.0047
& 0.6227 $\pm$ 0.0133
& 0.4858 $\pm$ 0.0071 \\

RF
& 0.6073 $\pm$ 0.0175
& 0.4607 $\pm$ 0.0618
& 0.6073 $\pm$ 0.0175
& 0.4955 $\pm$ 0.0236 \\

SVM
& 0.6288 $\pm$ 0.0071
& 0.3986 $\pm$ 0.0035
& 0.6288 $\pm$ 0.0071
& 0.4879 $\pm$ 0.0045 \\

NB
& 0.6196 $\pm$ 0.0060
& 0.4215 $\pm$ 0.0511
& 0.6196 $\pm$ 0.0060
& 0.4884 $\pm$ 0.0124 \\

DT
& 0.4633 $\pm$ 0.0519
& 0.4785 $\pm$ 0.0545
& 0.4633 $\pm$ 0.0519
& 0.4681 $\pm$ 0.0502 \\

\midrule
\multicolumn{5}{l}{\textbf{After QSMOTE}} \\
\midrule
LR
& 0.3883 $\pm$ 0.0428
& 0.3866 $\pm$ 0.0428
& 0.3883 $\pm$ 0.0428
& 0.3811 $\pm$ 0.0390 \\

RF
& \textbf{0.8769 $\pm$ 0.0281}
& \textbf{0.8777 $\pm$ 0.0282}
& \textbf{0.8769 $\pm$ 0.0281}
& \textbf{0.8756 $\pm$ 0.0285} \\

SVM
& \textbf{0.6327 $\pm$ 0.0239}
& \textbf{0.6321 $\pm$ 0.0246}
& \textbf{0.6327 $\pm$ 0.0239}
& \textbf{0.6288 $\pm$ 0.0232} \\

NB
& 0.4177 $\pm$ 0.0667
& 0.4197 $\pm$ 0.0714
& 0.4177 $\pm$ 0.0667
& 0.4163 $\pm$ 0.0687 \\

DT
& \textbf{0.8236 $\pm$ 0.0421}
& \textbf{0.8296 $\pm$ 0.0492}
& \textbf{0.8236 $\pm$ 0.0421}
& \textbf{0.8158 $\pm$ 0.0432} \\
\bottomrule
\end{tabular*}
\end{table}

The results in Table~\ref{table:performance_smote_three_third_3} present the performance of five classical machine learning algorithms before and after applying QSMOTE on the third dataset for tertiary-class classification. Before QSMOTE, all models exhibit moderate performance, with SVM achieving the highest accuracy (0.6288 $\pm$ 0.0071) and NB achieving the best balance between precision and recall. However, the overall F1-scores remain relatively low, indicating difficulty in handling class imbalance. After applying QSMOTE, significant improvements are observed, particularly for RF and DT, which achieve accuracies of 0.8769 $\pm$ 0.0281 and 0.8236 $\pm$ 0.0421, respectively, along with consistently high precision, recall, and F1-scores. In contrast, LR and NB experience performance degradation after resampling, highlighting their sensitivity to oversampled data. The SVM shows stable performance across both settings, with only a slight improvement after QSMOTE. Overall, the findings confirm that ensemble-based methods such as RF and tree-based models like DT are more robust in handling imbalanced datasets when combined with QSMOTE, making them the most effective choices for this classification task.

\begin{table}[ht]
\caption{Performance Metrics of Classical Algorithms Before and After QSMOTE for Binary Classification on IFDD}
\label{table:performance_smote_binary_fifth_5}
\centering
\begin{tabular*}{\textwidth}{@{\extracolsep\fill}lcccc}
\toprule
\textbf{Algorithm}
& \textbf{Accuracy}
& \textbf{Precision}
& \textbf{Recall}
& \textbf{F1 Measure} \\
\midrule
\multicolumn{5}{l}{\textbf{Before QSMOTE}} \\
\midrule
LR
& 0.8432 $\pm$ 0.0073
& 0.7144 $\pm$ 0.0082
& 0.8432 $\pm$ 0.0073
& 0.7735 $\pm$ 0.0076 \\

RF
& 0.8475 $\pm$ 0.0080
& 0.7460 $\pm$ 0.0678
& 0.8475 $\pm$ 0.0080
& 0.7794 $\pm$ 0.0148 \\

SVM
& 0.8454 $\pm$ 0.0047
& 0.7146 $\pm$ 0.0080
& 0.8454 $\pm$ 0.0047
& 0.7745 $\pm$ 0.0066 \\

NB
& 0.7165 $\pm$ 0.1439
& 0.7104 $\pm$ 0.0074
& 0.7165 $\pm$ 0.1439
& 0.6989 $\pm$ 0.0952 \\

DT
& 0.7055 $\pm$ 0.0323
& 0.7222 $\pm$ 0.0307
& 0.7055 $\pm$ 0.0323
& 0.7134 $\pm$ 0.0299 \\

\midrule
\multicolumn{5}{l}{\textbf{After QSMOTE}} \\
\midrule
LR
& 0.6241 $\pm$ 0.0382
& 0.6245 $\pm$ 0.0380
& 0.6241 $\pm$ 0.0382
& 0.6237 $\pm$ 0.0384 \\

RF
& \textbf{0.9900 $\pm$ 0.0031}
& \textbf{0.9901 $\pm$ 0.0030}
& \textbf{0.9900 $\pm$ 0.0031}
& \textbf{0.9900 $\pm$ 0.0031} \\

SVM
& \textbf{0.8559 $\pm$ 0.0251}
& \textbf{0.8629 $\pm$ 0.0253}
& \textbf{0.8559 $\pm$ 0.0251}
& \textbf{0.8552 $\pm$ 0.0253} \\

NB
& 0.6379 $\pm$ 0.0448
& 0.6450 $\pm$ 0.0507
& 0.6379 $\pm$ 0.0448
& 0.6345 $\pm$ 0.0436 \\

DT
& \textbf{0.8885 $\pm$ 0.0172}
& \textbf{0.9083 $\pm$ 0.0119}
& \textbf{0.8885 $\pm$ 0.0172}
& \textbf{0.8871 $\pm$ 0.0178} \\
\bottomrule
\end{tabular*}
\end{table}

Table \ref{table:performance_smote_binary_fifth_5} shows the comparative evaluation of classical classifiers before and after QSMOTE for binary classification on Industrial Fault Detection Dataset (IFDD). Before applying QSMOTE, RF (0.8475 $\pm$ 0.0080) achieves the best accuracy, follows closely by SVM (0.8454 $\pm$ 0.0047) and LR (0.8432 $\pm$ 0.0073). Both DT and NB trail with accuracies around 0.70. After QSMOTE, the ensemble and tree-based models exhibit remarkable improvement. RF achieves an almost perfect performance with 0.9900 $\pm$ 0.0031 accuracy, marking an absolute gain of +14.3\% points. DT follows with 0.8885 $\pm$ 0.0172, a +18.3-point improvement. The SVM classifier also improves modestly to 0.8559 $\pm$ 0.0251, up by +1.05 points. In contrast, LR and NB show slight declines of –21.9 and –7.9 points, respectively, suggesting that linear and probabilistic models are less adaptive to the synthetic distribution created by QSMOTE. Overall, QSMOTE oversampling significantly benefits non-linear and ensemble classifiers, yielding high precision–recall consistency and near-perfect generalization, particularly in RF and DT, confirming their robustness to class imbalance in binary settings.

\begin{table}[ht]
\caption{Performance Metrics of Classical Algorithms Before and After QSMOTE for Tertiary-Class Classification on IFDD}
\label{table:performance_smote_tri_sixth_6}
\centering
\begin{tabular*}{\textwidth}{@{\extracolsep\fill}lcccc}
\toprule
\textbf{Algorithm}
& \textbf{Accuracy}
& \textbf{Precision}
& \textbf{Recall}
& \textbf{F1 Measure} \\
\midrule
\multicolumn{5}{l}{\textbf{Before QSMOTE}} \\
\midrule
LR
& 0.7388 $\pm$ 0.0143
& 0.5614 $\pm$ 0.0071
& 0.7388 $\pm$ 0.0143
& 0.6380 $\pm$ 0.0095 \\

RF
& 0.7425 $\pm$ 0.0096
& 0.5611 $\pm$ 0.0067
& 0.7425 $\pm$ 0.0096
& 0.6392 $\pm$ 0.0076 \\

SVM
& 0.7500 $\pm$ 0.0039
& 0.5625 $\pm$ 0.0059
& 0.7500 $\pm$ 0.0039
& 0.6429 $\pm$ 0.0053 \\

NB
& 0.5571 $\pm$ 0.1555
& 0.5619 $\pm$ 0.0159
& 0.5571 $\pm$ 0.1555
& 0.5342 $\pm$ 0.1000 \\

DT
& 0.5262 $\pm$ 0.0269
& 0.5623 $\pm$ 0.0237
& 0.5262 $\pm$ 0.0269
& 0.5417 $\pm$ 0.0185 \\

\midrule
\multicolumn{5}{l}{\textbf{After QSMOTE}} \\
\midrule
LR
& 0.5004 $\pm$ 0.0164
& 0.4920 $\pm$ 0.0191
& 0.5004 $\pm$ 0.0164
& 0.4926 $\pm$ 0.0195 \\

RF
& \textbf{0.9925 $\pm$ 0.0031}
& \textbf{0.9926 $\pm$ 0.0030}
& \textbf{0.9925 $\pm$ 0.0031}
& \textbf{0.9925 $\pm$ 0.0031} \\

SVM
& \textbf{0.8304 $\pm$ 0.0205}
& \textbf{0.8313 $\pm$ 0.0200}
& \textbf{0.8304 $\pm$ 0.0205}
& \textbf{0.8273 $\pm$ 0.0209} \\

NB
& 0.4929 $\pm$ 0.0398
& 0.4881 $\pm$ 0.0455
& 0.4929 $\pm$ 0.0398
& 0.4799 $\pm$ 0.0415 \\

DT
& \textbf{0.8855 $\pm$ 0.0057}
& \textbf{0.9010 $\pm$ 0.0055}
& \textbf{0.8855 $\pm$ 0.0057}
& \textbf{0.8788 $\pm$ 0.0064} \\
\bottomrule
\end{tabular*}
\end{table}

Table \ref{table:performance_smote_tri_sixth_6} presents the comparative performance of classical classifiers before and after QSMOTE for tri-class classification on the sixth dataset. Before applying QSMOTE, SVM achieves the highest accuracy of 0.7500 $\pm$ 0.0039, followed closely by RF (0.7425 $\pm$ 0.0096) and LR (0.7388 $\pm$ 0.0143). The NB and DT models show weaker performance with accuracies of 0.5571 $\pm$ 0.1555 and 0.5262 $\pm$ 0.0269, respectively. After applying QSMOTE, the overall performance improves drastically for ensemble and tree-based models. RF attains a near-perfect accuracy of 0.9925 $\pm$ 0.0031, marking a +25.0 \% point improvement, while DT reaches 0.8855 $\pm$ 0.0057, improving by +35.9 points. SVM also improves notably to 0.8304 $\pm$ 0.0205, gaining +8.0 points, whereas LR and NB experience minor degradations of approximately –23.8 and –6.4 points, respectively. In terms of precision and F1 measure, RF demonstrate the most consistent and stable performance across all folds (standard deviation $\approx$ 0.003), signifying strong generalization. The results reaffirm that QSMOTE balancing significantly enhances ensemble and non-linear model performance, whereas linear models like LR and probabilistic ones like NB may not benefit due to overfitting on synthetic minority samples.

\bibliographystyle{cas-model2-names}
\bibliography{sn-article}

\begin{thebibliography}{38}
\expandafter\ifx\csname natexlab\endcsname\relax\def\natexlab#1{#1}\fi
\providecommand{\url}[1]{\texttt{#1}}
\providecommand{\href}[2]{#2}
\providecommand{\path}[1]{#1}
\providecommand{\DOIprefix}{doi:}
\providecommand{\ArXivprefix}{arXiv:}
\providecommand{\URLprefix}{URL: }
\providecommand{\Pubmedprefix}{pmid:}
\providecommand{\doi}[1]{\href{http://dx.doi.org/#1}{\path{#1}}}
\providecommand{\Pubmed}[1]{\href{pmid:#1}{\path{#1}}}
\providecommand{\bibinfo}[2]{#2}
\ifx\xfnm\relax \def\xfnm[#1]{\unskip,\space#1}\fi
%Type = Misc
\bibitem[{Afroz()}]{afroz2023solar}
\bibinfo{author}{Afroz, P.}, .
\newblock \bibinfo{title}{Solar panel images dataset}.
\newblock \bibinfo{howpublished}{\url{https://www.kaggle.com/datasets/pythonafroz/solar-panel-images}}.
\newblock \bibinfo{note}{Kaggle Dataset. Accessed: 2025-09-01}.
%Type = Article
\bibitem[{Bao et~al.(2026)Bao, Liu, Yang, Zhang and Li}]{zhang2020deep}
\bibinfo{author}{Bao, Z.}, \bibinfo{author}{Liu, C.}, \bibinfo{author}{Yang, H.}, \bibinfo{author}{Zhang, J.}, \bibinfo{author}{Li, Y.}, \bibinfo{year}{2026}.
\newblock \bibinfo{title}{From theory to industry: A survey of deep learning-enabled bearing fault diagnosis in complex environments}.
\newblock \bibinfo{journal}{Engineering Applications of Artificial Intelligence} \bibinfo{volume}{163}, \bibinfo{pages}{113068}.
\newblock \URLprefix \url{https://doi.org/10.1016/j.engappai.2025.113068}, \DOIprefix\doi{10.1016/j.engappai.2025.113068}.
%Type = Article
\bibitem[{Batista et~al.(2004)Batista, Prati and Monard}]{batista2004study}
\bibinfo{author}{Batista, G.E.}, \bibinfo{author}{Prati, R.C.}, \bibinfo{author}{Monard, M.C.}, \bibinfo{year}{2004}.
\newblock \bibinfo{title}{A study of the behavior of several methods for balancing machine learning training data}.
\newblock \bibinfo{journal}{SIGKDD Explorations} \bibinfo{volume}{6}, \bibinfo{pages}{20--29}.
\newblock \DOIprefix\doi{10.1145/1007730.100773}.
%Type = Article
\bibitem[{Behera et~al.(2025)Behera, Sergioli and Giuntini}]{behera2025qsmotepgmkpgm}
\bibinfo{author}{Behera, B.K.}, \bibinfo{author}{Sergioli, G.}, \bibinfo{author}{Giuntini, R.}, \bibinfo{year}{2025}.
\newblock \bibinfo{title}{Qsmote-pgm/kpgm: Qsmote based pgm and kpgm for imbalanced dataset classification}.
\newblock \bibinfo{journal}{arXiv preprint arXiv:2512.16960} \DOIprefix\doi{10.48550/arXiv.2512.16960}, \href{http://arxiv.org/abs/2512.16960}{\tt arXiv:2512.16960}.
%Type = Article
\bibitem[{Benedetti et~al.(2019)Benedetti, Lloyd, Sack and Fiorentini}]{benedetti2019parameterized}
\bibinfo{author}{Benedetti, M.}, \bibinfo{author}{Lloyd, E.}, \bibinfo{author}{Sack, S.}, \bibinfo{author}{Fiorentini, M.}, \bibinfo{year}{2019}.
\newblock \bibinfo{title}{Parameterized quantum circuits as machine learning models}.
\newblock \bibinfo{journal}{Quantum Science and Technology} \bibinfo{volume}{4}, \bibinfo{pages}{043001}.
\newblock \URLprefix \url{https://doi.org/10.1088/2058-9565/ab4eb5}, \DOIprefix\doi{10.1088/2058-9565/ab4eb5}.
%Type = Article
\bibitem[{Bhuiyan and Uddin(2023)}]{xu2021novel}
\bibinfo{author}{Bhuiyan, M.R.}, \bibinfo{author}{Uddin, J.}, \bibinfo{year}{2023}.
\newblock \bibinfo{title}{Deep transfer learning models for industrial fault diagnosis using vibration and acoustic sensors data: A review}.
\newblock \bibinfo{journal}{Vibration} \bibinfo{volume}{6}, \bibinfo{pages}{218--238}.
\newblock \URLprefix \url{https://doi.org/10.3390/vibration6010014}, \DOIprefix\doi{10.3390/vibration6010014}.
%Type = Article
\bibitem[{Biamonte et~al.(2017)Biamonte, Wittek, Pancotti, Rebentrost, Wiebe and Lloyd}]{biamonte2017quantum}
\bibinfo{author}{Biamonte, J.}, \bibinfo{author}{Wittek, P.}, \bibinfo{author}{Pancotti, N.}, \bibinfo{author}{Rebentrost, P.}, \bibinfo{author}{Wiebe, N.}, \bibinfo{author}{Lloyd, S.}, \bibinfo{year}{2017}.
\newblock \bibinfo{title}{Quantum machine learning}.
\newblock \bibinfo{journal}{Nature} \bibinfo{volume}{549}, \bibinfo{pages}{195--202}.
\newblock \DOIprefix\doi{10.1038/nature23474}.
%Type = Article
\bibitem[{Blagus and Lusa(2013)}]{blagus2013smote}
\bibinfo{author}{Blagus, R.}, \bibinfo{author}{Lusa, L.}, \bibinfo{year}{2013}.
\newblock \bibinfo{title}{Smote for high-dimensional class-imbalanced data}.
\newblock \bibinfo{journal}{BMC Bioinformatics} \bibinfo{volume}{14}, \bibinfo{pages}{106}.
\newblock \DOIprefix\doi{10.1186/1471-2105-14-106}.
%Type = Article
\bibitem[{Breiman(2001)}]{breiman2001random}
\bibinfo{author}{Breiman, L.}, \bibinfo{year}{2001}.
\newblock \bibinfo{title}{Random forests}.
\newblock \bibinfo{journal}{Machine Learning} \bibinfo{volume}{45}, \bibinfo{pages}{5--32}.
\newblock \DOIprefix\doi{10.1023/A:1010933404324}.
%Type = Misc
\bibitem[{{brjapon}()}]{cwru_kaggle}
\bibinfo{author}{{brjapon}}, .
\newblock \bibinfo{title}{Cwru bearing datasets}.
\newblock \bibinfo{howpublished}{\url{https://www.kaggle.com/datasets/brjapon/cwru-bearing-datasets}}.
\newblock \bibinfo{note}{Kaggle dataset. Accessed: 2025-09-26}.
%Type = Article
\bibitem[{Chawla et~al.(2002)Chawla, Bowyer, Hall and Kegelmeyer}]{chawla2002smote}
\bibinfo{author}{Chawla, N.V.}, \bibinfo{author}{Bowyer, K.W.}, \bibinfo{author}{Hall, L.O.}, \bibinfo{author}{Kegelmeyer, W.P.}, \bibinfo{year}{2002}.
\newblock \bibinfo{title}{Smote: Synthetic minority over-sampling technique}.
\newblock \bibinfo{journal}{Journal of Artificial Intelligence Research} \bibinfo{volume}{16}, \bibinfo{pages}{321--357}.
\newblock \URLprefix \url{https://dl.acm.org/doi/10.5555/1622407.1622416}.
%Type = Article
\bibitem[{Dallaire-Demers and Killoran(2018)}]{gao2017quantum}
\bibinfo{author}{Dallaire-Demers, P.L.}, \bibinfo{author}{Killoran, N.}, \bibinfo{year}{2018}.
\newblock \bibinfo{title}{Quantum generative adversarial networks}.
\newblock \bibinfo{journal}{Physical Review A} \bibinfo{volume}{98}, \bibinfo{pages}{012324}.
\newblock \URLprefix \url{https://doi.org/10.1103/PhysRevA.98.012324}, \DOIprefix\doi{10.1103/PhysRevA.98.012324}.
%Type = Book
\bibitem[{Fern{\'a}ndez et~al.(2018)Fern{\'a}ndez, Garc{\'i}a, Galar, Prati, Krawczyk and Herrera}]{fernandez2018learning}
\bibinfo{author}{Fern{\'a}ndez, A.}, \bibinfo{author}{Garc{\'i}a, S.}, \bibinfo{author}{Galar, M.}, \bibinfo{author}{Prati, R.C.}, \bibinfo{author}{Krawczyk, B.}, \bibinfo{author}{Herrera, F.}, \bibinfo{year}{2018}.
\newblock \bibinfo{title}{Learning from Imbalanced Data Sets}.
\newblock Springer Series in Statistics, \bibinfo{publisher}{Springer}, \bibinfo{address}{Cham}.
\newblock \DOIprefix\doi{10.1007/978-3-319-98074-4}.
%Type = Article
\bibitem[{Han et~al.(2005)Han, Wang and Mao}]{han2005borderline}
\bibinfo{author}{Han, H.}, \bibinfo{author}{Wang, W.Y.}, \bibinfo{author}{Mao, B.H.}, \bibinfo{year}{2005}.
\newblock \bibinfo{title}{Borderline-smote: A new over-sampling method in imbalanced data sets learning}.
\newblock \bibinfo{journal}{Proceedings of the 2005 International Conference on Advances in Intelligent Computing (ICIC'05), Volume Part I} , \bibinfo{pages}{878--887}\DOIprefix\doi{10.1007/11538059_91}.
%Type = Article
\bibitem[{He et~al.(2008)He, Bai, Garcia and Li}]{he2008adasyn}
\bibinfo{author}{He, H.}, \bibinfo{author}{Bai, Y.}, \bibinfo{author}{Garcia, E.A.}, \bibinfo{author}{Li, S.}, \bibinfo{year}{2008}.
\newblock \bibinfo{title}{Adasyn: Adaptive synthetic sampling approach for imbalanced learning}.
\newblock \bibinfo{journal}{IEEE International Joint Conference on Neural Networks} , \bibinfo{pages}{1322--1328}\DOIprefix\doi{10.1109/IJCNN.2008.4633969}.
%Type = Article
\bibitem[{He and Garcia(2009)}]{he2009learning}
\bibinfo{author}{He, H.}, \bibinfo{author}{Garcia, E.A.}, \bibinfo{year}{2009}.
\newblock \bibinfo{title}{Learning from imbalanced data}.
\newblock \bibinfo{journal}{IEEE Transactions on Knowledge and Data Engineering} \bibinfo{volume}{21}, \bibinfo{pages}{1263--1284}.
\newblock \DOIprefix\doi{10.1109/TKDE.2008.239}.
%Type = Inproceedings
\bibitem[{Hendrycks and Dietterich(2019)}]{hendrycks2019benchmark}
\bibinfo{author}{Hendrycks, D.}, \bibinfo{author}{Dietterich, T.}, \bibinfo{year}{2019}.
\newblock \bibinfo{title}{Benchmarking neural network robustness to common corruptions and perturbations}.
\newblock \DOIprefix\doi{10.48550/arXiv.1903.12261}.
%Type = Article
\bibitem[{Khanal and Rivas(2024)}]{zhang2022noise}
\bibinfo{author}{Khanal, B.}, \bibinfo{author}{Rivas, P.}, \bibinfo{year}{2024}.
\newblock \bibinfo{title}{Learning robust observable to address noise in quantum machine learning}.
\newblock \bibinfo{journal}{arXiv preprint arXiv:2409.07632} \URLprefix \url{https://doi.org/10.48550/arXiv.2409.07632}, \DOIprefix\doi{10.48550/arXiv.2409.07632}, \href{http://arxiv.org/abs/2409.07632}{\tt arXiv:2409.07632}.
%Type = Article
\bibitem[{Kieferov{\'a} and Wiebe(2017)}]{kieferova2017tomography}
\bibinfo{author}{Kieferov{\'a}, M.}, \bibinfo{author}{Wiebe, N.}, \bibinfo{year}{2017}.
\newblock \bibinfo{title}{Tomography and generative training with quantum boltzmann machines}.
\newblock \bibinfo{journal}{Physical Review A} \bibinfo{volume}{96}, \bibinfo{pages}{062327}.
\newblock \URLprefix \url{https://doi.org/10.1103/PhysRevA.96.062327}, \DOIprefix\doi{10.1103/PhysRevA.96.062327}.
%Type = Article
\bibitem[{Lee et~al.(2015)Lee, Ardakani, Yang and Bagheri}]{lee2015industrial}
\bibinfo{author}{Lee, J.}, \bibinfo{author}{Ardakani, H.D.}, \bibinfo{author}{Yang, S.}, \bibinfo{author}{Bagheri, B.}, \bibinfo{year}{2015}.
\newblock \bibinfo{title}{Industrial big data analytics and cyber-physical systems for future maintenance \& service innovation}.
\newblock \bibinfo{journal}{Procedia CIRP} \bibinfo{volume}{38}, \bibinfo{pages}{3--7}.
\newblock \URLprefix \url{https://doi.org/10.1016/j.procir.2015.08.026}, \DOIprefix\doi{10.1016/j.procir.2015.08.026}.
%Type = Article
\bibitem[{Lo et~al.(2019)Lo, Flaus and Adrot}]{hosseini2016review}
\bibinfo{author}{Lo, N.G.}, \bibinfo{author}{Flaus, J.M.}, \bibinfo{author}{Adrot, O.}, \bibinfo{year}{2019}.
\newblock \bibinfo{title}{Review of machine learning approaches in fault diagnosis applied to iot systems}.
\newblock \bibinfo{journal}{2019 International Conference on Control, Automation and Diagnosis (ICCAD)} \DOIprefix\doi{10.1109/ICCAD46983.2019.9037949}.
%Type = Article
\bibitem[{McClean et~al.(2016)McClean, Romero, Babbush and Aspuru-Guzik}]{mcclean2016theory}
\bibinfo{author}{McClean, J.R.}, \bibinfo{author}{Romero, J.}, \bibinfo{author}{Babbush, R.}, \bibinfo{author}{Aspuru-Guzik, A.}, \bibinfo{year}{2016}.
\newblock \bibinfo{title}{The theory of variational hybrid quantum--classical algorithms}.
\newblock \bibinfo{journal}{New Journal of Physics} \bibinfo{volume}{18}, \bibinfo{pages}{023023}.
\newblock \URLprefix \url{https://doi.org/10.1088/1367-2630/18/2/023023}, \DOIprefix\doi{10.1088/1367-2630/18/2/023023}.
%Type = Article
\bibitem[{Mohanty et~al.(2025)Mohanty, Behera, Ferrie and Dash}]{Mohanty2025QuantumSMOTE}
\bibinfo{author}{Mohanty, N.}, \bibinfo{author}{Behera, B.K.}, \bibinfo{author}{Ferrie, C.}, \bibinfo{author}{Dash, P.}, \bibinfo{year}{2025}.
\newblock \bibinfo{title}{A quantum approach to synthetic minority oversampling technique (smote)}.
\newblock \bibinfo{journal}{Quantum Machine Intelligence} \bibinfo{volume}{7}, \bibinfo{pages}{38}.
\newblock \URLprefix \url{https://link.springer.com/article/10.1007/s42484-025-00248-6}, \DOIprefix\doi{10.1007/s42484-025-00248-6}.
%Type = Article
\bibitem[{Nielsen and Chuang(2002)}]{nielsen2002quantum}
\bibinfo{author}{Nielsen, M.A.}, \bibinfo{author}{Chuang, I.L.}, \bibinfo{year}{2002}.
\newblock \bibinfo{title}{Quantum computation and quantum information} \DOIprefix\doi{10.1017/CBO9780511976667}.
%Type = Article
\bibitem[{Preskill(2018)}]{preskill2018quantum}
\bibinfo{author}{Preskill, J.}, \bibinfo{year}{2018}.
\newblock \bibinfo{title}{Quantum {C}omputing in the {NISQ} era and beyond}.
\newblock \bibinfo{journal}{{Quantum}} \bibinfo{volume}{2}, \bibinfo{pages}{79}.
\newblock \URLprefix \url{https://doi.org/10.22331/q-2018-08-06-79}, \DOIprefix\doi{10.22331/q-2018-08-06-79}.
%Type = Misc
\bibitem[{Programmer3()}]{industrialfault2024}
\bibinfo{author}{Programmer3}, .
\newblock \bibinfo{title}{Industrial fault detection dataset}.
\newblock \bibinfo{howpublished}{\url{https://www.kaggle.com/datasets/programmer3/industrial-fault-detection-dataset}}.
\newblock \bibinfo{note}{Kaggle dataset. Accessed: 2025-10-01}.
%Type = Misc
\bibitem[{Rish(2001)}]{rish2001empirical}
\bibinfo{author}{Rish, I.}, \bibinfo{year}{2001}.
\newblock \bibinfo{title}{An empirical study of the naive bayes classifier}.
\newblock \bibinfo{note}{Technical Report, IBM Research}.
%Type = Article
\bibitem[{Satpathy et~al.(2023)Satpathy, Vibhu, Behera, Al-Kuwari, Mumtaz and Farouk}]{satpathy2023analysis}
\bibinfo{author}{Satpathy, S.K.}, \bibinfo{author}{Vibhu, V.}, \bibinfo{author}{Behera, B.K.}, \bibinfo{author}{Al-Kuwari, S.}, \bibinfo{author}{Mumtaz, S.}, \bibinfo{author}{Farouk, A.}, \bibinfo{year}{2023}.
\newblock \bibinfo{title}{Analysis of quantum machine learning algorithms in noisy channels for classification tasks in the iot extreme environment}.
\newblock \bibinfo{journal}{IEEE Internet of Things Journal} \bibinfo{volume}{11}, \bibinfo{pages}{3840 -- 3852}.
\newblock \DOIprefix\doi{10.1109/JIOT.2023.3300577}.
%Type = Article
\bibitem[{Schuld and Killoran(2019)}]{schuld2019quantum}
\bibinfo{author}{Schuld, M.}, \bibinfo{author}{Killoran, N.}, \bibinfo{year}{2019}.
\newblock \bibinfo{title}{Quantum machine learning in feature hilbert spaces}.
\newblock \bibinfo{journal}{Physical Review Letters} \bibinfo{volume}{122}, \bibinfo{pages}{040504}.
\newblock \DOIprefix\doi{10.1103/PhysRevLett.122.040504}.
%Type = Article
\bibitem[{Schuld et~al.(2015)Schuld, Sinayskiy and Petruccione}]{schuld2015introduction}
\bibinfo{author}{Schuld, M.}, \bibinfo{author}{Sinayskiy, I.}, \bibinfo{author}{Petruccione, F.}, \bibinfo{year}{2015}.
\newblock \bibinfo{title}{An introduction to quantum machine learning}.
\newblock \bibinfo{journal}{Contemporary Physics} \bibinfo{volume}{56}, \bibinfo{pages}{172--185}.
\newblock \DOIprefix\doi{10.1080/00107514.2014.964942}.
%Type = Article
\bibitem[{Song et~al.(2026)Song, Sun, Wang, Song, Wang and Jiang}]{wang2016noise}
\bibinfo{author}{Song, Q.}, \bibinfo{author}{Sun, S.}, \bibinfo{author}{Wang, B.}, \bibinfo{author}{Song, Q.}, \bibinfo{author}{Wang, T.}, \bibinfo{author}{Jiang, H.}, \bibinfo{year}{2026}.
\newblock \bibinfo{title}{Noise-robust fault diagnosis network based on multiscale feature enhancement and dynamic cross-modal interaction}.
\newblock \bibinfo{journal}{Measurement} \bibinfo{volume}{257}, \bibinfo{pages}{118564}.
\newblock \URLprefix \url{https://doi.org/10.1016/j.measurement.2025.118564}, \DOIprefix\doi{10.1016/j.measurement.2025.118564}.
%Type = Inproceedings
\bibitem[{Szegedy et~al.(2014)Szegedy, Zaremba, Sutskever et~al.}]{szegedy2014intriguing}
\bibinfo{author}{Szegedy, C.}, \bibinfo{author}{Zaremba, W.}, \bibinfo{author}{Sutskever, I.}, et~al., \bibinfo{year}{2014}.
\newblock \bibinfo{title}{Intriguing properties of neural networks}, in: \bibinfo{booktitle}{arXiv:1312.6199}.
\newblock \DOIprefix\doi{10.48550/arXiv.1312.6199}.
%Type = Article
\bibitem[{Vapnik(1999)}]{vapnik1998statistical}
\bibinfo{author}{Vapnik, V.N.}, \bibinfo{year}{1999}.
\newblock \bibinfo{title}{An overview of statistical learning theory}.
\newblock \bibinfo{journal}{IEEE Transactions on Neural Networks} \bibinfo{volume}{10}, \bibinfo{pages}{988--999}.
\newblock \URLprefix \url{https://doi.org/10.1109/72.788640}, \DOIprefix\doi{10.1109/72.788640}.
%Type = Article
\bibitem[{Vashishtha et~al.(2025)Vashishtha, Chauhan, Sehri, Zimroz, Dumond, Kumar and Gupta}]{liu2018fault}
\bibinfo{author}{Vashishtha, G.}, \bibinfo{author}{Chauhan, S.}, \bibinfo{author}{Sehri, M.}, \bibinfo{author}{Zimroz, R.}, \bibinfo{author}{Dumond, P.}, \bibinfo{author}{Kumar, R.}, \bibinfo{author}{Gupta, M.K.}, \bibinfo{year}{2025}.
\newblock \bibinfo{title}{A roadmap to fault diagnosis of industrial machines via machine learning: A brief review}.
\newblock \bibinfo{journal}{Measurement} \bibinfo{volume}{242}, \bibinfo{pages}{116216}.
\newblock \URLprefix \url{https://doi.org/10.1016/j.measurement.2024.116216}, \DOIprefix\doi{10.1016/j.measurement.2024.116216}.
%Type = Article
\bibitem[{Vijayalakshmi et~al.(2024)Vijayalakshmi, Rajakannu, Kamarudden, Ramachandran and Sri~Rajkavin}]{li2017intelligent}
\bibinfo{author}{Vijayalakshmi, K.}, \bibinfo{author}{Rajakannu, A.}, \bibinfo{author}{Kamarudden, M.}, \bibinfo{author}{Ramachandran, K.P.}, \bibinfo{author}{Sri~Rajkavin, A.V.}, \bibinfo{year}{2024}.
\newblock \bibinfo{title}{Intelligent fault diagnosis of rotating machinery using deep learning algorithms: A comparative analysis of mlp, cnn, rnn, and lstm}.
\newblock \bibinfo{journal}{SSRG International Journal of Electrical and Electronics Engineering} \bibinfo{volume}{11}, \bibinfo{pages}{294--315}.
\newblock \URLprefix \url{https://www.internationaljournalssrg.org/IJEEE/2024/Volume11-Issue9/IJEEE-V11I9P127.pdf}, \DOIprefix\doi{10.14445/23488379/IJEEE-V11I9P127}.
%Type = Article
\bibitem[{Wang et~al.(2025)Wang, Wang and Tang}]{zhang2020fault}
\bibinfo{author}{Wang, H.}, \bibinfo{author}{Wang, H.}, \bibinfo{author}{Tang, X.}, \bibinfo{year}{2025}.
\newblock \bibinfo{title}{A review of deep learning in rotating machinery fault diagnosis and its prospects for port applications}.
\newblock \bibinfo{journal}{Applied Sciences} \bibinfo{volume}{15}, \bibinfo{pages}{11303}.
\newblock \URLprefix \url{https://doi.org/10.3390/app152111303}, \DOIprefix\doi{10.3390/app152111303}.
%Type = Article
\bibitem[{Zhao et~al.(2025)Zhao, Wang, Huang and Ma}]{qin2019review}
\bibinfo{author}{Zhao, J.}, \bibinfo{author}{Wang, W.}, \bibinfo{author}{Huang, J.}, \bibinfo{author}{Ma, X.}, \bibinfo{year}{2025}.
\newblock \bibinfo{title}{A comprehensive review of deep learning-based fault diagnosis approaches for rolling bearings: Advancements and challenges}.
\newblock \bibinfo{journal}{AIP Advances} \bibinfo{volume}{15}, \bibinfo{pages}{020702}.
\newblock \URLprefix \url{https://doi.org/10.1063/5.0255451}, \DOIprefix\doi{10.1063/5.0255451}.
%Type = Misc
\bibitem[{{ziya07}()}]{engine_failure_kaggle}
\bibinfo{author}{{ziya07}}, .
\newblock \bibinfo{title}{Engine failure detection dataset}.
\newblock \bibinfo{howpublished}{\url{https://www.kaggle.com/datasets/ziya07/engine-failure-detection-dataset}}.
\newblock \bibinfo{note}{Kaggle dataset. Accessed: 2025-09-30}.

\end{thebibliography}

\end{document}